\documentclass{ijmpbstyle}

\usepackage{bbm}
\usepackage{pstricks,pst-plot,pst-node}
\psset{unit=1cm,linewidth=1pt,arrowscale=1.5}
\newrgbcolor{purple}{0.8 0 1}
\newrgbcolor{pink}{1 0.6 0.6}

\let\draftnote\relax

\newcommand\be{\begin{equation}}
\newcommand\ee{\end{equation}}

\newcommand\Z{{\mathbb Z}}
\newcommand\R{{\mathbb R}}

\def\vev#1{\left\langle#1\right\rangle}

\newcommand\tr{\mathop{\rm tr}\nolimits}
\newcommand\E{{\cal E}}

\def\CF{\mathcal{F}}

\def\bx{{\bf x}}

\begin{document}

\markboth{V.~Kaplunovsky, D.~Melnikov, J.~Sonnenschein}
{Holographic Baryons and Instanton Crystals}

%
\catchline{}{}{}{}{}
%

\title{HOLOGRAPHIC BARYONS AND INSTANTON CRYSTALS\footnote{Prepared for the special issue \emph{Skyrmion} of the IJMPB, ed. by M.~Rho and I.~Zahed.}  }

\author{VADIM KAPLUNOVSKY}

\address{Physics Theory Group and Texas Cosmology Center\\
University of Texas, Austin, TX 78712, USA\\
vadim@physics.utexas.edu }

\author{DMITRY MELNIKOV\footnote{also at the Institute for Theoretical and Experimental Physics, B.~Cheremushkinskaya 25, 117218 Moscow, Russia.}}

\address{International Institute of Physics, Federal University of Rio Grande do Norte,\\
Av. Odilon Gomes de Lima 1722, Natal, RN 59078-400, Brazil\\
dmitry@iip.ufrn.br}

\author{JACOB SONNENSCHEIN}

\address{The Raymond and Beverly Sackler School of Physics and Astronomy,\\
Tel Aviv University, Ramat Aviv 69978, Israel\\
cobi@post.tau.ac.il}

\maketitle


\begin{abstract}
In a wide class of holographic models, like the one proposed by Sakai and Sugimoto, baryons can be approximated by instantons of non-abelian gauge fields that live on the world--volume of flavor D--branes. In the leading order, those are just the Yang-Mills instantons, whose solutions can be constructed from the celebrated ADHM construction. This fact can be used to study various properties of baryons in the holographic limit. In particular, one can attempt to construct a holographic description of the cold dense nuclear matter phase of baryons. It can be argued that holographic baryons in such a regime are necessarily in a solid crystalline phase. In this review we summarize the known results on the construction and phases of crystals of the holographic baryons.
\end{abstract}


\section{Introduction}

The 1980's witnessed a revival of interest in the Skyrme model of baryons.\cite{Skyrme} At the time there was a growing evidence that the correct low-energy description of baryons is that of solitons of meson fields.\cite{Witten:1979kh} The chiral Lagrangian, unknown at the time of the Skyrme's original work, provided a natural framework for the realization of this idea.\cite{Balachandran:1982dw} Starting with the work of Adkins, Nappi and Witten\cite{Adkins:1983ya}, many static properties of baryons were computed using the Skyrme model. The approach produced many new exciting results (for reviews see \emph{e.g.} Refs.~\refcite{Balachandran:1985pa}--\refcite{Brown:2010}), though remained far from being a systematic accurate method to describe baryons.

A branch of applications of the Skyrme model, which partially motivated our work and the current review, emerged in the analysis of the properties of cold dense baryonic matter. In 1985 Klebanov\cite{Klebanov:1985qi} analyzed skyrmions arranged in a three--dimensional simple cubic lattice and found a ground state of such a system. However, the true ground state of skyrmions at finite density, is not the simple cubic, but rather the face--centered cubic lattice. This was later found by Kugler and Shtrikman.\cite{KuglerShtrikman,Kugler:1989uc} Even before that it was noticed by Goldhaber and Manton\cite{Goldhaber:1987pb}, that at large density --- essentially when the skyrmions start to overlap --- skyrmion lattices undergo a transition to configurations with a higher degree of symmetry. In the new state, the appropriate description of the system seemed to be not the one of the skyrmion, but rather of the half--skyrmion lattice. In particular, the elementary Wigner--Seitz cell of the high density configuration contained a half of the skyrmion's topological charge. Kugler and Shtrikman found that the low--energy ground state, the face--centered cubic lattice of skyrmions, at high densities turn into a simple cubic lattice of half--skyrmions.

An important property of the half-skyrmion phase is the partial restoration of chiral symmetry, which occurs when skyrmion fields are averaged over an elementary cell. Such behavior can be compared with baryons at finite density. The $N_c=3$ QCD at the chemical potential slightly larger than the baryon mass is believed to be in a deconfined color superconductor phase with the chiral symmetry restored. For large $N_c$ there is no color superconductivity and no deconfinement, for reasonably large values of the quark chemical potential $\mu_q\gg \Lambda_{\rm QCD}$. In this regime, the phase is some kind of a quark liquid, with quark--like excitations in the interior of the Fermi sea, but with hadronic ones close to the Fermi surface.\cite{McLerran:2007qj} The interior of the Fermi sea is chirally symmetric, but the symmetry is broken close to the surface by the chiral density waves. In other words the chiral symmetry is locked with translational symmetry, exactly like in the case of skyrmions.

Another part of the motivation to study baryonic crystals came from the advances in the holographic correspondence, or simply holography, in the 2000's. Shortly after the original proposal of the AdS/CFT correspondence,\cite{Maldacena:1997re} it was understood how to put baryons on the holographic setting, \emph{e.g.} in Refs.~\refcite{WittenBaryons}--\refcite{Callan:1999zf}. However, the most popular setup to study holographic hadron physics came a bit later, with a proposal of Sakai and Sugimoto in 2004.\cite{Sakai:2004cn} Remarkably, the latter model naturally incorporates in it the original Skyrme model with some modifications. Specifically, in terms of the pion fields, one can present the effective action as the chiral Lagrangian plus the Skyrme term, plus an infinite series of the couplings to vector meson fields.

For a reader less acquainted with the developments in the holographic correspondence, let us list the main milestones as far as applications to hadron physics is concerned. The correspondence provides a very powerful and useful tool to study strongly coupled systems by mapping them to a weakly coupled  string theory, equivalently low--curvature gravity. The most natural physical system in which it could be and was implemented is QCD and hadron physics.

It is well known that the gauge theory dual of the type IIB string theory on the $AdS_5\times S^5$ space is the ${\cal N}=4$ super Yang-Mills (SYM) theory, which is conformal, with the maximal amount of supersymmetries. In order to study purely non-supersymmetric YM theory one has to deform the gauge theory so that all supersymmetries are broken and the system turn into a confining rather than conformal phase. Associated with such a deformation the background string theory has to be deformed accordingly.

A prototype holographic model of this nature is the so called Witten's model,\cite{Witten:1998zw} where one starts with the near horizon limit of the background of large number $N_c$ of D4 branes (instead of D3 branes in the $AdS_5\times S^5$ scenario) and compactifies one space coordinate in such a way that together with the radial direction it has a cigar--like shape. Imposing anti-periodic boundary conditions for the fermions breaks all the supersymmetries so that in the limit of a small compactification radius the geometry is dual to the low--energy regime of the four--dimensional YM theory contaminated with Kaluza--Klein modes.

This background admits a Wilson loop with an area law, namely confining, behavior as well as a gapped glueball spectrum. To incorporate quark degrees of freedom one introduces additional $N_f$ ``flavor" branes, typically in the probe $N_f\ll N_c$ regime. This prescription was originally introduced by Karch and Katz,\cite{Karch:2002sh} and first implemented for a system in the Coulomb phase. It was later considered in a confining background in Ref.~\refcite{Sakai:2003wu}, and finally, Sakai and Sugimoto\cite{Sakai:2004cn} proposed a model that up to date provides the most useful holographic dual model of a system in the same ``universality class" as QCD with $N_f<<N_c$ flavors.

The model in based on extending the Witten's model by placing stacks of $N_f$ D8 and anti-D8 flavor branes that merge with each other at the tip of the cigar, so that the construction exhibits a ``geometrical spontaneous breakdown" of chiral symmetry. This model was later generalized by Aharony \emph{et al.}\cite{Aharony:2006da} by allowing the flavor branes to merge at any point along the radial direction and not necessarily at the tip of the cigar. (See a more detailed discussion below.)

A Wilson line is described in holography by a string that stretches out and ends on the boundary of the holographic background. The endpoints of the string can be viewed as an infinitely heavy quark and anti-quark pair. In this picture a meson takes the form of a string whose endpoints are flavored and of finite mass. In holography this maps into a string that starts and ends on probe flavor branes. In the limit of large string tension and small string coupling, which corresponds in the gauge theory side to the limits of large number of colors $N_c$ and large 't Hooft coupling $\lambda\equiv g_{\rm YM}^2 N_c$, one can describe the degrees of freedom of the meson by those of fields (scalars or vectors) that reside on the flavor branes.\cite{Mintakevich:2008mm} On the other hand a description of realistic mesons, and in particular their Regge spectrum, requires the use of the stringy holographic picture.\cite{Sonnenschein:2014jwa}

What is the holographic dual of a baryon? Since a quark corresponds to an endpoint of a string, a baryon should include a structure to which $N_c$ strings are connected. It was found out that in the context of the $AdS_5\times S^5$ string theory such a construction takes the form of a ``baryonic vertex", which is a D5 brane that wraps the $S^5$ with $N_c$ strings that connect it to the boundary.\cite{WittenBaryons} This in fact corresponds to an external baryon with infinitely heavy quarks. Such constructions of stringy baryons were proposed also in the context of confining backgrounds\cite{Brandhuber:1998xy,Callan:1999zf}.

In a similar manner to the passage from a Wilson line to a meson, to get a dynamical baryon, and not an external one, the strings  have to connect the baryonic vertex to flavor branes and not to the boundary. It was shown in Ref.~\refcite{Seki:2008mu} that in the generalized Sakai-Sugimoto (gSS) model the baryonic vertex is immersed in the flavor brane itself. Recall that in the gSS model the flavor branes are D8 branes that wrap an $S^4$ and hence the baryonic vertex, which in this model is a D4 brane wrapping the same cycle, is a soliton to the five dimensional flavor gauge theory that resides on the flavor branes. It is also an instanton in the Euclidean four dimensions spanned by the three ordinary space directions and the radial holographic direction.\cite{Hata:2007mb}  Thus, in the large $N_c$ and large $\lambda$ limit, baryons can be described as flavor instantons.

Similarly to the Skyrme model, the Sakai-Sugimoto model can be analyzed to extract the static properties of baryons. In certain aspects it may do better then the Skyrme model, but overall its predictive power is comparable.\cite{Hashimoto:2008zw} What is interesting about the holographic realization, is that static baryons can be described by solitons (instantons) of the non-abelian gauge (Yang-Mills) fields. Curiously, similar approximate description was ingenuously anticipated by Atiyah and Manton for skyrmions somewhat a decade before holography.\cite{AtiyahManton}

The main problems one encounters when describing baryons holographically are the following. First, the size of the baryon turns out to be of the order ${1}/{\sqrt{\lambda}}$ and hence in the $\lambda\to\infty$ limit it becomes comparable to the string scale and stringy corrections are not necessarily negligible. Second, there are two different scales, off by a factor $\sim 2$, one has to use to describe the mesonic and baryonic spectra. The latter problem, but not the former, can be overcome when one uses the gSS rather than the original SS model.\cite{Seki:2008mu}

Another severe problem in using the SS model is the fact that the interaction between the baryons of this model is repulsive at any separation distance. For long distances between the baryons the repulsion is due to the fact that the lightest isoscalar vector, whose exchange yields repulsion, is lighter than the lightest scalar that yields attraction. In the near and intermediate zones the interaction between two instantons of the SS model is purely repulsive. By using the gSS instead of the SS model, the severeness of the problem can be reduced. As was shown in Ref.~\refcite{Kaplunovsky:2010eh} in the gSS (and not in the SS) model there is, in addition to the repulsive force, also an attractive one due to an interaction of the instantons with a scalar field that associates with the fluctuation of the embedding, though the ratio of the attractive to repulsive potential can never exceed ${1}/{9}$. Thus in both the SS and the gSS nuclei will not be formed.

In Ref.~\refcite{Dymarsky:2010ci} it was shown that there is another holographic model\cite{Kuperstein:2004yf} with  a dominance of the attraction at long distances, but at the same time, a tiny $\sim 1.7\%$ binding energy. In that model the lightest scalar is in fact a pseudo--Goldstone boson associated with the spontaneous breaking of the scale symmetry. This meson can be made to be lighter, but not much lighter, than the lightest isoscalar vector and hence the emerging system is that of small attraction dominance.\cite{Dymarsky:2010ci} The fact that in the gSS the interaction between the baryons is for any separation distance repulsive will be important, but not crucial for this review, as we shall see in more detail later.

Motivated by the interesting behavior of skyrmions at high density and being equipped with the new methodology of holography, in Refs.~\refcite{Kaplunovsky:2012gb} and~\refcite{Kaplunovsky:2013iza} the authors of this review decided to look at the problem of the cold nuclear matter from the holographic perspective. Conversely to the real nuclear matter, in the large $N_c$ limit of holography the latter is always a solid (crystal).\cite{Kaplunovsky:2010eh} Thus, it appeared interesting to look at the realization of the crystal of baryons in the instanton description of the Sakai--Sugimoto--like setup. Moreover, related studies appeared in papers~\refcite{Kim:2006gp}--\refcite{Rho:2009ym}. In particular, Refs.~\refcite{Bergman} and~\refcite{Rozali:2007rx} have analyzed the phase diagram of configurations with finite baryon density, approximating the density of instantons by a mean value. Refs.~\refcite{Kim:2007vd} and~\refcite{Rho:2009ym} searched for the manifestation of the half-skyrmion transition in a holographic setting.  The research program, to be described here, of studying the instanton crystals as holographic duals of the cold nuclear matter  stemmed from these and related results.

In brief, the program assumes construction of instanton solutions that may approximate the crystals of the holographic baryons. We use the Sakai-Sugimoto model as a prototypical example, but any other model where the baryons are realized as instantons would be just as good for our purposes, for example the model of Ref.~\refcite{DKS}, for some seven--brane geometries,\cite{Dymarsky:2010ci} or the $\rm AdS_5\times S^1$ model of Ref.~\refcite{Kuperstein:2004yf} (the baryons of that model are studied in Sec.~6 of Ref.~\refcite{Seki:2008mu}). For this large class of models one can introduce a general energy functional that will need to be minimized over various instanton configurations to find the true ground state for a given set of external parameters.

Holographic study implies certain limits for the parameters of the underlying gauge theory, such as large number of colors $N_c$  and large 't~Hooft coupling $\lambda$. On one hand this simplifies the problem and allows to use perturbation theory in the construction of instanton solutions. One the other hand it introduces new problems, such as additional suppression of the baryon radii $r\sim1/M_{\rho}\sqrt{\lambda}$, which is problematic not only for the phenomenology of the model, but also for its general consistency. In the leading order (LO) of the perturbation theory the solutions are just the well--known instantons of the $SU(N)$ Yang-Mills. However all these instantons are LO degenerate and one has to minimize the energy in the next--to--leading order (NLO), selecting from a large moduli space of the LO solutions. The problem could be simplified by appropriately reducing the moduli space, but the experience shows that the our naive intuition does not always work well in selecting the appropriate configurations.

The review is supposed to summarize our findings and expectations in the realization of this program. We will assume that the cold nuclear matter phase of the holographic baryons exist and is realized by an appropriate configuration of the Yang-Mills instantons. We will discuss examples of instanton configurations and try to select an appropriate ones to describe the ground states.

The review is organized as follows. In section~\ref{sec:hQCD} we review how baryons can be realized in the Sakai-Sugimoto and a class of similar holographic models. In section~\ref{sec:SSmodel} we review the model of Sakai--Sugimoto, while in section~\ref{sec:SSBaryons} we discuss baryons, their properties and further adaptation and generalization of the model, suitable for the subsequent study.

In section~\ref{sec:densebaryons} we discuss the consequences of the holographic limit (large number of colors and large 't~Hooft coupling) for the physics of hadrons at finite density. In subsections~\ref{sec:LargeNQCD} and~\ref{sec:Largelambda} we consider the general implications of these limits. In particular, we explain why we expect to find baryon crystals at large $N_c$ and outline the main features of such a phase. In section~\ref{sec:multibaryons} we review what the holographic limits mean specifically in the context of the Sakai-Sugimoto and similar models. We outline the main steps needed for the instanton description of the baryon crystal. In section~\ref{sec:pointcharge} we illustrate our expectations in a simplest example using the lattice of point charges. This example summarizes the behavior of the holographic lattices, when the density is increased, and demonstrates what we believe is the onset of the transition to the quark liquid phase.

In section~\ref{sec:exactsolns}, we review available exact solutions for the instanton lattices. These are only known for the (quasi--) one--dimensional configurations, a straight periodic chain and a (abelian) zigzag. Naively the zigzag is the first configuration to which the straight chain can break, when the pressure is increased. However, only the straight chain can be realized as a ground state in our setup. For the efficient study of more complicated configurations it seems necessary to rely on approximations. This is implemented in section~\ref{sec:secondpaper}.

In section~\ref{sec:twobody} the two-body force approximation is discussed. This approximation can be used in our setup and it allows numerical simulations to be used for the search of available energy minima. It is employed in section~\ref{sec:phasediagram} where the possible quasi--one--dimensional configurations are considered. It is assumed that at densities not too high possible configurations are either linear arrays (straight chains) of instantons, or zigzag--like configurations. Section~\ref{sec:newresults} summarizes the results of a more recent study in progress of the two--dimensional lattices, both infinite and finite size. It demonstrates that if the zigzag assumption of Sec.~\ref{sec:phasediagram} is relaxed, there exist other configurations that can be favored over the (abelian) zigzag.

Section~\ref{sec:summary} summarizes the main results and problems.


\section{Holographic QCD}
\label{sec:hQCD}

\subsection{Sakai-Sugimoto Model}
\label{sec:SSmodel}

Some gauge theories, \emph{e.g.} ${\cal N}=4$ supersymmetric Yang-Mills (SYM) theory, have exact holographic duals, where both sides of the duality follow as IR limits of the same string-theoretical construction, while all the undesirable degrees of freedom are superheavy. This is unfortunately not the case of Quantum Chromo Dynamics (QCD), not even of its large $N_c$ limit: it either does not have an exact holographic limit, or we have not found it yet. Instead, there is a large number of ``holographic QCD'' (hQCD) models, which are rather dual to some QCD-like theories with a lot of extra stuff that the real QCD does not have. Our hope is that such models are good enough to get qualitative understanding, though they cannot probably be too accurate to make useful numerical predictions.

One of the most popular hQCD models on the market is the Sakai-Sugimoto model\cite{Sakai:2004cn}, which we are going to review momentarily. The predictions of this model are at best qualitative, although it can successfully compete with conventional hadron physics models like Skyrme model as far as certain quantitative predictions are concerned.

The construction of the Sakai-Sugimoto model starts with $N_c$ coincident D4 branes, spanning the Minkowski space times a circle of radius $R$ with antiperiodic boundary conditions for the fermions. Such a configuration breaks the ${\cal N}=4$ supersymmetry down to ${\cal N}=0^*$. For weak 't~Hooft coupling $\lambda=g_{\rm YM}^2N_c\sim g_sN_c \ll1$ (field theory limit, $g_s$ --- string coupling), the open strings between the branes give rise to the gluons of the pure $U(N_c)$ Yang-Mills (YM) theory. However in the holographic limit $\lambda\gg1$ everything happens right at the Kaluza--Klein scale $\Lambda_{\rm QCD}\sim M_{\rm KK}\equiv 1/R$, so the YM glueballs end up with similar $O(M_{\rm KK})$ masses to a lot of non-YM stuff. This scenario is known as the Witten's model.\cite{Witten:1998zw}

On the gravity side of the duality, at $\lambda\gg 1$, the D4 branes merge into a black brane which warps the 10D metric. All we see outside the horizon is a warped space-time geometry and the Ramond--Ramond flux induced by the conserved charge of the D-branes. Specifically we have a warped product of $\R^{3,1}$ Minkowski space, the $S^4$ sphere (originally surrounding the D4 branes), and a two-dimensional cigar spanning the radial direction perpendicular to the branes and the $S^1$ circle. The radial coordinate, here denoted $u$, runs from $u_\Lambda>0$ to infinity. At $u_\Lambda$ the $S^1$ shrinks to a point, hence the cigar.

Altogether, we have warped metric, the four--form flux, and the running dilaton according to
\begin{align}
ds^2\ &
=\ \left( \frac{u}{R_{D4}}\right)^{3/2}
\Bigl[-d t^2+\delta_{ij}d x^i d x^j+f(u)d x_4^2\Bigr]\,
+\,\left( \frac{R_{D4}}{u}\right)^{3/2}
\left[\frac{d u^2}{f(u)}+u^2d\Omega_4^2\right],\nonumber\\[7pt]
F_4\ &
=\ 3\pi\ell_s^3 N_c\times\mbox{vol}(S^4)\,,\qquad
e^{\phi}\ =\ g_s\left( \frac{u}{R_{D4}}\right)^{3/4},
\label{D4background}
\end{align}
where $x_4$ is the coordinate along the $S^1$ or the polar angle on the cigar,
\be
R_{D4}^3\ =\ \pi g_s\ell_s^3 N_c\,,\qquad
f(u)\ =\ 1\,-\,\left( \frac{u_{\Lambda}}{u}\right)^3,
\ee
$\ell_s=\sqrt{\alpha'}$ is the string length scale. The $u_\Lambda$ --- the minimal value of the radial coordinate $u$ at the tip of the cigar --- is related to the original radius $R$ of the $S^1$ circle as
\be
\label{KKscale}
2\pi R\ =\ \frac{4\pi}{3}\left( \frac{R_{D4}^3}{u_{\Lambda}}\right)^{1/2}.
\ee
The same radius $R$ also controls the 4D Yang--Mills coupling and hence the 't~Hooft's
coupling $\lambda$.
Analytically continuing from $\lambda\ll1$ to $\lambda\gg1$, we have
\be
\lambda\ =\ g^2_{\rm 4D} N_c\
=\ \frac{g^2_{5D}}{2\pi R}\, N_c\
=\ \frac{2\pi g_s\ell_s}{R}\, N_c\,.
\label{'tHooftconst}
\ee

To add the flavor degrees of freedom to the model, Sakai and Sugimoto added $N_f$ D8 and $N_f$ anti--D8 (flavor) branes. The D8 (anti--D8) branes span all space coordinates except the $x_4$ coordinate along the $S^1$ circle. At weak $\lambda$, the open strings connecting the color branes to the flavor branes give rise to the quarks and the anti-quarks. The massless quarks are localized at the intersections of D8 and D4 branes; likewise, at the intersections of the anti-D8 branes and the D4 branes we get massless antiquarks. The open strings between the flavor branes yield $N_f^2$ vector and scalar fields living on those branes. The four-dimensional modes of these vector and scalar fields are dual to the QCD \emph{mesons.} In the meantime the YM instantons of the vector fields are dual to the QCD \emph{baryons}, see Sec.~\ref{sec:SSBaryons} for more details.

On the holographic side of the duality ($N_c\to\infty$, $\lambda\to\infty$) the exact solution for the flavor branes interacting with the warped metric and fluxes is not known, but for $N_f\ll N_c$ and $g_s N_f\ll1$ we may use the probe approximation: the flavor branes seek the lowest-action configuration in the background metric~\eqref{D4background}, while their back-reaction upon the metric is neglected. The flavor branes remains D-branes, rather than merge into a black brane of their own.

\begin{figure}[bt]
\centerline{\psfig{file=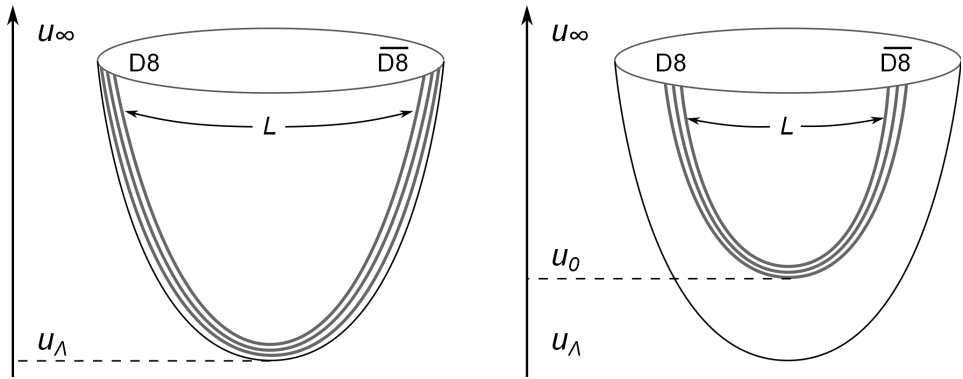,width=\linewidth}}
\vspace*{8pt}
\caption{The figure on the right is the generalized  non-antipodal configuration. The figure on the left describes the limiting antipodal case $L=\pi R$, where the branes connect at $u_0=u_{\Lambda}$.}
\label{Ushape}
\end{figure}

Consequently, at low temperatures (below the deconfinement transition)\footnote{ At higher temperatures --- above the deconfining transition --- the background metric has different topology, and the flavor branes also have different shapes, see Ref.~\refcite{Aharony:2006da} for details.}, the flavor branes span a product of the Minkowski space, four-sphere $S^4$ and a one-dimensional curve on the cigar; the exact shape of this curve follows from minimizing the branes' action, but its topology follows from the cigar itself: since the  D8 and the anti-D8 branes cannot continue all the way to the origin of the geometry $u=0$, they must reconnect to each other and form a U-shaped configuration as shown on figure~\ref{Ushape}.

The reconnection is the geometric realization of the spontaneous chiral symmetry breaking: the separate stacks of $N_f$ D8 and $N_f$ anti-D8 probe branes give rise to the $U(N_f)_L\times U(N_f)_R$ gauge symmetry, which corresponds to the $U(N_f)_L\times U(N_f)_R$ global chiral symmetry in the dual four-dimensional theory. But when the D8 and anti-D8 reconnect, only a single stack of $N_f$ U-shaped branes remains and hence only one unbroken $U(N_f)$ symmetry. Thus, the chiral symmetry is spontaneously broken,
\be
U(N_f)_L\times U(N_f)_R\to U(N_f).
\ee

As shown on figure~\ref{Ushape}, the U-shaped profiles of reconnected branes depend on an additional parameter --- the asymptotic separation $L$ of the D8 and the anti-D8 branes along the $S^1$ circle for $u\to\infty$. For $L=\pi R$ the branes form the \emph{antipodal configuration} in which the branes remain at opposite points on the circle for all $u$: this is the original configuration of Sakai and Sugimoto. In the more general version of the model\cite{Aharony:2006da} we allow for the $L<\pi R$ \emph{non-antipodal configurations.} In such configurations, the distance between the branes in the $x_4$ direction depends on $u$ --- it becomes smaller for smaller $u$ --- and eventually the branes reconnect at $u_0$ before they reach the bottom of the cigar. The $\zeta=u_0/u_\Lambda$ ratio may be used to parameterize the non-antipodal configurations instead of the $L/R$.

The $\zeta$ or $L/R$ parameter of the Sakai--Sugimoto model does not correspond to any known adjustable parameters of the real-life QCD. Unfortunately, this parameter affects many physical properties of the model. For example, for $(L/R)>0.97$ the deconfinement and the restoration of chiral symmetry happen at the same temperature, but for $(L/R)<0.97$ they happen at different temperatures and the model has an intermediate deconfined but chirally broken phase\cite{Aharony:2006da}. Also, in the antipodal model the central nuclear forces are purely repulsive, while the non-antipodal models give rise to both repulsive and attractive nuclear forces\cite{Kaplunovsky:2010eh}, though the net force remains repulsive at all distances.

The low-energy dynamics of the flavor degrees of freedom living on the D8 branes is governed by the effective action comprised of the Dirac--Born--Infeld (DBI) and Chern--Simons (CS) terms,
\be
S=S_{\rm DBI} + S_{\rm CS}\,.
\ee
The DBI action is
\be
\label{DBIaction}
S_{\rm DBI}\ =\ - T_8\!\int\limits_{D8+\overline{D8}}\! d^9x\, e^{-\phi}
\mathop{\rm Str}\nolimits\left(\sqrt{-\det(g_{mn} + 2\pi\alpha' \CF_{mn})}\right),
\ee
where $T_8 = (2\pi)^{-8}\ell_s^{-9}$ is the D8-brane tension, $g_{mn}$ is the nine-dimensional induced metric on the branes, $\mathcal{F}_{mn}$ is the $U(N_f)$ gauge field strength, and Str denotes the symmetrized trace over the flavor indices. The lowercase arabic letters $m$, $n$ denote the nine coordinates of the D8 branes' world-volume.

In the limit of fixed brane geometry and weak gauge fields, the DBI action reduces to that of Yang--Mills,
\be
S_{\rm DBI}[{\cal F}]\ =\ {\rm const}\ +\ S_{\rm YM}[{\cal F}]\
+\ O({\cal F}^4).
\ee
Furthermore, the low-energy field modes we are interested in are constant along the sphere $S^4$ and only contain the components of vector fields  perpendicular to $S^4$. Thus we are going to dimensionally reduce the flavor gauge theory down to five dimensions: the four Minkowski dimensions $x^{0,1,2,3}$, plus one coordinate $z$ along the U-shaped line on the cigar. We find it convenient to choose a particular $z$ coordinate that makes the five-dimensional metric conformal
\be
ds^2\ =\ \left(\frac{u(z)}{R_{D4}}\right)^{3/2}\bigl( -dt^2\,+\,d\bx^2\,+\,dz^2\bigr)\,.
\ee

In the $x^M=(x^0,x^1,x^2,x^3,z)$ coordinates, the five-dimensional YM action for the flavor gauge fields becomes
\be
S_{\rm YM}\ \approx \int\!\!d^4x\!\int\!\!dz\,\frac{1}{2g^2_{\rm YM}(z)}\,\tr\bigl({\cal F}_{MN}^2\bigr)\,,
\label{actionz0}
\ee
where
\be
\label{5Dcoupling}
\frac{1}{2g^2_{\rm YM}(z)}\
=\ \frac{N_c\lambda M_{\rm KK}}{216\pi^3}\times\frac{u(z)}{u_\Lambda}\,.
\ee

Near the bottom of the U-shaped flavor branes one finds
\be
\frac{1}{2g^2_{\rm YM}(z)}\ =\ \frac{N_c\lambda M_{\rm KK}}{216\pi^3}\left( \zeta\,+\,\frac{8\zeta^3-5}{9\zeta}\,M_{\rm KK}^2 z^2\, +\,O(M_{\rm KK}^4 z^4)\right).
\label{Uexpansion}
\ee

The Chern--Simons term arises from the coupling of the gauge fields on the D8 brane to the bulk Ramond--Ramond field. In nine dimensions
\be
\label{CSterm0}
S_{\rm CS} \ =\ T_8\!\!\int\limits_{\rm D8+\overline{D8}} \!\! C_3\wedge \tr e^{2\pi\alpha' \CF}\,,
\qquad \text{where}\qquad F_4=d C_3.
\ee
After integrating over the $S^4$ and dimensionally reducing to five dimensions, the CS term becomes
\be
\label{CSterm1}
S_{\rm CS}\ =\ \frac{N_c}{24\pi^2} \int_{\rm 5D}\tr\left(\mathcal{A}\mathcal{F}^2\, -\,\frac{i}{2}\,\mathcal{A}^3\mathcal{F}\, -\,\frac{1}{10}\,\mathcal{A}^5\right) .
\ee
In a particularly interesting case of two flavors, it is convenient to separate the $U(2)$ gauge fields ${\cal A}_M$ into their $SU(2)$ components $A_M$ and the $U(1)$ components $\hat A_M$. In terms of these components, the CS action becomes
\be
\label{CSterm2}
S_{\rm CS}\ =\ \frac{N_c}{16\pi^2}\int\! \hat A\wedge {\rm tr} F^2\
+\ \frac{N_c}{96\pi^2}\int\! \hat{A}\wedge \hat{F}^2\,.
\ee

We shall see in a moment that the baryons and the multi-baryon systems have strong self-dual $SU(2)$ magnetic fields $F_{\mu\nu}$.
\footnote{%
	In our notations, the space-time indices $0,1,2,3,z$ of the effective five-dimensional theory are labeled
     $M,N,\ldots=0,1,2,3,z$ while the space indices $1,2,3,z$ of the same theory are labeled $\mu,\nu,\ldots$
     When we need the 9D indices for the whole D8 brane, we use $m,n,\ldots$
     }
Thanks to the first term in this CS action, the instanton number density,
\be
\label{idensity}
I(x,z)\ =\ \frac{1}{32\pi^2}\,\epsilon^{\kappa\lambda\mu\nu} F^a_{\kappa\lambda}F^a_{\mu\nu}\,,
\ee
acts as electric charge density for the abelian field $\hat A_0$; the net electric charge of an instanton is $Q_{\rm el}=N_c/2$.

Besides the $U(N_f)$ gauge fields, the effective low-energy five-dimensional theory also contains the scalar fields $\Phi^a(x,z)$ describing the small fluctuations of the D8 branes in the transverse directions. For $N_f$ branes, the scalars form the adjoint multiplet of the $U(N_f)$ gauge symmetry. The action for the scalar fields follows from the DBI action for the induced metric $g_{mn}$ of the fluctuating branes. For the $\Phi(x,z)$ fields normalized to have similar kinetic energies to the vector fields, the scalar action looks like
\begin{align}
S_{\rm scalar}\ &
=\int\!\!d^4x\!\int\!\!dz\,\frac{1}{2g^2_{\rm YM}(z)}\,\tr\Bigl(
    (D_M\Phi)^2\,+\,V(\Phi)\Bigr)
\nonumber\\
&\quad+\ \frac{N_c}{16\pi^2}\int\!\!d^4x\!\int\!\!dz\ C(z)\times \tr\bigl(\Phi{\cal F}_{MN}{\cal F}^{MN}\bigr)\ +\ \cdots.
\label{ScalarAction}
\end{align}
The details of the scalar potential $V(\Phi)=m^2(z)\Phi^2+a(z)\Phi^4+\cdots$ need not concern us here, what is important is the second term describing the backreaction of the gauge fields on the brane geometry. In the antipodal Sakai--Sugimoto model $C(z)=0$ and there is no backreaction because of a geometric symmetry, but in the non-antipodal models $C(z)\neq0$ and the scalar fields $\Phi$, induced by the vector fields of the baryons, lead to  attractive nuclear forces\cite{Kaplunovsky:2010eh}. The ratio of these attractive forces $F_a$ to the repulsive forces $F_r$ mediated by the abelian electric fields depends of the $C(z)$ profile of the interaction term~(\ref{ScalarAction}).
For the Sakai--Sugimoto models
\be
\frac{F_a}{F_r}\
=\ C^2(z)\ =\ \frac{1-\zeta^{-3}}{9}\left(\frac{u_0}{u(z)}\right)^8\ \le\ \frac19\ <\ 1,
\label{ScalarToVector}
\ee
so the net force is always repulsive.

To see how that works, let us focus on the baryons in the Sakai--Sugimoto and other models of the holographic QCD.


\subsection{Baryons in holographic QCD models}
\label{sec:SSBaryons}

In the old hadronic string, the baryons were made out of Y-shaped configurations of three open strings connected to each other at one end;  the other end of each string was connected to a quark. To realize this picture in holographic QCD, we need a \emph{baryon vertex} (BV) --- some object connected to $N_c$ open strings. The other ends of the strings must be connected to the flavor branes and act like the quarks; this would give the baryon its flavor quantum numbers. Witten had constructed the baryon vertex for the ${\rm AdS}_5\times S^5$ model from a D5 brane wrapping the five-sphere\cite{WittenBaryons}; the generalized versions of this construction in Refs.~\refcite{Brandhuber:1998xy} and~\refcite{Callan:1999zf} use ${\rm D}p$ branes wrapping compact cycles carrying $O(N_c)$ Ramond--Ramond fluxes.

In the Sakai--Sugimoto version of this construction, the baryon vertex is realized as a D4 brane wrapped around the $S^4$ sphere (but localized in all other dimensions except the time). The $S^4$ carries $N_c$ units of the $F_4$ Ramond--Ramond flux,
\be
\frac{1}{(2\pi)^3l_s^3}\int_{S^4} F_4 = N_c\,,
\ee
so the Chern--Simons coupling of this flux to the $U(1)$ gauge field $\mathcal{B}$ living on the D4 brane acts a $N_c$ units of the net electric charge for the component $\mathcal{B}_0$:
\be
T_4\int_{\rm D4} C_3\wedge e^{2\pi\alpha'd \mathcal{B}} = N_c \int\mathcal{B}_0\,dx^0\,.
\label{D4BraneCharge}
\ee
In a compact space like $S^4$, the net electric charge must vanish. To cancel the charge~\eqref{D4BraneCharge} we need to connect the D4 brane to open strings. The back end of an oriented open string has electric charge minus one, so we must connect the D4 brane with $N_c$ such strings; their front ends connect to the D8 flavor branes (since the strings do not have any other place to end) and act as $N_c$ quarks.

We may put the D4 brane anywhere in space and anywhere on the cigar. However, the $S^4$ volume, equivalently mass, increases with the $u$ coordinate, so the lowest-energy location of the D4 is the cigar's tip $u=u_\Lambda$. At other locations, the brane feels a gravity-like force pulling it down to the tip. However, the strings connected to the BV pull it towards the flavor branes; in the non-antipodal models the D8 branes do not reach the cigar's tip, so the strings pull the baryonic vertex up from the tip towards the lowest point $u_0$ of the flavor branes. The competition between the upward and downward forces on the BV determines its ultimate location. In some models, the forces reach equilibrium for the BV hanging on strings below the flavor branes\cite{Dymarsky:2010ci}, while in many other models, including the non-antipodal Sakai--Sugimoto, the string forces win and the D4 sticks to the lowest point $u_0$ of the flavor branes\cite{Seki:2008mu}.

In all such models, the BV is a ${\rm D}p$ brane completely embedded in a stack of ${\rm D}({p+4})$ flavor branes. As explained in Refs.~\refcite{Witten:1994tz}--\refcite{Lee:1997vp} such ${\rm D}p$ branes are equivalent to zero-radius Yang--Mills instantons of the $U(N_f)$ gauge symmetry on the flavor branes, and for $N_f>1$ it may be smoothly
inflated to a finite-radius instanton. In $p+5$ dimensions of the flavor branes, this instanton is a fat ${\rm D}p$ brane wrapping some compact cycle, but once we dimensionally reduce to five dimensions, the instanton becomes a finite-size particle. Thus, in the low-energy effective five-dimensional theory of the holographic QCD, a baryon is realized
as a finite-size instanton of the $U(N_f)$ gauge theory.

In the BV picture, each of the $N_c$ strings connecting the vertex to the flavor branes has electric charge $1/N_f$ under the abelian $U(1)$ subgroup of the $U(N_f)$, so the whole baryon has abelian charge $N_c/N_f$. In the instanton picture, the same electric charge is obtained from the CS coupling between the abelian electric field and the non-abelian magnetic fields of the instanton.
\begin{align}
S_{\rm CS}\ =\ \frac{N_c}{24\pi^2}\int \tr\Bigl(
	{\cal AF}^2\,-\,\frac{i}{2}{\cal A}^3{\cal F}\,-\,\frac{1}{10}{\cal A}^5\Bigr)\ &
\to\ \frac{N_c}{ N_f}\int\hat A\wedge \frac{1}{8\pi^2} \tr\bigl(F\wedge F\bigr)
\nonumber\\
&\to\ \frac{N_c}{N_f}\int d^5x\, \hat A_0(x)\times I(x)\,,
\label{CSterm3}
\end{align}
where $I(x)$ is the instanton number density of the magnetic fields~(\ref{idensity}). For $N_f\ge3$, the CS couplings also endow instantons with non-abelian electric charges.

The like-sign electric charges, both abelian and non-abelian, repel each other. It is the Coulomb repulsion between different parts of the same instanton that prevents it from collapsing to a point-like D-brane. However, the instantons do not grow large because the five-dimensional gauge coupling~(\ref{5Dcoupling}) decreases away from the $z=0$ hyperplane: a large instanton would spread into regions of space where the coupling is weaker, and that would increase the instanton's energy. Instead, the equilibrium radius of the instanton scales like
\be
a_{\rm inst}\ \sim\ \frac{1}{\sqrt{\lambda}M_{\rm KK}}\,.
\label{RBscaling}
\ee

For a holographic model of a baryon, this radius is unrealistically small. Indeed, using the $\rho$ meson's mass as a unit, the real-life baryon radius $R_b\sim 3.4 M^{-1}_\rho$, while in holography $a\ll M_\rho^{-1}\sim M_{\rm KK}^{-1}$. Moreover it raises the question of whether we may adequately describe such a small instanton using the $\rm DBI+CS$ action, or perhaps higher-order stringy corrections need to be included.\footnote{
    In string theory, the  $\rm DBI+CS$ action for the gauge fields on a D-brane
    is exact for \emph{constant} tension fields ${\cal F}_{mn}$, however strong.
    But for the variable tension fields, the DBI action includes all powers of the
    ${\cal F}_{mn}$ but neglects their derivatives $D_k{\cal F}_{mn}$,
    $D_kD_p{\cal F}_{mn}$, \emph{etc.}
    It is not clear what effect (if any) such higher-derivative terms would have on
    a small instanton.
    In a supersymmetric background, the instanton is BPS and its net mass is protected
    against stringy corrections, so the DBI action --- or even the Yang--Mills action ---
    gives the exact value.
    But what happens to small instanton in non-supersymmetric backgrounds is an open question.
    }

On the other hand, assuming the $\rm DBI+CS$ description is valid, the small radius~(\ref{RBscaling}) of the instanton allows for consistent expansion of the instanton related quantities in powers of $1/\lambda$.\cite{Hata:2007mb} In particular, the leading contribution to the instanton's mass $M_I$ is $O(\lambda N_cM_{\rm KK})$ while the corrections due to $z$-dependent gauge coupling and due to Coulomb self-repulsion are both $O(N_cM_{\rm KK})\sim M_I/\lambda$.

To see how that works, consider a static instanton --- a time-independent configuration of $SU(N_f)$ magnetic fields, plus the electric fields induced by the CS couplings, and the scalar fields induced by the $\tr(\Phi FF)$ coupling to the magnetic fields. Since the canonically normalized couplings in five dimensions are $O(\lambda^{-1/2})$, the leading contribution to the instanton's energy comes from a purely magnetic field configuration. In the DBI approximation
\be
E_{\rm DBI}\ =\int\!\!d^3\bx\!\int\!dz\,\frac{1}{2 g_{\rm YM}^2(z)}\,\mathop{\rm Str}\nolimits\left(\sqrt{\det\bigl(K(z)\delta_{\mu\nu}+F_{\mu\nu}\bigl)}\,-\,K^2(z)\right),
\label{DBIactionK}
\ee
where $K(z)=(2\pi\alpha')^{-1}g_{11}(z)\sim1/(\lambda M_{KK}^2\ell_s^4)$. For both $g_{\rm YM}(z)$ and $K(z)$ the dependence on the $z$ coordinate is important only on the distance scale of order$1/M_{\rm KK}$, so for instantons of much smaller size we may start with the approximation of constant $K$ and constant five--dimensional coupling. In this approximation, the DBI energy is minimized by the magnetic fields that are exactly self-dual (with respect to the four-dimensional space of $(x^1,x^2,x^3,z)$); moreover, the DBI energy of an instanton is equal to its Yang--Mills energy:
\be
E_{\rm DBI}({\rm instanton})\ =\ E_{\rm YM}({\rm instanton})\ =\ \frac{8\pi^2}{g^2_{\rm YM}(z=0)}\,,
\ee
regardless of its radius and of the $K$ parameter of the DBI action.

The equilibrium radius of an instanton is determined by its (self-) interaction energy, in the next order in $1/\lambda$ expansion. To this order, we assume the magnetic fields to be exactly self-dual --- which allows us to use the YM action instead of DBI --- but the gauge coupling is $z$--dependent, and we also account for the electric and the scalar fields. For small instantons, we may approximate the five-dimensional gauge coupling as
\be
\frac{8\pi^2}{g_{\rm YM}^2(z)}\ =\ N_c\lambda M_{\rm KK}\left(
	B\,+\,DM_{\rm KK}^2\, z^2\,+\,O(M_{\rm KK}^4z^4)\right)
\ee
for some numerical constants $B$ and $D$. For the Sakai--Sugimoto model $B=\zeta/27\pi$ and $D=(8\zeta^3-5)/9\zeta^2$, while other models may have different values. Consequently, the YM energy of the non-abelian magnetic fields evaluates to
\be
E_{\rm NA}\ =\ \lambda N_c M_{\rm KK}\left(
	B\,+\,DM_{\rm KK}^2 Z^2\,+\,DM_{\rm KK}^2\frac{a^2}{2}\right)
\label{ENAsingle}
\ee
where $a$ is the instanton's radius and $Z$ is is the $z$ coordinate of its center.

The electric potentials $A_0^a$ couple to the $F_{\mu\nu}\tilde F^{\mu\nu}$ products of the magnetic fields while the scalar potentials $\Phi^a$ couple to the $F_{\mu\nu}F^{\mu\nu}$. For the self-dual magnetic fields, both potentials couple to the same source $I(\bx,z) Q^a_{\rm el}$, with $Q^a_{\rm el}$ --- the non-abelian electric charge,  and the only difference is the coupling strength ratio $C(z)$ (\emph{cf.}\ Eqs.~(\ref{ScalarAction}) and (\ref{ScalarToVector})). For small instantons we may neglect the $z$--dependence of this ratio and let $C(z)\approx C(0)\equiv C$, and as long as we do not go very far from the instanton (for distances $r\ll 1/M_{\rm KK}$), we may also neglect the $z$--dependence of the gauge coupling. Consequently, both electric and scalar potentials become the four-dimensional Coulomb potentials
\be
\Phi(\bx,z)\ =\ C\, A_0(\bx,z)\
=\ \frac{2Q_{\rm el}}{B N_c\lambda M_{\rm KK}}\int d^3\bx'\!\int dz'\,
	\frac{I(\bx',z')}{(\bx-\bx')^2+(z-z')^2}\,.
\ee

The electric fields lead to repulsive forces between the charges while the scalar forces lead to the attractive forces. Altogether, the net Coulomb energy amounts to
\begin{align}
E_C\ &
=\ \frac{B}{8\pi^2}\,N_c\lambda M_{\rm KK} \int d^3\bx\,dz\,\tr\bigl( (\nabla A_0)^2\,-\,(\nabla\Phi)^2\bigr)\\
&=\ \frac{(1-C^2)N_c}{B \lambda M}\int d^3\bx\,d^3\bx'\,dz\,dz'\,
	\frac{I(\bx,z)\times I(\bx',z')}{(\bx-\bx')^2+(z-z')^2}\,,\nonumber
\end{align}
and for a single instanton of radius $a$ it  evaluates to
\be
E_C\ =\ \frac{(1-C^2)N_c}{5B\lambda M_{\rm KK} a^2}\,.
\label{ECsingle}
\ee

Let us note that for $C<1$ the electric fields are stronger than the scalar fields and the net Coulomb energy of the instanton is positive --- which makes for a net self-repulsive force that prevents the instanton from shrinking to zero radius. In models with $C>1$ (assuming they exist), the scalar fields would be stronger, the net Coulomb energy would be negative, which means a self-attractive force rather than self-repulsive. In such a model, the instanton would shrink to zero radius and our approximations would not be valid.

Collecting the non-abelian~(\ref{ENAsingle}) and the Coulomb~(\ref{ECsingle}) self-interaction energies together we find the net energy of an instanton to be
\be
E_{\mbox{1-inst}}\ =\ \lambda N_c M_{\rm KK}\left(B\,+\,DM_{\rm KK}^2\,Z^2\,+\,DM_{\rm KK}^2\,\frac{a^2}{2}\,
	+\,\frac{1-C^2}{5B\lambda^2 M_{\rm KK}^2 a^2}\right).
\ee
Indeed, for $a^2\sim1/(\lambda M_{\rm KK}^2)$, both radius-dependent terms here are  $O(1/\lambda)$ corrections to the leading term. Minimizing the net energy, we find the equilibrium value of the instanton radius
\be
a\ =\ \frac{1}{M_{\rm KK}\sqrt{\lambda}}\left(
 \frac{2(1-C^2)}{5BD}\right)^{1/4}\,.
\label{GenRadius}
\ee
The instanton center is in equilibrium at $Z=0$ (the bottom of the U-shaped flavor branes), and the instanton's mass is
\be
M_{\rm 1-inst} =\ N_c M_{\rm KK}\left(
	\lambda B\,+\,\sqrt{\frac{2D(1-C^2)}{5B}}\,+\,O(1/\lambda)\right).
\label{GenMass}
\ee

For the antipodal Sakai--Sugimoto model
\be
a\ =\ \frac{(162\pi/5)^{1/4}}{M_{\rm KK}\sqrt{\lambda}}\quad
{\rm and}\quad M_{\rm 1-inst} =\ N_c M_{\rm KK}\left(
	\frac{\lambda}{27\pi}\,+\,\sqrt{\frac{18\pi}{5}}\,+\,O(1/\lambda)\right),
\ee
while other models should have different $O(1)$ numeric factors.

We may absorb two out of three model-dependent parameters $B$, $C$, $D$ into a redefinition of the $\lambda$ and $M_{\rm KK}$ parameters of the effective five-dimensional theory, for example
\begin{align}
\lambda\ &\to\ \frac{\lambda}{\sqrt{DB^3}}\,,\quad
M_{\rm KK}\ \to M\,=\,M_{\rm KK}\sqrt{\frac{D}{B}}\,,\\
\intertext{thus}
\frac{8\pi^2}{g_{\rm YM}^2(z)}\ &
=\ N_c\lambda M\Bigl(1\,+\,M^2 z^2\,+\,\cdots\Bigr).
\label{HoloGZ}
\end{align}
For \emph{static} instanton or multi-instanton systems we may also get rid of the third model-dependent parameter $C$. Indeed, for the static systems $\Phi(\bx,z)=C A_0(\bx,z)$, so the only effect of the scalar fields is to reduce the net Coulomb force by a constant factor $(1-C^2)$. We may simulate this effect without any scalar fields by using different gauge couplings for the electric and magnetic fields, $g^2_{\rm el}=(1-C^2)g^2_{\rm mag}$, or equivalently by rescaling the time dimension $x^0$ relative to the space dimensions $(x^1,x^2,x^3,z)$,
\be
t\ \to\ \frac{t}{\sqrt{1-C^2}}\quad{\rm but}\quad \bx\ \to\ \bx,\quad z\ \to\ z.
\ee
Consequently, the static instanton's energy becomes
\begin{align}
E_{\rm net}\ &
=\ N_c\lambda M\ +\ N_c\lambda M^3 \int d^3\bx\,dz\,I(\bx,z)\times z^2 \nonumber\\
&\qquad+\ \frac{N_c}{4\lambda M}\int d^3\bx\,dz\,d^3\bx'\,dz'\,
\frac{I(\bx,z)\times I(\bx',z')}{(\bx-\bx')^2+(z-z')^2}\
+\ {\rm subleading}\\
&=\ N_c\lambda M\ +\ N_c\lambda M^3\left(Z^2+\frac{a^2}{2}\right)\
+\ \frac{N_c}{5\lambda M a^2}\ +\ {\rm subleading,}\nonumber
\end{align}
hence in equilibrium $Z=0$ and
\be
\label{equiradius}
a\ =\ \frac{(2/5)^{1/4}}{M\sqrt{\lambda}}\,.
\ee

Besides the radius $a$ and $Z$, the instanton has other moduli --- the $X^{1,2,3}$ coordinates of the center (which corresponds to the baryon's coordinates in three dimensions) and $4N_f-5$ orientation moduli in the $SU(N_f)$ gauge algebra. The net energy is degenerate in these moduli to all orders in $1/\lambda$. For finite $N_c$ --- even if it's very large --- one should quantize the motion of the instanton in those moduli directions. Consequently, a holographic baryon acquires definite spin $J$ and isospin $I$ quantum numbers: for $N_f=2$ the baryons have $I=J=0,1,2,\ldots,\frac{N_c}{2}$ for even $N_c$ or $I=J=\frac12,\frac32,\frac52,\ldots,\frac{N_c}{2}$ for odd $N_c$\cite{Hata:2007mb,Seki:2008mu}, and there are similar spin-to-flavor multiplet relations for $N_f>2$.

However, in multi-baryon systems interactions between the baryons break the rotational and flavor symmetries of individual baryons, and only the overall
$SO(3)$ and $SU(N_f)$ symmetries remain unbroken. In holography, the multi-instanton systems suffer from the same problem: the magnetic fields of multiple instantons interfere with each other, which spoils the degeneracy of the net energy with respect to orientations moduli of the individual instantons. In the large $N_c$ limit, this effect becomes more important than the quantum motion in the moduli space.

Consequently, in this review we stick to classical static instantons with definite classical orientations in space and in $SU(N_f)$. From the quantum point of view, such instantons are superpositions of states with different spins and isospins (or rather $SO(4)$ and $SU(N_f)$ quantum numbers). At this level there is no difference between instantons being bosons or fermions. Since minimizing the classical energy of a classical multi-instanton system with respect to classical positions and orientations of all the instantons is already a very hard problem, going into the details of the quantization of moduli is beyond the scope of the review.


\section{Holographic Baryons at High Densities}
\label{sec:densebaryons}



\subsection{Nuclear matter in the large $N_c$ limit}
\label{sec:LargeNQCD}

\begin{figure}[bt]
\centerline{\psfig{file=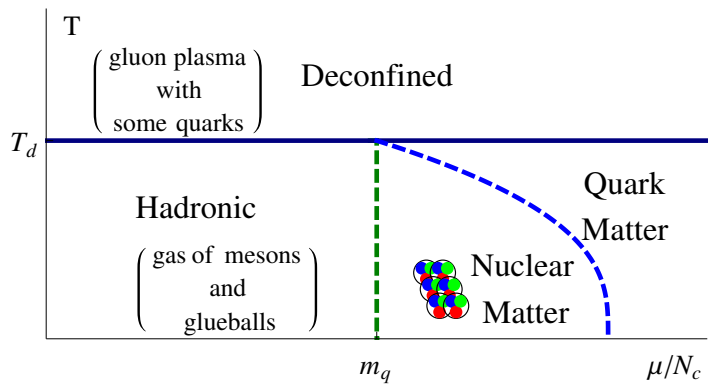,width=3.65in}}
\vspace*{8pt}
\caption{Conjectured phase diagram of large $N_c$ QCD.\protect\cite{McLerran:2007qj}}
\label{figLargeNQCD}
\end{figure}

Let us first focus on the 't Hooft $N_c\to\infty$ limit keeping the number of quark flavors small, $N_f/N_c\ll1$. In such a case QCD perturbation theory is dominated by the planar gluon diagrams while contributions of the non-planar diagrams and of the quark loops are suppressed. In such a regime the $T-\mu$ phase diagram is believed to look like Fig.~\ref{figLargeNQCD}. (For a review see Ref.~\refcite{McLerran:2007qj} and references therein,  also Ref.~\refcite{Blake:2012dp}.)

Since the dynamics of the theory is dominated by the gluons, the quarks are sensitive to the gluonic background, but the backreaction from them to the gluons is suppressed by $N_f/N_c$. Consequently, at lower temperatures there is confinement, but for increasing temperature there is a first order transition to the deconfined phase. The transition temperature $T_d$ is at the $\Lambda_{\rm QCD}$ scale and it is almost independent of the quark chemical potential $\mu_q=\mu_b/N_c$ (as long as $\mu_q$ does not grow with $N_c$).

It is not clear whether for $N_c\to\infty$ the deconfining phase transition coincides with the chiral symmetry restoration for the light quarks. Several field theory arguments, for example in Refs~\refcite{Casher:1979vw}, \refcite{Banks:1979yr} and~\refcite{McLerran:2007qj}, suggest that for $\mu_q=0$ the two transitions should happen at the same point. However, these arguments do not work for $\mu_q>0$,\cite{McLerran:2007qj} and there are other arguments for the existence of confined but chirally restored phases, or deconfined phases where the chiral symmetry remains broken. In particular, some holographic models, \emph{e.g.} in Ref.~\refcite{Aharony:2006da}, have deconfined but chirally broken phases even at zero chemical potential.

For temperatures below the deconfining transition $T_d$ and baryon chemical potentials below the baryon mass, equivalently for $\mu_q\lesssim m_q\equiv M_b/N_c$, the thermal state of the theory is a dilute gas of glueballs and mesons, whose interactions are suppressed by powers of $N_c$, and almost no baryons or antibaryons: the masses of mesons and glueballs are $O(\Lambda_{\rm QCD})$, and they scale as $N_c^0$; on the other hand, the  baryons, which are made out of $N_c$ quarks, have masses of order $N_c\Lambda_{\rm QCD}$, so their relative abundance in thermal equilibrium is exponentially suppressed. The interaction energy between baryons  also scales as $N_c$. Hence at $\mu_q\approx m_q$ there is an abrupt phase transition to the bulk nuclear matter with finite baryon density.

Unlike the ordinary nuclear matter, which is in the quantum liquid state, the large $N_c$ nuclear matter is crystalline solid since the ratio of kinetic energy to potential energy decreases with $N_c$. Indeed, the potential energy of baryon-baryon forces scales like $N_c$; more precisely,\cite{Kaplan:1996rk} in the large $N_c$ limit the two-baryon potential becomes
\begin{multline}
\label{Potential}
V\ \sim \  N_c\times A_C(r) \ + \ N_c\times A_S(r) \left({\bf I}_1\cdot {\bf I}_2\right)\left({\bf J}_1\cdot {\bf J}_2\right) \ +
\\  + \  N_c\times A_T(r) \left({\bf I}_1\cdot {\bf I}_2\right)\left(3  \left({\bf n}\cdot {\bf J}_1\right) \left({\bf n}\cdot {\bf J}_2\right)- \left({\bf J}_1\cdot {\bf J}_2\right)\right) \ + \
 O\left(1/N_c \right),
\end{multline}
for some $N_c$--independent profiles $A_C(r)$, $A_S(r)$, and $A_T(r)$ for the central, spin-spin, and tensor forces; their overall magnitudes are $A\sim\Lambda_{\rm QCD}$ for $r\sim 1/\Lambda_{\rm QCD}$. Classically, this potential tries to organize the baryons into some kind of a crystal, where the distances between neighboring baryons do not depend on the $N_c$, while the binding energy (per baryon) scales like $N_c\Lambda_{\rm QCD}$. Quantum--mechanically, the baryons in such a crystal behave like atoms in ordinary crystals: they oscillate in their potential wells with zero-point kinetic energies
\be
\label{Kinetic}
K\  \sim \ \frac{\pi}{2m_B\,d^2}\ \sim \ \frac{\Lambda_{\rm QCD}}{N_c}\,\frac{1}{d^2}\,,
\ee
where $d\sim 1/\Lambda_{\rm QCD}$ is the $N_c$--independent diameter of the potential well. Therefore, at zero temperature the ratio of kinetic energy to the potential energy scales like
\be
\frac{K}{V}\ \sim \  \frac{1}{N_c^2}
\label{KVratio}
\ee
and becomes very small for large $N_c$.

At higher temperatures the kinetic energies of baryons become larger, $K\sim T$, but in the confined phase we are limited to $T<T_d\sim\Lambda_{\rm QCD}$. Consequently, the kinetic to potential energy ratio scales at most like
\be
\frac{K}{V}\ \sim\ \frac{1}{N_c}\,,
\ee
which is larger than (\ref{KVratio}) but still small in the large $N_c$ limit. Consequently, for large $N_c$ neither zero-point quantum motion, nor thermal motion of baryons can destroy the baryon crystal, so the nuclear matter remains solid all the way to the deconfining temperature.

Before holography, the best models for the large--$N_c$ nuclear crystals were lattices of skyrmions.\cite{Klebanov:1985qi} In this framework, Goldhaber and Manton had found a curious phase transition from a lattice of whole skyrmions at low chemical potential to a denser lattice of half-skyrmions at higher potential.\cite{Goldhaber:1987pb} (According to Refs~\refcite{KuglerShtrikman} and~\refcite{Kugler:1989uc}, the whole-skyrmion lattice at low potentials has face-centered cubic (fcc) arrangement while the half-skyrmion lattice at higher potential is simple cubic.) In the half-skyrmion-lattice phase, the order parameter for the chiral symmetry breaking vanishes after space averaging, so in QCD terms the transition is interpreted as chiral symmetry restoration at high $\mu_q$.

In QCD with $N_c=3$ and two massless flavors there is a similar chirally-symmetric phase at low $T$ and high $\mu_q$. This phase is a quark liquid rather than a baryon liquid (the quarks are no longer confined to individual baryons) and there is a condensate of quark pairs making this liquid a color superconductor. But for large $N_c$ the situation is more complicated: there is no color superconductivity, and there is no deconfinement for $T<T_d$. Instead, the dense cold nuclear matter forms a phase which combines the features of the baryonic and quark phases: the quarks fill up a Fermi sea, but the interactions near the Fermi surface are strong, so the excitations are not free quarks or holes but rather meson--like quark--hole pairs or baryon-like states of $N_c$ quarks.

For $\mu_q\gg\Lambda_{\rm QCD}$, the interior of the Fermi see is chirally symmetric, but near the Fermi surface the symmetry is broken by the chiral density waves.\cite{Rubakov} (Although Son and Shuster\cite{SonShuster} argue that such waves develop only for very large $N_c>10^3$.) To be precise, the chiral density waves mix the chiral symmetry of the quark phase with the translational symmetry  rather than simply break it. Averaging over space restores the chiral symmetry, just like it happens for the lattice of half-skyrmions.

On the other hand, for $\mu_q$ just above $m_q=M_b/N_c$, the baryonic crystal has a completely broken chiral symmetry. Thus, at some critical $\mu_q^{(c)}=O(m_q)$, there should be a chiral symmetry restoring phase transition from the baryonic crystal to a distinct quark phase. In holography, this transition is believed to be dual to the ``popcorn transition'' from a three dimensional to a four dimensional instanton lattice.\cite{Rozali:2007rx,Kaplunovsky:2012gb}


\subsection{Effect of the large $\lambda$ limit on the holographic nuclear matter}
\label{sec:Largelambda}
\begin{figure}[bt]
\centerline{\psfig{file=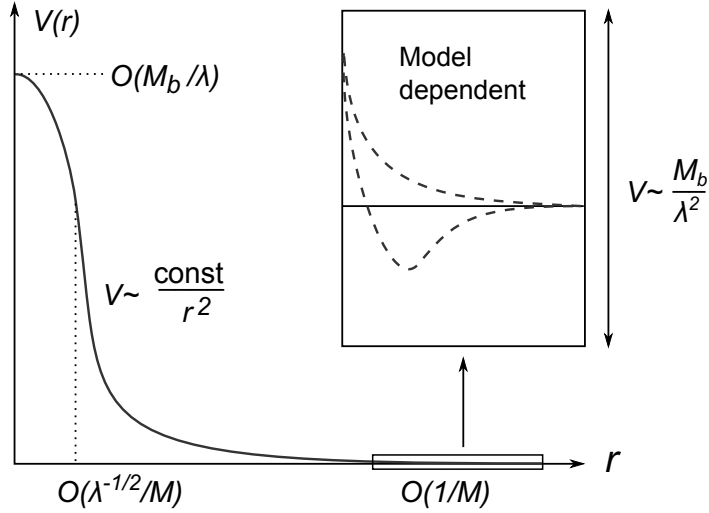,width=3.65in}}
\vspace*{8pt}
\caption{Two-body nuclear potential in holographic QCD.}
\label{figPotential}
\end{figure}

In holography, the semiclassical description of the gravity side of the gauge/gravity duality in terms of metric, fluxes, and branes requires the limits of large $N_c$ and also large 't~Hooft coupling $\lambda=g_{\rm YM}^2N_c$. In the large $\lambda$ limit, the baryons become very heavy: in units of the mesonic mass scale ($M\sim\Lambda_{\rm QCD}$) the baryon mass is $M_b\sim\lambda N_c M$. However, the interactions between the baryons do not grow with $\lambda$: even for two baryons right on top of each other, the repulsive potential between them is only $V\sim N_c M\sim M_b/\lambda$. At larger distances, the forces are even weaker since the hard-core radius of a holographic baryon shrinks with $\lambda$ as $R_b\sim M^{-1}\lambda^{-1/2}$. Outside this radius, the repulsive potential decreases as $1/r^2$ until $r\sim M^{-1}$, at which point it becomes dependent on the meson mass spectrum of a specific holographic model: In some models, the potential becomes attractive for $r\gtrsim M^{-1}$ while in others it remains repulsive at all distances. The overall picture is shown in figure~\ref{figPotential}.

Since the nuclear forces are so weak in the holographic QCD, all transitions between different phases of the cold nuclear matter happen at chemical potentials $\mu$ very close to  the baryon mass: just below $M_b$ we have glueball/meson gas (or vacuum for $T=0$), while just above $M_b$ we have dense quark matter. To see the baryonic matter phase (or any other intermediate phases) we need to zoom into the $\mu\approx M_b$ region.

\begin{figure}[bt]
\centerline{\psfig{file=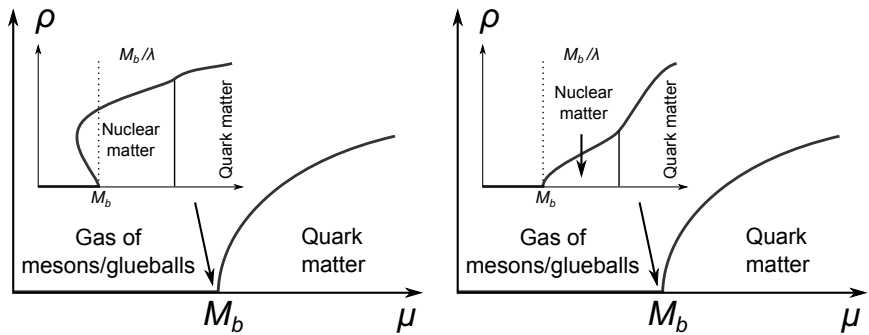,width=0.9\linewidth}}
\vspace*{8pt}
\caption{Density as a function of the chemical potential in the large $\lambda$ regime
	for attracting~(left) and repelling~(right) baryons.
	The nuclear matter phase is confined to a narrow window of the order $\Delta\mu\sim M_b/\lambda$.
	In the naive diagram the transition occurs directly from the no-baryon to a quark phase.}
\label{figAttractiveRepulsive}
\end{figure}

Figure~\ref{figAttractiveRepulsive} illustrates this point: to see the baryonic matter phase between the vacuum and the quark-matter phases on the plot of baryon density $\rho$ as a function of the chemical potential $\mu$, we need to zoom into the narrow range $\mu-M_b=O(M_b/\lambda)$. The figure also shows that the thermodynamic order of the phase transition between the vacuum (or the meson/glueball gas for $T>0$) and the baryonic matter depends on the sign of the long-distances nuclear force. If the force becomes attractive at long distances, then bulk baryonic matter exists at zero external pressure and has $\mu\geq M_b-E_{\rm bind}$, where $E_{\rm bind}$ is the binding energy. Consequently, the transition from the vacuum (or gas) to the nuclear matter (in the form of a baryonic crystal) is  first order as shown on figure~\ref{figAttractiveRepulsive}(left). On the other hand, if the nuclear forces are repulsive at all distances, then the bulk nuclear matter does not exists except at positive external pressures, and its chemical potential must be $\mu>M_b$. Moreover, $\mu$ raises monotonically with the pressure and the density, so the transition from the vacuum to the bulk nuclear matter is second order as shown on figure~\ref{figAttractiveRepulsive}(right).

Phase diagram of the Sakai-Sugimoto model was studied, for example, in Refs.~\refcite{Horigome:2006xu}--\refcite{Yamada:2007} and~\refcite{Bergman}, where the narrow nuclear matter phase was ignored. In the latter examples the scenario on the right part of figure~\ref{figAttractiveRepulsive} applies. Here we make no assumptions about the long-distance nuclear forces. Consequently, we cannot say anything specific about the transition from the vacuum to the nuclear matter phase. Instead, we focus on the transition from that phase to the quark liquid phase, which correspond in the holographic picture to changing the configuration of the instanton crystal, from a three-dimensional lattice, through a sequence of intermediate steps, to a four--dimensional one. Or rather, in this article, we review a simplified problem, namely the transitions between one-dimensional and two-dimensional instanton lattices.


\subsection{Multi--Baryon Systems in holographic QCD}
\label{sec:multibaryons}
In the large $N_c$ limit, nuclear forces between the baryons are dominated by the static potentials. Holographically, a static system of $A$ baryons corresponds to a time-independent configuration of the non-abelian magnetic flavor fields $F^a_{\mu\nu}(\bx,z)$ ($\mu,\nu=1,2,3,z$) of net instanton number $A$,
\be
\int\!\!d^3\bx\,dz\,\frac{\epsilon^{\kappa\lambda\mu\nu}}{16\pi^2}\,\tr(F_{\kappa\lambda}F_{\mu\nu})\
=\ A\,,
\label{InstantonNumber}
\ee
accompanied by the Coulomb electric $A_0^a(\bx,z)$ and scalar $\Phi^a(\bx,z)$ potentials induced by their Chern--Simons and $\Phi FF$ couplings to the magnetic fields (see section~\ref{sec:SSBaryons} and Ref.~\refcite{Kaplunovsky:2010eh} for the details). The whole configuration should minimize the net $\rm DBI+CS$ energy of the system subject to the constraint~\eqref{InstantonNumber}.

In the $\lambda\to\infty$ limit, the DBI energy of the magnetic fields is $O(\lambda)$ while the net effect of the electric and scalar fields is only $O(1)$. Moreover, the magnetic fields are concentrated within $O(\lambda^{-1/2})$ distance from the $z=0$ hyperplane, so to the leading order we may approximate the $4+1$ dimensional spacetime as flat. Thus Eq.~(\ref{DBIactionK}) becomes
\be
E_{\rm net}\ \approx\ E_{\rm DBI}\ \approx\ \frac{1}{g^2_{YM}}
\int d^3\bx\,dz\,\mathop{\rm Str}\nolimits\left[
	\sqrt{\det\bigl(K_0\delta_{\mu\nu}+F_{\mu\nu}\bigr)}\,
	-\,K_0\right]
\label{DBIenergy}
\ee
where we neglect the $z$ dependence of the $g^2_{\rm YM}$ and $K_0=K(0)$. Similar to the Yang--Mills energy of an $A$--instanton system, this leading--order DBI energy is minimized by the self-dual configurations of the magnetic fields $F^a_{\mu\nu}(\bx,z)$. In fact, all such self-dual configurations (of the same instanton number $A$) have the same leading-order energies
\be
E_{\rm LO}\ =\ \frac{8\pi^2}{g_{\rm YM}^2}\times A \ =\ A \lambda N_c M\,,
\label{LeadingOrder}
\ee
and $M$ was introduced in~(\ref{HoloGZ}). The self-dual configurations form a continuous family parameterized by $4N_fA$ moduli, which correspond to the locations, radii, and $SU(N_f)$ orientations of the $A$ instantons. But the leading-order energy~\eqref{LeadingOrder} does not depend on any of these moduli.

Fortunately, the sub-leading corrections to the net energy lift the degeneracy of the leading order, which provides for the $O(\lambda^0)$ interactions between the baryons. To work out such interactions we need the degenerate perturbation theory for the magnetic field configurations and their energies. Let us outline the formal procedure.

At  first order of the perturbation theory, one (a) limits the $F^a_{\mu\nu}(\bx,z)$ configurations to the degenerate minima of the leading-order energy function, \emph{i.e.} to the self-dual magnetic fields; (b) calculates the $O(\lambda^0)$ corrections $\Delta E$ for the energies of these configurations; (c) minimizes $\Delta E$ among the self-dual configurations. At the next order, to compute the $O(\lambda^{-1})$ corrections to the energy, one needs to find the $O(\lambda^{-1})$ corrections $\Delta F^a_{\mu\nu}(\bx,z)$ to the magnetic fields, which would no longer be self-dual. This is however beyond the aim of this review.

In more detail, to get $O(\lambda^0)$ interactions between $A$ baryons, we proceed as follows:
\begin{enumerate}
\item
Use Atiyah--Drinfeld--Hitchin--Manin (ADHM) construction\cite{ADHM} to obtain general self-dual magnetic field configurations, for a review see \emph{e.g.} Ref.~\refcite{Corrigan:1983sv}. Those are encoded in terms of $A\times A$ and $A\times N_f$ matrices obeying certain quadratic (ADHM) constraints. The first task is to solve these constraints and write down the ADHM matrices in terms of the instantons' locations, radii, and orientations.

\item
Given the ADHM matrices, work out the instanton number density profile
\be
I(\bx,z)\ =\ \frac{\epsilon^{\kappa\lambda\mu\nu}}{16\pi^2}\,\tr(F_{\kappa\lambda}F_{\mu\nu}),
\ee
for $N_f>2$ one would also need the non-abelian adjoint density
\be
I^a(\bx,x)\ =\ \frac{\epsilon^{\kappa\lambda\mu\nu}}{16\pi^2}\,d^{abc} F^b_{\kappa\lambda}F^c_{\mu\nu}\,.
\ee

\item
Next, calculate the $O(\lambda^0N_c)$ corrections to the net energy of the system, which include the effect of the $z$--dependence of the five--dimensional gauge coupling and the Coulomb electric and scalar potentials induced by the Chern--Simons and the $\Phi FF$ couplings, thus
\be
\Delta E\ =\ \Delta E_{NA}\ +\ \Delta E_C\,.
\ee
The $z$--dependent gauge coupling changes the DBI energy of the magnetic fields by
\be
\Delta E_{\rm NA}\ =\ N_c M\int d^3x\,dz\,\lambda M^2 z^2\times I(\bx,z),
\label{ENA}
\ee
while the Coulomb energy depends on the $N_f$. For $N_f=2$, the  $U(2)$ CS and  $\Phi FF$ terms couple the $SU(2)$ magnetic fields to the $U(1)$ Coulomb fields only. Consequently, the $A_0^a$ and the $\Phi^a$ fields are abelian and couple to the instanton density $I(\bx,z)$. Their net energy is simply $4+1$ dimensional Coulomb energy
\be
\Delta E_C\ =\ \frac{N_c M}{4}\int d^3\bx_1\,dz_1 \int d^3\bx_2\,dz_2\,
\frac{I(\bx_1,z_1)\times I(\bx_2,z_2)}{\lambda M^2( (\bx_1-\bx_2)^2+(z_1-z_2)^2)}\,.
\label{ECoulomb}
\ee
For $N_f>2$, the $U(N_F)$ CS and $\Phi FF$ terms couple the $SU(N_F)$ magnetic fields to both abelian and non-abelian electric and scalar fields; the abelian Coulomb fields are sourced by the instanton density $I(\bx,z)$ while the non-abelian fields are sourced by the adjoint density $I^a(\bx,z)$.
Altogether, the net energy of these Coulomb fields is
\begin{multline}
\Delta E_C  =\,\frac{N_c M}{4}\int d^3\bx_1\,dz_1\int d^3\bx_2\,dz_2\,
\left\{\frac{2I(\bx_1,z_1)\times I(\bx_2,z_2)}{\lambda N_f M^2( (\bx_1-\bx_2)^2+(z_1-z_2)^2)} \right.
\\  \left. + \, \frac{4I^a(\bx_1,z_1)\times I^a(\bx_2,z_2)}{\lambda M^2( (\bx_1-\bx_2)^2+(z_1-z_2)^2)}\right\}.
\end{multline}

\item
The previous three steps give $\Delta E$ as a function of baryons' locations, radii, and $SU(N_f)$ orientations. In the final step minimize the $\Delta E$ with respect to all these moduli.
\end{enumerate}

This four--step procedure is fairly straightforward for a few baryons --- \emph{cf.} calculations of the two--body nuclear forces by Kim and Zahed\cite{Kim:2008iy} and by Hashimoto \emph{et~al.}\cite{Hashimoto:2009ys} However, it becomes prohibitively difficult for large numbers of baryons\cite{Hashimoto:2009as,Hashimoto:2010ue} and outright impossible for infinite baryon crystals. At best, one can survey a small subspace of the $A$-instanton moduli space and try to minimize the $\Delta E$ over that subspace. For example, we may assume that all the instantons have the same radius, that their centers form a periodic lattice of some particular symmetry, and that the orientations of the instantons also form some kind of a periodic pattern. This gives us an ansatz for all the $A\times 4N_f$ moduli in terms of just a few overall parameters, and we can try to calculate and minimize the $\Delta E$ as a function of these parameters. However, any such ansatz is likely to miss the true lowest-energy configuration of the system. Indeed, in condensed matter guessing the crystalline symmetry of some substance from the properties of the individual atoms is a game of chance with poor odds, and there is no reason why the instanton crystals should be any simpler. Moreover, even if we could somehow guess all the symmetries of the instanton crystal, actually working through above four steps is impossible without additional approximations (besides $\lambda\gg1$).

In the examples we consider in the remainder of this review we will use several approximations. In section~\ref{sec:pointcharge} we will illustrate our main expectations about the high density properties of the holographic nuclear matter by considering point-like instantons. In section~\ref{sec:StraightChain}, where the low--density will be presented, we will need the sparse lattice approximation (the size of the instanton is much smaller than the unit cell size) in order to compute the Coulomb energy. In section~\ref{sec:zigzagsoln} this approximation will be necessary to solve for the ``high--density" phase. In fact, this is as far as the analytical methods have allowed us so far.\cite{Kaplunovsky:2012gb} In section~\ref{sec:secondpaper} we will further neglect the contribution of the many--body forces to the instantons' interactions. This allows for an efficient numerical analysis and has lead to a plethora of high-density phases.\cite{Kaplunovsky:2013iza} In both cases, however, we will restrict only to the one--dimensional crystals.

\begin{figure}[bt]
\centerline{\psfig{file=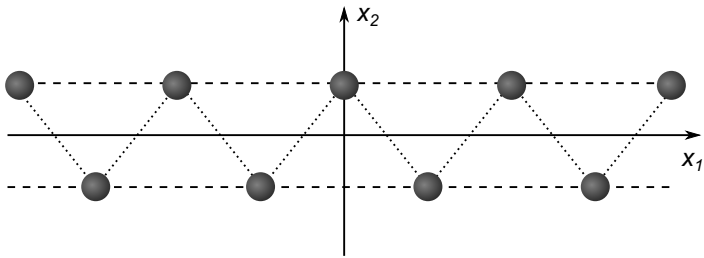,width=3.65in}}
\vspace*{8pt}
\caption{The zigzag configuration is the leading order small deformation of the one--dimensional chain.}
\label{fig:zigzag}
\end{figure}

To make the baryons form a one--dimensional lattice instead of spreading out in three dimensions, we curve the two transverse dimensions of the flavor branes, say $x_2$ and $x_3$, similar to the curvature of the $z$ coordinate. In terms of the effective $4+1$ dimensional theory, this corresponds to the five--dimensional flavor gauge coupling depending on the $x_2$ and the $x_3$ as well as the $x_4\equiv z$,
\be
\frac{8\pi^2}{g_{\rm YM}^2(x)}\ =\ \lambda N_c M\Bigl(
	1\,+\,M_2^2x_2^2\,+\,M_3^2x_3^2\,+\,M_4^2x_4^2\,+\,O(M^4x^4)\Bigr).
\label{GeneralCoupling}
\ee
This gauge coupling acts as a harmonic potential for the instantons which pulls them towards the $x_1$ axis, so at low densities the instantons form a one-dimensional lattice along the $x_1$. At higher densities, the instantons push each other away from the $x_1$ axis and form
more complicated two-dimensional or three-dimensional lattices, starting with the zigzag, the configuration shown in figure~\ref{fig:zigzag}. To make sure the transition from the straight chain to the zigzag happens for lattice spacings much larger than the instanton radius (which is required by the two-body force approximation), we will assume $M_2\ll M_3,M_4$.

\subsection{Point charge approximation}
\label{sec:pointcharge}

In order to illustrate our expectation about the phase structure of holographic baryons at high density let us consider a drastic simplification. Let us take the limit in which the size of instantons goes to zero, in which they become like point charges. In this limit the only surviving moduli of the instantons are their positions. As explained in the previous section, by tuning a potential in the transverse directions $x_2$ and $x_3$, see Eq.~(\ref{GeneralCoupling}), we will confine the lattice of instantons--charges to a one--dimensional chain along $x_1$. We will assume that potential in one of the directions, say $x_2$, is weaker than in the others, specifically $M_2\ll M_3, M_4$. Later, in section~\ref{sec:secondpaper}, we will observe that keeping the ration $M_3/M_4$ non-trivial allows for a number of new high--density phases.

In the point--charge limit there are two competing forces exerted on instantons. One comes from the curvature--generated ``non-abelian" potential, which confines them to the $x_1$ dimension, and the second is the four-dimensional Coulomb repulsion, which dislikes instantons sitting too close to each other. At low densities the equilibrium configuration must be a straight periodic chain of instantons. The one--dimensional density is $1/D$, where $D$ is the lattice spacing.

When $D$ goes to zero, equivalently density grows high, the Coulomb repulsion will force the instantons to leave the one--dimensional alignment and to expand in the transverse directions at the expense of increasing their non--abelian energy. The leading displacement mode is the zigzag shown in figure~\ref{fig:zigzag}, in which each pair of nearest neighbors goes in the opposite directions away from the chain. Recall that from the point of view of the flavor branes the neighboring instantons in the zigzag move to the opposite branches of the U-shaped configuration. This was called the \emph{popcorn} transition in Ref.~\refcite{Kaplunovsky:2012gb}.

Let us now study this transition quantitatively. We replace the instanton density $I(x)$ by the sum of delta-functions
\be
I({\bf x}) = \sum\limits_{n=-\infty}^{\infty}\delta^{(4)}({\bf x}-n {\bf D})\,,
\ee
where ${\bf D}$ is a 4-vector generating translations from one site of the chain to a neighboring one, here chosen to be along $x_1$. For the straight chain the non-abelian part of the energy, calculated per instanton, gives
\begin{multline}
\label{nonAbelianEnergy}
E_{\rm NA}= N_c\lambda M\int\limits_0^d d x_1\int d^3 x \ I(x)\left(1 + M_2^2x_2^2+ M_3^2x_3^2+ M^2 x_4^2\right) =
\\ = N_c\lambda M\left(1+M_2^2x_2^2+M_3^2x_3^2 + M^2 x_4^2\right).
\end{multline}
This energy is minimized for the choice $x_2=x_3=x_4=0$. From Eq.~(\ref{ECoulomb}) the Coulomb energy per instanton is given by the sum
\be
\label{AbelianEnergy}
E_{\rm C} =\frac{N_c}{4\lambda M}\sum\limits_{n\neq 0}\frac{1}{(nD)^2}=\frac{N_c}{\lambda M}\,\frac{\pi^2}{12 D^2}\,.
\ee

Let us investigate stability of the charges in the straight chain against the transition to the zigzag. The preferred zigzag deformation is in the $x_2$ direction as we have frozen $x_3=x_4=0$ by requiring $M_2\ll M_3,M_4$. For every charge in the zigzag $\Delta x_2=\pm\epsilon$. As a result the averaged change in the non-abelian energy (per instanton) will be
\be
\label{NAshift}
\Delta E_{\rm NA} = N_c\lambda M M_2^2\epsilon^2\,.
\ee
It is also straightforward to evaluate the Coulomb energy per instanton in the zigzag phase:
\begin{multline}
\label{AbelianEnergy2}
E_{\rm C} = \frac{N_c}{4\lambda M}\left(\sum\limits_{{\rm even}~n\neq 0 }\frac{1}{(nD)^2}+\sum\limits_{{\rm odd}~n}\frac{1}{(nd)^2+(2\epsilon)^2}\right) \\= \frac{N_c}{\lambda M}\left(\frac{\pi^2}{48 D^2}+\frac{\pi}{16\epsilon D}\,\tanh\frac{\pi\epsilon}{D}\right).
\end{multline}

For small $\epsilon$ let us expand the energy difference between the zigzag and the straight phases:
\be
\label{1dChargesEn}
\Delta E=N_c\lambda MM_2^2\epsilon^2 + \frac{N_c}{\lambda M}\left(-\frac{\pi^4\epsilon^2}{48 D^4} + \frac{\pi^6\epsilon^4}{120 D^6}+ O(\epsilon^6)\right).
\ee
The sign of the $\epsilon^2$-term depends on the density (lattice spacing). It is positive for small densities and negative for large. Therefore there is a second order phase transition, in which $\epsilon$ acquires a non-trivial vev, \emph{i.e.} the straight chain turns into a zigzag. The critical density corresponds to the point, where $\epsilon^2$-term changes sign:
\be
\label{PointChargeDc}
D=D_c \equiv \frac{\pi}{2\cdot 3^{1/4}\sqrt{MM_2\lambda}}\,.
\ee
For the spacing slightly smaller than $D_c$ the order (zigzag) parameter has the mean field behavior:
\be
\label{Trans1DCharge}
\langle\epsilon\rangle\simeq \pm \frac{\sqrt{5}}{\pi}\,\sqrt{D_c(D_c-D)}\,.
\ee
More generally $\epsilon$ is given by the solution to
\be
\label{Esolution}
\frac{(\pi\epsilon/D)^3\cosh^2(\pi\epsilon/D)}{\sinh(2\pi\epsilon/D)\,-\, (2\pi\epsilon/D)}\
=\ \frac{3}{4}\left(\frac{D_{c}}{D}\right)^4.
\ee
The zigzag amplitude grows monotonously with the density. However at larger densities, that is smaller spacing between instantons in $x_1$, it may become favorable for instantons to organize themselves in a more complicated configuration with respect to $x_2$ in order to minimize the Coulomb energy. Indeed, as was shown in Ref.~\refcite{Kaplunovsky:2012gb}, the zigzag first turns into a double zigzag (four layers in $x_3$), then to a three--layer configuration \emph{etc.} Gradually the original one--dimensional lattice turns into a two-dimensional one. Except for the first transition to the zigzag, which is second order, the remaining ones are of first order.

A similar analysis can be done for two-- and three--dimensional lattices. For the three-dimensional case one does not need additional parameters $M_2$, $M_3$, there is only the curvature of the holographic $z$--dimension, $M_4\equiv M$. If one considers a simple cubic (sc) lattice, it is likely that at large densities it will split along the $z$ direction into a three-dimensional analog of the zigzag. In such a transition every even site will displace in the $\pm z$ direction while the every odd one will displace in the opposite. In this second order phase transition the original sc lattice will split into two fcc layers.

The transition ${\rm sc}\to 2{\rm fcc}$ can be studied quantitatively. The only difference is that in two and three dimensions the Coulomb energy $E_{\rm C}$ per instanton will diverge. However, what matter for the the question of stability is the energy difference of the two configurations, which is finite. Overall, one finds
\be
\label{3dChargesEn}
\Delta E = N_c\lambda M^3\epsilon^2 + \frac{N_c}{\lambda M}\left(-\frac{\Delta\mu^2\epsilon^2}{D^4} + \frac{4\ell\epsilon^4}{D^6}+ O(\epsilon^6)\right),
\ee
where the following dimensionless quantities have been introduced
\be
\Delta \mu^2 = \sum\limits_{\text{odd}}\frac{1}{\left(n_1^2+n_2^2+n_3^2\right)^2}\simeq 10.0\,, \qquad \ell = \sum\limits_{\text{odd}}\frac{1}{\left(n_1^2+n_2^2+n_3^2\right)^3}\simeq 6.60\,.
\ee
This gives the critical density corresponding to the lattice spacing
\be
\label{scCritSpacing}
D_c= \frac{1}{M}\sqrt{\frac{\Delta\mu}{\lambda}} \simeq\ 0.69\,\frac{1}{M\sqrt{\lambda}}\,.
\ee

The sc lattice however is unlikely the ground state of the system of point charges. The minimum energy configuration of in this case is unknown, but we believe it to close packing, that is the largest interatomic distance between the nearest neighbors for a given density, which is achieved in the fcc lattice. Stability analysis of the fcc lattice is trickier because for the fcc lattice there is no natural way to split into two sublattices. One can expect two things to happen when the density is increased. First there can be a first order transition to a multi--layer four-dimensional lattice. Second the splitting into two sublattices can occur through a breaking of the cubic symmetries (half of nearest neighbors will go one way and another half the other way). The latter will be restored as soon as the separation between the sublattices will be large enough.

The example of the point charges demonstrates the main expectations about the phase structure of the holographic baryon crystal. When squeezed hard enough the lattices of baryons-instantons will expand in transverse dimensions to form higher--dimensional structures. In particular, the three-dimensional lattice of instantons will turn into a four--dimensional one with a non-zero thickness in the holographic dimension. As the density grows high the number of possible phases (lattice configurations) grows rapidly. This number is even higher if one goes beyond the point--charge approximation, as one has to take into account other instantons' moduli, such as orientation. In the remaining sections we will review the current results on the phase diagram of the nuclear matter phase in the generalized Sakai-Sugimoto model, so far obtained for the one-dimensional instanton lattices.


\section{Exact solutions}
\label{sec:exactsolns}

In this section we will discuss some analytical $N_f=2$ instanton solutions, which describe holographic one--dimensional baryon crystals close to the threshold chemical potential, \emph{i.e.} around the nuclear--to--quark matter phase transition. As we discussed in the previous section for analytical treatment to be viable, we need some assumptions about the instanton moduli. In this section we will assume a certain periodicity of the instantons' orientations. In order to show that the solutions, considered here, are true ground states it is desirable to expand the available moduli space. This will be done in section~\ref{sec:secondpaper}, where a larger class of solutions will be scanned numerically based on the further two--body force approximation.

\subsection{Straight chain}
\label{sec:StraightChain}

Let us start from reviewing the straight chain of instantons, which is the appropriate solution for the low--density baryon crystal. The solution for an infinite periodic chain of $SU(2)$ instantons was first obtained by Harrington and Shepard in Ref.~\refcite{Harrington:1978ve}. That solution corresponded to parallel--oriented instantons. Two decades later, in Ref.~\refcite{Kraan:1998pm}, Kraan and van Baal derived a generalization of the Harrington--Shepard solution, which described an instanton chain with a constant relative orientation shift between any pair of neighboring instantons. In this subsection we will review the result of Ref.~\refcite{Kraan:1998pm} in the context of holographic QCD.

One reason why it took a while to generalize the solution of Ref.~\refcite{Harrington:1978ve} to the solution of Ref.~\refcite{Kraan:1998pm} was that the former can be obtained through the simple 't Hooft ansatz, while the case of instantons of variable orientation required the details of the ADHM construction. ADHM data are matrices, which contain the information about the locations of the instanton centers, their radii and orientations. For an infinite chain the matrices are infinite--dimensional. Specifically, $Sp(k)$ $A$--instanton solution is encoded by two quaternionic matrices: an $A\times A$ symmetric matrix $X$ and a $k\times A$ vector $Y$, which satisfy the constraint
\be
\label{ADHMeqs}
X^\dagger X + Y^\dagger Y \qquad \text{is real symmetric.}
\ee

In the case of $Sp(1)\simeq SU(2)$ the solution can be reformulated in terms of four real symmetric matrices $\Gamma^\mu$ and real vectors $Y^\mu$:
\be
\label{GammaDef}
X=\Gamma^\mu\tau^\mu\equiv \Gamma^4+ i\tau^j \Gamma^j\,, \qquad Y=Y^\mu\tau^\mu\equiv Y^4 + i\tau^jY^j\,,
\ee
where $\tau^j$, $j=1,2,3$ are Pauli matrices. The real matrices satisfy the constraint
\be
[\Gamma^\mu,\Gamma^\nu]\,+\,Y^\mu\otimes Y^\nu\,-\,Y^\nu\otimes Y^\mu\
=\ \frac{\epsilon^{\mu\nu\kappa\lambda}}2
\left([\Gamma^\kappa,\Gamma^\lambda]\,+\,Y^\kappa\otimes Y^\lambda\,-\,Y^\lambda\otimes Y^\kappa\right),
\label{GammaRel}
\ee
where $\bigl(Y^\mu\otimes Y^\nu\bigr)_{mn}=Y^\mu_m\times Y^\nu_n$. Physically, the diagonal matrix elements $\Gamma^\mu_{nn}$ are the four--dimensional coordinates of the instantons' centers. The components $Y_{n}^\mu$ combine the radii and the $SU(2)$ orientations of the instantons. In particular, we introduce $a_n=|Y_n|$ for the real radii, and $A$ $SU(2)$ matrices $y_n$ (equivalent to unimodular quaternions $Y_n/a_n$) parameterizing the instanton's
orientations. Equations~(\ref{GammaRel}) can be equivalently written as
\be
\eta^a_{\mu\nu}\bigl[\Gamma^\mu,\Gamma^\nu\bigr]_{mn}\
+\ a_ma_n\times\tr\bigl(y_m^\dagger y_n^{}(-i\tau^a)\bigr) =\ 0\,,
\label{ADHM}
\ee
where $\eta^a_{\mu\nu}$ is the 't~Hooft's symbol mapping between the $SU(2)_{\rm gauge}$ and the $SU(2)_L$ inside ${\rm Spin}(4)=SU(2)_L\times SU(2)_R$,
\be
\eta_{44}^a=0,\quad
\eta^a_{4i}=-\delta^a_i\,,\quad
\eta^a_{i4}=+\delta^a_i\,,\quad
\eta^a_{ij}=\epsilon^{aij},\quad
a,i,j=1,2,3.
\ee

For a given ansatz, with selected radii, positions and orientations of instantons, the off-diagonal matrix elements $\Gamma^\mu_{m\neq n}\equiv \alpha^\mu_{mn}$ are determined through the ADHM conditions~(\ref{GammaRel}). The ADHM data are somewhat redundant --- an $O(A)$ symmetry acting on all the $\Gamma^\mu_{mn}$ and $Y_n=a_ny_n$ does not change any physical properties of the multi-instanton data. This symmetry includes $\Z_2^A$ which flips the instanton orientations $y_n\to-y_n$ (independently for each $n$). It also includes small $SO(A)$ rotations that change the off-diagonal elements by $\delta\alpha^\mu_{mn}=\epsilon_{mn}(X_m-X_n)^\mu+O(\epsilon^2)$. To eliminate these rotations, the ADHM equations~(\ref{ADHM}) for the off-diagonal elements should be combined with additional constrains (one for each $m\neq n$), for example
\be
\forall m\neq n:\
(X_m-X_n)^\mu \alpha^\mu_{mn}\ =\ 0.
\label{ADHMsols}
\ee

To find the self-dual vector--potential one also needs to solve an additional set of matrix equations. Luckily, for the kind of analysis we do here what we need is just the expression for the instanton density. In terms of the matrices $\Gamma^\mu$ and $Y^\mu$ it can be constructed as follows. Provided that~(\ref{GammaRel}) is satisfied define a real $A\times A$ symmetric matrix
\be
L_{mn}(x)\
=\,\sum_\ell\bigl(\Gamma^\mu_{\ell m}-x^\mu\delta_{\ell m}\bigr)
	\bigl(\Gamma^\mu_{\ell n}-x^\mu\delta_{\ell n}\bigr)\
+\ \frac12\, a_m a_n\tr\bigl(y_m^\dagger y_n^{}\bigr).
\label{Ldef}
\ee
The instanton density is then given by\cite{Osborn}
\be
I(x)\ =\ -\frac{1}{16\pi^2}\,\square\square\log\det(L(x))\,.
\label{Idensity}
\ee

We will look for a solution, which corresponds to an array of equal size instantons arranged in a straight periodic lattice. Periodicity means discrete translational symmetry ${\bf S}:x^\mu\to x^\mu+D^\mu$ of the ADHM solution; in the language of the $\Gamma^\mu_{mn}$ and $Y^\mu_m$, this symmetry acts as
\begin{align}
\Gamma^\mu\ &\to\ S^{-1}\Gamma^\mu S\ =\ \Gamma^\mu\ +\ D^\mu\mathbbm{1}\qquad
\text{to keep the $x^\mu\mathbbm{1}-\Gamma^\mu$ invariant}
\label{SymAct1}
\\
(Y^\mu_n\tau^\mu)\ &\to\,\sum_m G(Y^\mu_m\tau^\mu)S_{mn}\ =\ (Y^\mu_n\tau^\mu)\,,
\label{SymAct}
\end{align}
for some $O(A=\infty)$ matrix $S_{mn}$ and $SU(2)$ matrix $G$.

Physically, $G$ rotates the orientation of an instanton relative to its immediate neighbor. The more distant neighbors are related by $G^{n-m}$ rotations, which generate a $U(1)$ subgroup of the $SU(2)$. Without loss of generality, we take $G=\exp\bigl(i\phi\tau_3/2\bigr)$ for some ``twist'' angle $\phi$ between 0 and $2\pi$.

In this section we will take the direction of the instanton chain to be the $x^4$, while the transverse directions are $x^1$, $x^2$, and $x^3$. We will also make $x_3$ the preferred direction for the zigzag deformation.\footnote{Notice that this is different from the convention used in the previous section, where the chain was along $x^1$ and the zigzag was formed in $x^2$. We will return again to that convention in the next section.} In terms of equation~(\ref{SymAct1}) this means $D^\mu=(0,0,0,D)$. Finally, we take $S_{m,n}=\delta_{m,n+1}$ (shifts from the $n-{\rm th}$ instanton to the $(n+1)-{\rm st}$) and $Y_0^\mu=(0,0,0,a)$ (where $a$ is the radius of the instantons).  Consequently, the translational symmetry~(\ref{SymAct1})--(\ref{SymAct}) requires
\be
i\tau^\mu Y^\mu_n\ =\ a\,\exp\left(in\tfrac\phi2\tau_3\right)
\quad\Longleftrightarrow\quad
Y^\mu_n\ =\ \bigl(0,0,a\sin(n\phi/2),a\cos(n\phi/2)\bigr)
\ee
and
\be
\Gamma^\mu_{mn}\
=\ D\,\delta^{\mu4}\times n\,\delta_{mn}\ +\ \hat\Gamma^\mu(m-n)\,,
\ee
where the
$\hat\Gamma^\mu(m-n)$ do not have separate dependence on $m$ and $n$ but only on $m-n$. Combining these symmetry conditions with the ADHM constraint (\ref{GammaRel}), we get
\begin{align}
\label{GammaStraight1}
Y^\mu_m\otimes Y^\mu_n\ &=\ a^2\,\cos\left[(m-n)\phi/2\right],\\
\label{GammaStraight2}
\Gamma^4_{mn}\ &=\ Dn\,\delta_{mn}\,,\\
\label{GammaStraight3}
\Gamma^1_{mn}\ &=\ \Gamma^2_{mn}\ =\ 0,\\
\label{GammaStraight4}
\Gamma^3_{mn}\ &=\ \frac{a^2}{d}\,\frac{\sin\left[(m-n)\phi/2\right]}{m-n}\quad
{\rm for}\ m\neq n,\ {\rm but\ 0\ for}\ m=n.
\end{align}

To calculate the instanton density of the periodic chain we need the determinant of the infinite matrix $L$~(\ref{Ldef}). This determinant is badly divergent, but we may obtain it up to an overall infinite-but-constant factor from the derivatives
\be
\partial_\mu\log\det(L(x))\ =\ 2\tr\left((x^\mu\mathbbm{1}-\Gamma^\mu)L^{-1}(x)\right).
\label{Derivatives}
\ee

For three of these derivatives ($\mu=1,2,3)$ the trace converges, while for $\mu=4$ the trace diverges, but can be regularized using symmetry $x_4\to-x_4$, $n\to-n$. To evaluate those traces, it is natural to use the Fourier transform from infinite matrices to linear operators acting on periodic functions of $\theta~(\rm mod~2\pi)$.\footnote{This operation is also known as Nahm transform, \emph{cf.} Ref.~\refcite{Nahm}.} Consequently, $L$ becomes
\begin{align}
L\ &=\ x_1^2\ +\ x_2^2\ +\ \bigl(x_3-\Gamma^3(\theta)\bigr)^2\
+\ \left(x_4+iD\frac{d}{d\theta}\right)^2\ +\ T(\theta),\\
{\rm where}\ T(\theta)\ &=\ \pi a^2\delta\big(\theta-\frac{\phi}{2}\big)\ +\ \pi a^2\delta\big(\theta+\frac{\phi}{2}\big),
\label{Tantiferromagnetic}\\
{\rm and}\ \Gamma^3(\theta)\ &=\ \frac{\pi a^2}{D}\times\begin{cases}
	1-\frac\phi{2\pi} & {\rm for}\  0 <\theta<\frac{\phi}{2} \ \rm{and} \ 2\pi-\frac{\phi}{2}< \theta < 2\pi,\\
	\quad-\frac\phi{2\pi}& {\rm for}\ \frac{\phi}{2}< \theta < 2\pi - \frac{\phi}{2},
	\end{cases}
\end{align}
and $L^{-1}$ becomes the Green's function of this operator. Calculating this Green's function one obtains the following expression upon integration of the traces~(\ref{Derivatives}):
\begin{align}
\det(L)\ =\ &
\left(\cosh\frac{\phi r_1 }{D}\,+\,\frac{\pi a^2}{Dr_1}\,\sinh\frac{\phi r_1 }{D}\right)
\left(\cosh\frac{(2\pi-\phi) r_2 }{D}\,+\,\frac{\pi a^2}{Dr_2}\,\sinh\frac{(2\pi-\phi) r_2 }{D}\right)\nonumber \\
&+\ \frac{r_1^2+r_2^2-(\pi a^2/D)^2}{2r_1r_2}\,\sinh\frac{\phi r_1 }{D}\,\sinh\frac{(2\pi-\phi)r_2}{D}\nonumber\\
&-\ \cos \frac{2 \pi x_4}{D}\,,
\label{BigFla}
\end{align}
where
\begin{align}
r_1^2\ &=\ x_1^2\ +\ x_2^2\ +\ \left(x_3+\frac{a^2(\phi-2\pi)}{2D}\right)^2,\nonumber\\
r_2^2\ &=\ x_1^2\ +\ x_2^2\ +\ \left(x_3+\frac{a^2\phi}{2D}\right)^2.
\label{Radii}
\end{align}
This is precisely the result obtain by Kraan and van Baal in Ref.~\refcite{Kraan:1998pm}.



Let us now put this instanton solution in the context of holographic baryons. In line with the program discussed in section~\ref{sec:densebaryons} we are going to evaluate $O(\lambda^0)$ corrections to the leading order energy result. For that we will plug the density of the ``free" instanton solution into equations~(\ref{ENA}) and~(\ref{ECoulomb}). Then we will minimize the energy with respect to the relevant moduli, which in this case are the radius and the twist angle.

The expression for the instanton density follows from~(\ref{Idensity}) after plugging solution~(\ref{BigFla}). Although the expression derived from~(\ref{BigFla}) is way too complicated to print, several moments of the instanton density may be obtained via integrating by parts, thanks to the double D'Alembertian in Eq.~(\ref{Idensity}):
\begin{align}
\label{Moments0}
\int d^4 x\,I(x)\ &
=\ A,\\
\label{Moments1}
\int d^4 x\,I(x)\times x^\nu\ &
\equiv \langle\, x^\nu \,\rangle=\ \tr\bigl(\Gamma^\nu\bigr),\\
\label{Moments2}
\int d^4 x\,I(x)\times x^\mu x^\nu\ &
\equiv \langle\, x^\mu x^\nu \,\rangle =\ \tr\bigl(\Gamma^\mu\Gamma^\nu\bigr)\ +\ \frac{1}{2}\,\delta^{\mu\nu}\tr\bigl(T\bigr)\,,\\
\text{where}\ T_{mn}\ &
\equiv\ \frac{1}{2}\, a_m a_n\,\tr\bigl(y_m^\dagger y_n\bigr).
\end{align}
In particular,
\be
\vev{x_1^2}\ =\ \vev{x_2^2}\ =\ \frac{a^2}{2}
\label{Size12}
\ee
--- exactly as for a single stand-alone instanton of radius $a$, but
\be
\vev{x_3^2}\ =\ \frac{a^2}{2}\ +\ \frac{a^4}{4D^2}\,\phi(2\pi-\phi)\,,
\label{Size3}
\ee
where the second term is due to interference between the instantons. Curiously, the interference term vanishes for $\phi=0$, $i\,.e.$ for instantons with the same $SU(2)$ orientations.

The calculated moments of the instanton density yield the leading corrections to the non-abelian part of the energy~(\ref{ENA}). Taking the expansion of the five--dimensional gauge coupling from Eq.~(\ref{GeneralCoupling}) and using Eqs.~(\ref{Size12}) and~(\ref{Size3}), assuming small size instantons ($a\ll M^{-1}$), one obtains,
\begin{align}
\Delta E_{\rm NA}\
&=\ N_c\lambda M\left(\frac{a^2}2\left(M_1^2+M_2^2+M_3^2\right) \,
+\,\frac{M_3^2a^4}{4D^2}\,\phi(2\pi-\phi)\right).
\label{NAenergy}
\end{align}
Here $M_1\equiv M_4$ is appears since we stretch the chain along the $x_4$ dimension. Later in this section we will assume $M_1=M_2=M'$ and $M_3=M$.

Adapting Eq.~(\ref{ECoulomb}) one finds the abelian Coulomb energy per instanton to be
\be
E_{\rm C}\ = \ \frac{N_c}{256\pi^2 \lambda M}\int_0^D d x_4\int d^3x\,
(\partial_\mu\square\log\det(L))^2.
\label{Cenergy}
\ee
For generic lattice spacings $D$, this integral is too hard to take analytically. But it becomes much simpler in the $D\gg a$ limit of well-separated instantons.

For large lattice spacing $D\gg a$ the leading correction to the Coulomb energy~(\ref{Cenergy}) evaluates to
\be
\label{ECstraight}
\Delta E_{\rm C}\ \approx\ \frac{N_c}{\lambda M}\left[
\frac{1}{5a^2}\ +\ \frac{4\pi^2+3(\pi-\phi)^2}{30D^2}\ +\ O(a^2/D^4)\right].
\ee

Now we need to minimize the total energy with respect to the moduli: the instantons' radius $a$ and the twist angle $\phi$. Combining (\ref{NAenergy}) and (\ref{ECstraight}), we find the minimum at
\be
\label{phiastraight}
\phi_{\rm min}\ =\ \pi,\qquad  a_{\rm min}\ =\ a_0\ -\ \frac{\pi^2 a_0^3}{12D^2}\
+\ O(a_0^4/D^2)\,,
\ee
where $a_0$ is the equilibrium radius of a standalone instanton~(\ref{equiradius}). We see that in the ``low--density" regime the instantons in the straight chain prefer the anti-ferromagnetic order, \emph{i.e.} any two neighboring instantons have an opposite orientation.


\subsection{Zigzag}
\label{sec:zigzagsoln}

Let us now address the question of the stability (or instability) of the straight instanton chain considered in the previous subsection against transverse motion (in $x_{1,2,3}$ directions) of the instantons that would get them out of the linear alignment. If such a motion can decrease the chain's net energy, than the chain is unstable.

Following the conventions of this section, we set $M_1=M_2=M'$ and $M_3=M$ and assume $M<M'$, so that the leading instability would be in the $x_3$ direction.  The anisotropy also breaks the degeneracy between different directions $\vec n$ of the $SU(2)$ twist $\exp\bigl(i\phi(\vec n\cdot\vec\tau)/2\bigr)$ between adjacent instantons. The lowest-energy direction of the twist is now $\tau_3$ -- which is precisely what we have used in our formulae in the previous section. Specifically, while the Coulomb energy does not depend on the twist direction, the non-abelian energy is minimized when the largest $\vev{x^2}$ of the instantons is oriented in the lowest-potential direction $x_3$. According to Eqs.~(\ref{Size12}) and~(\ref{Size3}), the instantons are larger in the direction of the twist than in other directions, hence the non-abelian energy
\begin{multline}
E_{\rm NA}\ =\ N_c\lambda M\left(1\ +\ a^2 M^{\prime2}\ +\ \frac12\, a^2 M^2
\right.  \\
\left. + \ \frac{a^4\phi(2\pi-\phi)}{4D^2}\,
\left(M^{\prime2}(n_1^2+n_2^2)\,+\,M^2n_3^3\right)\
+\ O(a^4M^4)\right)
\end{multline}
is minimized for $\vec n=(0,0,\pm1)$, equivalently, $SU(2)$ twist in the $\tau_3$ direction.

Also, $M'>M$ reduces the equilibrium size of standalone or far-apart instantons from (\ref{equiradius}) to
\be
a'_0\ =\ \frac{(1/5)^{1/4}}{\lambda^2M^2(M^{\prime2}+\frac12\,M^2)^{1/4}}\
\approx\ \frac{(1/5)^{1/4}}{\sqrt{\lambda MM'}}\quad({\rm for}\ M'\gg M).
\label{InstSize}
\ee
Consequently the anisotropy, apart from stabilizing the chain in the $x_1$, $x_2$ directions,  provides an additional control over the equilibrium size of instantons. This will allow for an analytical analysis of the phase transitions below.

Before we start the stability analysis we must describe the expected instability mode in the proper ADHM terms, \emph{i.e.} in terms of $\Gamma^\mu_{mn}$ matrices and $Y^\mu_n$ vectors. Moving the instantons' centers in $x_{1,2,3}$ directions without changing their $x_4$ locations along the chain, or any radii, or $SU(2)$ orientations, as compared to the ansatz~(\ref{GammaStraight1})--(\ref{GammaStraight4}), means keeping
\be
\label{GammaZigzag1}
Y_n^\mu\ =\ (0,0,a\sin(n\phi/2),a\cos(n\phi/2)),\qquad
\Gamma^4_{mn}\ =\ D\,n\times \delta_{mn}\,,
\ee
exactly as for the straight chain but changing the $\Gamma^{i=1,2,3}_{mn}$ matrices
\be
\Gamma^3_{mn}\ \to\ \Gamma^3_{mn}[\mbox{straight chain}]\
+\ \delta\Gamma^3_{mn}\,,\qquad
\Gamma^1_{mn}\ \to\ \delta\Gamma^1_{mn}\,,\qquad
\Gamma^2_{mn}\ \to\ \delta\Gamma^2_{mn}\,,
\ee
in a manner that preserves the self-duality equations~(\ref{GammaRel}). The solution corresponding to the displacement in the $x_3$ direction reads
\be
\delta\Gamma^{1,2}_{mn}\ \equiv\ 0,\qquad
\delta\Gamma^3_{mn}\ =\ \delta_{mn}\times\delta X^3[n]\,,
\label{SD3mat}
\ee
and the leading instability is expected to be the zigzag (figure~\ref{fig:zigzag}):
\be
\delta X^3[n]\ =\ \epsilon\times(-1)^n.
\label{dX3}
\ee

After defining the ADHM matrices we make a Fourier (Nahm) transform to map the infinite-dimensional matrices to differential operators on a circle. In the following it will be natural to combine wave functions $\psi(\theta)$ and $\psi(\theta\pm\pi)$ on the circle into a two-component wave function. Two-component functions provide a natural description for a two-layered chain. For some particular cases of the twist angle $\phi=0$ or $\pi$ all expressions take particularly compact form. More generally for $n$ layers $n$--component functions should be used instead. Specifically, we choose two-component functions with the following boundary conditions
\be
\label{TwoComponent}
\Psi(\theta)\ =\,\begin{pmatrix} \psi(\theta)+\psi(\theta +\pi)\\ \psi(\theta)-\psi(\theta+\pi)\end{pmatrix},\quad
-\frac{\pi}{2}\le\theta\le\frac{\pi}{2},\quad \Psi(\pi/2)\,=\,\Sigma_3\Psi(-\pi/2),
\ee
where $\Sigma_1,\Sigma_2,\Sigma_3$ are Pauli matrices acting on the two components. In the space of such two--component wave functions, $\Gamma^3$ becomes the operator
\be
\label{Gamma3beta0}
\Gamma^3\ = \ \epsilon\times\Sigma_3 +\frac{\pi a^2}{2D}\Theta\bigl(-\frac{\phi}{2}<\theta<\frac{\phi}{2}\bigr)\times\Sigma_1 + \frac{\pi a^2}{2D}\left(\frac{\phi}{\pi}-\Theta\bigl(-\frac{\phi}{2}<\theta<\frac{\phi}{2}\bigr)\right)\times\mathbbm{1}\,,
\ee
where $\Theta$ is the step--function, \emph{i.e.} $\Theta=1$ if $- \phi/2<\theta<\phi/2$ and $\Theta=0$ otherwise; and we restrict $\phi$ to the values $0\leq\phi\leq\pi$. For $\phi=\pi$ the latter expression takes the form
\be
\Gamma^3\ =\ \frac{\pi a^2}{2D}\times\Sigma_1\ +\ \epsilon\times\Sigma_3\,.
\ee

Matrix $T\equiv Y^\mu\otimes Y^\mu$ in the two-component notation becomes
\be
\label{Tbeta0}
T(\theta)= \frac{\pi a^2}{2}\left(\mathbbm{1}+\Sigma_1\right)\left(\delta\bigl(\theta-\frac{\phi}{2}\bigr) + \delta\bigl(\theta+\frac{\phi}{2}\bigr)\right), \qquad 0\leq \phi<\pi\,.
\ee
For $\phi=\pi$ the latter can just be written as $T=\pi a^2\delta(\theta\pm \pi/2)\times\mathbbm{1}$, \emph{cf.} equation~(\ref{Tantiferromagnetic}).

Next we plug the data above in the formula for the operator $L$~(\ref{Ldef}). To calculate the determinant of this operator, one needs to calculate its Green's function $L^{-1}(\theta,\theta_0)$ and consequently the traces~(\ref{Derivatives}). As in the example of the straight chain the naive determinant is divergent and needs to be regularized, which is done by extracting an infinite but constant prefactor. One can reconstruct an analytic expression for $\det(L)$ with any value of $\phi$. However only for $\phi=0$ and $\pi$ the expression is compact enough to present in the paper. Accidentally $\phi=\pi$ is the minimal energy configuration of instantons for both the straight chain at low density and in the zigzag phase provided that the zigzag amplitude is not too large.\footnote{As we shall see in Sec.~\ref{sec:secondpaper} this is true only under the assumption $M'\gg M$. Furthermore, it turns out if the zigzag geometry is relaxed there are other configurations favored over zigzag. See section~\ref{sec:newresults}.} In the remainder of this part we discuss the case $\phi=\pi$. One finds that the regularized determinant reads
\begin{align}
\frac{\det(L)}{\rm const}\ =\ &
\left( \cosh\frac{\pi r_1}{D}\,+\,\frac{\pi a^2}{Dr_1}\,\sinh\frac{\pi r_1}{D}\right)
\left( \cosh\frac{\pi r_2}{D}\,+\,\frac{\pi a^2}{Dr_2}\,\sinh\frac{\pi r_2}{D}\right)\nonumber\\
&\qquad+\ \frac{r_1^2+r_2^2-(\pi a^2/D)^2}{2r_1 r_2}\,
	\sinh\frac{\pi r_1}{D}\,\sinh\frac{\pi r_2}{D}\
-\ \cos\frac{2\pi x_4}{D}\nonumber\\
&+\ \sin^2\nu \left(
	\cosh\frac{\pi r_1}{D}\,\cosh\frac{\pi r_2}{D}\,
	-\,\frac{r_1^2+r_2^2}{2r_1r_2}\sinh\frac{\pi r_1}{D}\,\sinh\frac{\pi r_2}{D}
	\right) \nonumber \\
&+\ \sin\nu\,\cos\frac{\pi x_4}{D}
	\left( 2\cosh\frac{\pi r_1}{D}\,+\,\frac{\pi a^2}{Dr_1}\,\sinh\frac{\pi r_1}{D}\right)\nonumber\\
&-\ \sin\nu\,\cos\frac{\pi x_4}{D}
	\left( 2\cosh\frac{\pi r_2}{D}\,+\,\frac{\pi a^2}{Dr_2}\,\sinh\frac{\pi r_2}{D}\right)
	\nonumber\\
&-\ \sin^2\nu\,  ,
\label{DetLmod}
\end{align}
where
\begin{align}
r_{1,2}^2\ &=\ x_1^2\ +\ x_2^2\ +\ (x_3\,\mp\,\tfrac12 b_\epsilon)^2,\\
\label{be}
b_\epsilon\ &=\ \sqrt{4\epsilon^2+(\pi a^2/D)^2},\\
\nu\ &=\ \arctan\frac{2\epsilon}{\pi a^2/D}\ =\ \arcsin\frac{2\epsilon}{b_\epsilon}\,.
\label{nu}
\end{align}

Naively, we would expect the zigzag deformation to have no effect on the width of the instanton chain in the $x_1$ and $x_2$ directions, while the width squared in the $x_3$ direction should increase by $\epsilon^2$. And indeed, this is precisely what happens for any instanton radius and lattice spacing: evaluating~(\ref{Moments2}), we obtain precisely
\be
\vev{x_1^2}\ =\ \vev{x_2^2}\ =\ \frac{a^2}{2},\qquad
\vev{x_3^2}\ =\ \frac{a^2}{2}\ +\ \frac{\pi^2 a^4}{4D^2}\ +\ \epsilon^2.
\ee
Consequently, the non-abelian energy of the zigzag is, {\it cf.} (\ref{NAshift}),
\begin{align}
E_{\rm NA}\ &
=\ N_c\lambda M\left( \bigl(\tfrac12 M^2+M^{\prime2}\bigr) a^2\
	+\ M^2\,\frac{\pi^2 a^4}{4D^2}\ +\ M^2\,\epsilon^2\right)\nonumber\\
&=\ E_{\rm NA}[\epsilon=0]\ +\ N_c\lambda M^3\,\epsilon^2.
\label{ENAmod}
\end{align}

To calculate the integral in Eq.~(\ref{ECoulomb}) and have workable expression for the Coulomb energy, we again resort to the approximation $D\gg a$, where analytical calculation is possible. For small, widely-separated instantons, $r_1\sim a\ll D$, but finite $\epsilon/D$, we approximate
\begin{align}
\det(L)\ \approx\ &
\frac{\pi^2}{D^2}\left( (a^2+r_1^2+x_4^2)\,+\,\frac{\pi^2}{12D^2}\,(r_1^4+2a^2r_1^2-x_4^2)\,
	+\,O(a^6/D^4)\right)\nonumber\\
&\qquad\qquad\qquad\times\left( \cos\frac{\pi x_4}{D}\,+\,\cosh\frac{\pi r_2}{D}\,
	+\,\frac{\pi a^2}{2r_2D}\,\sinh\frac{\pi r_2}{D}\right)\nonumber\\
&+\ \frac{\pi^2 a^4}{8D^2\epsilon^2}\times\frac{\pi r_2}{D}\,\sinh\frac{\pi r_2}{D}
\label{modDsmall}
\end{align}
and consequently obtain
\be
E_{\rm C}\ =\ \frac{N_c}{\lambda M}\left[ \frac{1}{5a^2}\ +\ \frac{3\pi^2}{80D^2}\
	+\ \frac{\pi^2}{80D^2}\,\frac{\tanh(\pi\epsilon/D)}{\pi\epsilon/D}\
	+\ O(a^4/D^6)\right].
\label{ECsmallMod}
\ee
Remarkably, the $\epsilon$--dependent part of this Coulomb energy is precisely five times smaller than the energy of point charges in a similar zigzag formation~(\ref{1dChargesEn}). Mathematically, this fivefold reduction stems from the last term in equation~(\ref{modDsmall}), which accounts for the interference between the instantons. It is not clear why the interference between well-separated small instantons
has such a drastic effect on the Coulomb energy of the zigzag. Anyhow, the net energy cost of a small zigzag deformation $\epsilon\ll D$ is
\begin{align}
\Delta E_{\rm net}\ & =\ \Delta E_{\rm NA}\ +\ \Delta E_{\rm C} \nonumber \\
 & =\ N_c\lambda M^3\,\epsilon^2\ +\ \frac{N_c}{\lambda M}\left[
	-\frac{\pi^4 \epsilon^2}{240 D^4}\
	+\ \frac{\pi^6\epsilon^4}{600 D^6}\
	+\ O(\epsilon^4/D^6)\right].
\label{ECsmallEdep}
\end{align}
This cost is positive --- and hence the straight chain is stable --- for all sufficiently large lattice spacings
\be
D\ >\ D_{c}\ \equiv\ \frac{\pi}{\root 4\of{240}}\,\frac{1}{M\sqrt{\lambda}}\,.
\label{Dcrit}
\ee
For smaller lattice spacings $D<D_{c}$, the energy function~(\ref{ECsmallEdep}) has a negative coefficient of $\epsilon^2$ but positive coefficient of $\epsilon^4$. Thus, for $D<D_c$ the straight chain becomes unstable and there is a second-order phase transition to the zigzag configuration. For lattice spacings just below critical, the zigzag parameter $\epsilon$ is
\be
\vev{\epsilon}\ \approx\ \pm\frac{\sqrt{5}}{\pi}\,\sqrt{D_c(D_c-D)}\,.
\label{SecondOrder}
\ee
For smaller lattice spacing (but larger than the instanton size), the zigzag parameter satisfies the same equation as in the point charge limit~(\ref{Trans1DCharge}) with the new $D_{\rm c}$.

In units of the equilibrium instanton size~(\ref{InstSize}), the critical lattice spacing~(\ref{Dcrit}) is
\be
\frac{d_{\rm crit}}{a\approx a_0'}\
=\ \frac{\pi}{\root 4\of{48}}\left(\frac{M^{\prime2}+\frac12M^2}{M^2}\right)^{1/4}\
\approx\ 1.2\sqrt\frac{M'}{M}\,,\quad{\rm for}\ M'\gg M.
\ee
For a highly anisotropic gauge coupling with $M'\gg M$, the critical spacing is much larger than the instantons, which justifies our approximations. However, for near-isotropic couplings with $M'\approx M$, the critical spacing is only $D_{\rm crit}\approx 1.32 a_0$, and in this regime we cannot be sure our analysis of the zigzag instability is even qualitatively correct. That is, we are not sure that there is a transition from a straight instanton chain to a zigzag for $M'\approx M$, never mind any details of such transitions.

In this section we have presented analytical solutions for two types of instanton lattices: the straight chain and the zigzag with anti-ferromagnetic orientation of the instantons. More specifically, we have spelled out the result for the determinant of the operator $L$, which can be readily used to evaluate the instanton density via Eq.~(\ref{Idensity}). The result allows to compute easily the non-abelian contribution to the energy of the configurations, however the integrals giving the abelian Coulomb part already require some approximations, such as the small instanton limit $a\ll D$.

The configurations considered here seem to approach the limits of what can be done exactly. More complicated instanton lattice configurations require simplifying assumptions already at the level of the ADHM ansatz. This will be done in the next section, where the two--body force approximation will be implemented. The approximations allow to find many new phases, including the ones with non-abelian twist angles. To see them however, one needs to further increase the anisotropy of the system. In all cases, however, we will assume the small instanton limit.


\section{Approximations and generalizations}
\label{sec:secondpaper}

In this section we will review other interesting phases of the holographic nuclear matter by exploring a shortcut around steps $(1)$, $(2)$, and $(3)$ in section~\ref{sec:multibaryons}. Following the idea of Ref.~\refcite{Kaplunovsky:2013iza} we will argue that when the distances between the instantons are much larger than their radii, the interactions between $A$ baryons are dominated by the two-body forces,
\be
\Delta E\ \approx\ \frac12\sum_{\textstyle{m,n=1,\ldots,A\atop m\neq n}}
F_2\bigl(X_m^\mu-X_n^\mu,{\bf m},{\bf n}\bigr)
\ee
for a manageably simple function  $F_2$ of the two instanton's positions $X^{\mu}_{m,n}$ and orientations ${\bf m}$ and ${\bf n}$. Consequently, we can minimize the net interaction energy over the entire moduli space of the multi-instanton system using a simple numerical simulation: Starting from a random set of instanton positions and orientations, use the steepest descent algorithm to find the nearest local minimum of the net energy; repeat this procedure for  different random starting points to find other local minima; eventually, find all the local minima, compare their energies, and identify the global minimum.

In the numerical simulation in Ref.~\refcite{Kaplunovsky:2013iza} the lattice geometry for the instanton positions was assumed, but the numerical simulation was used to find the lowest-energy pattern of their orientations. Once all the patterns are known, they can be used as ansatz's (with a few parameters) for which the net energy can be calculated as analytic functions of the parameters. In the boundaries between different orientation patterns are mapped much more accurately than just in the numerical approach. In this section we will review the main results of Ref.~\refcite{Kaplunovsky:2013iza}.

\subsection{Two-body forces}
\label{sec:twobody}

In real-life nuclear physics, besides the two-body nuclear forces due to meson exchanges, there are significant three-body forces, and presumably
also four-body forces, \emph{etc.},
\begin{multline}
\hat H_{\rm nucleus}\
=\,\sum_{n=1}^A\hat H^{\rm 1\,body}(n)\
+\,\frac12\sum_{\textstyle{{\rm different}\atop m,n=1,\ldots A }}\hat H^{\rm 2\,body}(m,n)\\
\ +\,\frac16\sum_{\textstyle{{\rm different}\atop \ell,m,n=1,\ldots A }}
	\hat H^{\rm 3\,body}(\ell,m,n)\
+\ \cdots
\end{multline}
where $n$ stands for  the quantum numbers of the $n-{\rm th}$ nucleon. Likewise, in the holographic nuclear physics interactions between multiple baryons include two-body forces and also three-body, four-body forces \emph{etc.} Even in the classical infinite--mass limit, where the holographic baryons become static instantons of the $SU(N_f)$ gauge fields in $4+1$ dimensions, the potential energy (due to five--dimensional curvature and due to Chern-Simons interactions) of an $A$--instanton system has form
\begin{multline}
\Delta E(1,2,\ldots,A)\
\equiv\ E^{\rm total}\ -\ A\, \lambda N_c M\\
=\,\sum_{n=1}^A  E^{(1)}(n)
+\,\frac12\!\!\!\sum_{\textstyle{{\rm different}\atop m,n=1,\ldots A }} \!\! E^{(2)}(m,n)
+\,\frac{1}{6}\!\!\!\sum_{\textstyle{{\rm different}\atop \ell,m,n=1,\ldots A }}
	\!\!E^{(3)}(\ell,m,n)\
+\ \cdots
\end{multline}
with significant three-body, \emph{etc.} terms.

What about the relative magnitudes of the two-body, three-body, four-body, \ldots, interaction terms? When the baryons are tightly packed so that their instanton cores overlap and merge, we expect all the $n$-body forces to have comparable strengths. But in the opposite low-density regime of baryons separated by distances much larger then their radii, the two-body forces dominate the interactions, while the multi-body forces are smaller by powers of $\it (radius/distance)^2$. In this subsection we briefly summarize the proof of this observation given in Ref.~\refcite{Kaplunovsky:2013iza}.

We review the simplest case of $N_f=2$, although the proof can be extended to the general $SU(N_f)$ case as well.\cite{Kaplunovsky:2013iza} First let us consider intermediate-range distances between the baryons,
\be
\left(a\sim\frac{1}{M\sqrt\lambda}\right)\ \ll\ |X_m-X_n|\ \ll\ \frac{1}{M}\,,
\ee
which allows us to treat the five-dimensional holographic space as approximately flat. The holographic baryons are instantons of the $SU(2)$ magnetic fields, which source the $U(1)$ electric and scalar fields via CS and $\tr(\Phi F_{\mu\nu}F^{\mu\nu}$) couplings. When those instantons are small and separated from each other by large distances $|X_m-X_n|\sim D\gg a$, their interactions  come from two sources:

\begin{enumerate}

\item Direct Coulomb repulsion (electric---scalar) between nearly-point-like abelian charges in $4+1$ dimensions,
\be
\Delta E^{\rm direct}\ =\ \frac{N_c}{4\lambda M}\sum_{m\neq n}\frac{1}{|X_m^\mu-X_n^\mu|^2}\,,
\label{Edirect}
\ee
which is a manifestly two-body interaction.

\item The interference between the instantons. The latter changes the distribution of the instanton number density in space,
\begin{align}
\label{DensityDef}
I(x)\ &
=\ \frac{\epsilon^{\mu\nu\rho\sigma}}{32\pi^2}\,
\tr\bigl(F_{\mu\nu}(x)F_{\rho\sigma}(x)\bigr),\\[5pt]
I^{A\,\rm instantons}(x)\ &
=\,\sum_{i=n}^A I^{\rm standalone}_n(x)\
+\,\Delta I^{\rm interference}(x),
\end{align}
which in turn changes the self-interaction energy (Coulomb and non-abelian) of each instanton by an amount comparable to~(\ref{Edirect}).

\end{enumerate}

Within an instanton, $i.\,e.$ at $O(a)$ distance from some instanton's center $X_n^\mu$, the interference from the other instantons should be relatively weak,
\be
\Delta I^{\rm interference}(x)\ \sim\ \frac{a^2}{D^2}\times I_n^{\rm standalone}(x),
\ee
so there should be some kind of a perturbation theory for it. At the first order of such perturbation theory, the $\Delta I$ arises from interference between the un-perturbed standalone-like instantons, so we expect it to be a sum of pair-wise interferences from the other instantons,
\be
\Delta I^{\rm 1st\,order}(x\,{\rm near}\,X_n)\
=\ I_n^{\rm standalone}(x)\,\sum_{m\neq n}{\cal F}^{(1)}_{n,m}(x)\,,
\label{DeltaIFirstOrder}
\ee
where
\be
{\cal F}^{(1)}_{n,m}(x)\ \sim\ O\left(\frac{a^2}{|X_n-X_m|^2}\right)
\ee
and depends only on the instantons $\#n$ and $\#m$, \emph{i.e.} on their positions, radii, and orientations, but not on any other instantons.

At the second-order, we expect to include the interference between the first-order $\Delta I$ and the additional instantons, so at this order we obtain three-body effects,
\be
\Delta I^{\rm 2nd\,order}(x\,{\rm near}\,X_n)\ =\ I_n^{\rm standalone}(x)\,\sum_{\ell,m\neq n}{\cal F}^{(2)}_{n,\ell,m}(x)
\ee
but
\be
{\cal F}^{(2)}\ \sim\ \frac{a^4}{D^4}\ \ll\ {\cal F}^{(1)}.
\label{DeltaLast}
\ee
Likewise, the higher orders may involve more and more instantons, but the magnitudes of such high-order interference effects are suppressed by the higher powers of $(a^2/D^2)$.

Now consider the Coulomb self-interaction of the instanton $\#n$,
\be
E_{\rm C}^{\,\rm self}(n)\ =\ \frac{N_c}{4\lambda M}\times\frac{1}{\rho_n^2}\,,
\ee
where $\rho_n$ is the instanton's effective charge radius,
\be
\frac{1}{\rho_n^2}\ =\! \mathop{\intop\!\!\!\!\intop}\limits_{x_1,x_2\,{\rm near}\,X_n}\!\!
d^4x_1\,d^4x_2\,\frac{I(x_1)I(x_2)}{|x_1-x_2|^2}\,.
\ee
A standalone instanton has
\be
\frac{1}{\rho_n^2}\ =\ \frac{4/5}{a_n^2}\,,
\ee
but interference from the other instantons should change this radius by small amount of similar relative magnitude to $\Delta I/I_n^{\rm standalone}$, thus
\be
\frac{1}{\rho_n^2}\ \to\ \frac{4/5}{a_n^2}\ +\ \Delta_n\,,\quad
\Delta_n\ \sim\ \frac{1}{D^2}\,,
\ee
which changes the instanton's Coulomb self-interaction energy by
\be
\Delta E_{\rm C}^{\,\rm self}(n)\ =\ \frac{N_c}{4\lambda M}\times\Delta_n\
\sim\ \frac{N_c}{\lambda M D^2}\,.
\ee
Note that this effect has a similar magnitude to the direct Coulomb repulsion~(\ref{Edirect}) between the instances.

Moreover, the charge radius correction $\Delta_n$ is linear in the $\Delta I^{\rm interference}$ at $x$ near the $X_n$, hence in light of Eq.~(\ref{DeltaIFirstOrder}), the leading-order contribution to the $\Delta_n$ is a sum of pair-wise interferences from the other instantons, thus
\be
\Delta_n\ =\,\sum_{m\neq n}\Delta^{(1)}_{n,m}\ +\ O(a^2/D^4)\,,
\ee
where each $\Delta_{n,m}^{(1)}$ depends only on the instantons $\#n$ and $\#m$. Consequently, the leading effect of the interference on the net Coulomb self-energy of all the instantons has form
\be
\label{ECtwobody}
\begin{split}
\Delta^{\rm int} E_{\rm C}^{\,\rm self}\ &
\equiv\ E_{\rm C}^{\,\rm self}[{\rm interfering}]\ -\ E_{\rm C}^{\,\rm self}[{\rm standalone}]\\
&=\ \frac{N_c}{4\lambda M}\sum_{\textstyle{n,m=1,\ldots,A\atop n\neq m}}\Delta_{n,m}^{(1)}\
+\ O\left(\frac{N_c}{\lambda M}\times\frac{a^2}{D^4}\right)\,,
\end{split}
\ee
where the leading terms act as two-body interactions between the instantons.

Besides the Coulomb energy, the non-abelian energy is also affected by the interference between the instantons,
\be
\Delta^{\rm int} E_{\rm NA}\
=\ N_c\lambda M^3 \int d^4 x\ {\bf x}^2\times  \Delta I^{\rm interference}(x),
\label{DENA}
\ee
but this time the integral should be taken over the whole four-dimensional space, including both the instantons and the inter-instanton space. Indeed, one can show that\cite{Kaplunovsky:2013iza}
\be
\label{DImagnitude}
\begin{array}{l}
\text{at $O(a)$ distance from an $X_n$,}\quad  \Delta I\ \sim\ \frac{1}{a^2 D^2}\,,\\
\\
\text{at $O(D)$ distances from {\it all} the $X_m$,}\quad
\Delta I\ \sim\ \frac{a^4}{D^8}\,,
\end{array}
\ee
and both kinds of places make $O(a^4/D^2)$ contributions to the integral~(\ref{DENA}). Moreover, in both kinds of places, the leading terms in the $\Delta I(x)$ is a sum of  independent two-instanton interference terms,
\be
\Delta I^{\rm interference}(x)\
=\ \frac12\sum_{n\neq m}{\cal I}^{(2)}_{n,m}(x)\ +\ {\rm subleading}.
\label{Idecomp}
\ee

Although our heuristic argument (based on $\Delta I^{\rm interference}\ll I^{\rm standalone}$) for such decomposition near instanton centers does not work in the inter-instanton space, we shall show later in this section that the decomposition~(\ref{Idecomp}) works everywhere in the four--dimensional space. Therefore, the non-abelian interactions between the instantons due to interference are also dominated by the two-body terms.

Let us now relax the un-necessary assumption of intermediate-range distances between the baryons and consider what happens at longer distances $D\sim1/M$. In this regime, the curvature of the fourth space dimension --- and especially the $z$--dependence of the five-dimensional gauge coupling~(\ref{HoloGZ}) --- can no longer be treated as a perturbation. Consequently, the magnetic fields of an instanton or a multi-instanton system
are no longer self--dual in the inter-instanton space. Therefore, the interference between very distant instantons is no longer governed by the
self-dual ADHM solutions. Instead, we must work it out the hard way: figure out how the magnetic fields of an instanton propagate through the curved $4+1$ dimensions towards the other instantons, and then find out how such fields disturb the other instantons' cores.

Fortunately, we do not need a hard calculation to see that at large distances from an instanton its magnetic fields are very weak. Indeed, even in flat five--dimensional space the fields weaken with distance as $A^a_\mu\sim a^2/r^3$ (in the IR-safe singular gauge), so at $r\gg a$ they are so weak that the field equations become effectively linear. In the curved space, we may decompose the weak five--dimesional gauge fields into four--dimensional mesonic fields, hence at large distances $r\gtrsim 1/M$ from an instanton, its fields become Yukawa--like
\be
A^a_\mu(r,z)\ \sim\ \frac{a^2}{r^3} \sum_k \Psi_k(z)\Psi_k(z_{\rm inst}) e^{-m_k r}\ \ll\ \frac{a^2}{r^3}\,.
\label{WeakFields}
\ee
Consequently, at the location $X_n^\mu$ of any particular instanton, the background fields from the other instantons are very weak, and their effect on the instanton $\#n$ itself can be adequately accounted by the first-order perturbation theory. In other words, the effects of other instantons $\#m\neq n$ on the instanton $\#n$ are weak and add up linearly. For the self-interaction energy of the $n^{\rm th}$ instanton, this means
\be
E(n)\ \approx\ E({\rm standalone})\ +\,\sum_{m\neq n}\Delta_m E(n)\,,
\ee
where the second term gives rise to two-body interaction energies
\be
E^{(2)}_{\rm interference}(n,m)\ =\ \Delta_m E(n)\ +\ \Delta_n E(m).
\ee

We expect the two-body terms to be rather small: in addition to the usual $1/|X_m-X_n|^2$ factors from five dimensions they should carry Yukawa-like exponentials $e^{-mr}$ (or rather sums of such exponentials), but these are the leading interactions due to interference. The three-body or multi-body interactions follow from higher-order perturbation by very weak fields~(\ref{WeakFields}), so they are much smaller than the two-body interactions.

As to the direct Coulomb interactions between the holographic baryons via electric or scalar fields, at large distances, $|X_m-X_n|\sim1/M$, they also
decompose into sums of Yukawa forces. Moreover, since the scalar mesons generally have different masses from the vector mesons, the attractive potential due to scalars may have a different $r$ dependence from the repulsive potential due to vectors. Thus, for a right model, the net two-body force between two holographic baryons may become attractive at large enough distances between them. But regardless of the model, the direct Coulomb interactions are always manifestly two-body for $N_f=2$, while for $N_f\ge3$ the multi-body terms exist but become very small at large distances between the baryons.

Thus,
\be
E_{\rm direct}\ =\ \frac12\sum_{m\neq n} \mathop{\rm overlap}(m,n)\times
V^{(2)}(|X_m-X_n|)\,,
\ee
where the precise form of the potential $V^{(2)}(r)$ is model-dependent, but the two-body form of the direct interactions is quite universal.

To conclude this part, let us demonstrate how the multi-body expansion works by evaluating the non-abelian interaction energy of an $A$--instanton system. Due to the nice tricks with integration by parts of the instanton density in Eqs.~(\ref{Moments0})--(\ref{Moments2}) the expansion can be worked out quite easily. This is not the case with the Coulomb energy and we refer the reader to Ref.~\refcite{Kaplunovsky:2013iza} for the full details of the proof.

We would like to implement the large distance expansion for the solution of the ADHM equations for the off-diagonal elements $\Gamma_{mn}^\mu\equiv \alpha^{\mu}_{mn}$, see section~\ref{sec:StraightChain}. Indeed for large distances $|X_m-X_n|\sim D$ between the instantons, $D\gg a_n$, we may solve the ADHM equations, together with constraints~(\ref{ADHMsols}) as a power series in $a^2/D^2$:

\be
\label{AlphaExpansion}
\alpha^\mu_{mn}\ =\ \alpha^{(1)}_{\mu mn}\ +\ \alpha^{(2)}_{\mu mn}\ +\ \alpha^{(3)}_{\mu mn}\
+\ \cdots,
\ee
where
\begin{align}
\alpha^{(1)}_{\mu mn}\ &
=\ \frac{\eta^a_{\mu\nu}\,(X_m-X_n)_\nu}{|X_m-X_n|^2}\times
	\frac12\, a_ma_n\,\tr\bigl(y_m^\dagger y_n^{}(-i\tau^a)\bigr)\
=\ O(a^2/D),\\
\alpha^{(2)}_{\mu mn}\ &
=\ -\frac{\eta^a_{\mu\nu}\,(X_m-X_n)_\nu}{|X_m-X_n|^2}\times
	\sum_{\ell\neq m,n}\eta^a_{\kappa\lambda}
		\alpha^{(1)}_{\kappa \ell m}\alpha^{(1)}_{\lambda \ell n}\
=\ O(a^4/D^3),\\
\alpha^{(3)}_{\mu mn}\ &
=\ -2\frac{\eta^a_{\mu\nu}\,(X_m-X_n)_\nu}{|X_m-X_n|^2}\times
	\sum_{\ell\neq m,n}\eta^a_{\kappa\lambda}
		\alpha^{(1)}_{\kappa \ell m}\alpha^{(2)}_{\lambda \ell n}\
=\ O(a^6/D^5),
\end{align}
and so on. Note that for each off-diagonal matrix element, the leading term $\alpha^{(1)}_{\mu mn}$ in this expansion depends only on the instantons $\#m$ and $\#n$ (that is on the positions, radii, and orientations of only these two instantons), while the subleading terms $\alpha^{(2)}_{\mu mn},\alpha^{(3)}_{\mu mn},\ldots$ involve additional instantons.

Let us plug this expansion in the expression for the second moment of the instanton number density, which computes the correction to the non-abelian energy
\be
\Delta E_{\rm NA} \ = \  \sum\limits_{\mu=1}^4\Delta E_{\rm NA}^{(\mu)}\,,
\ee
where the sum goes over the contributions of all dimensions and
\begin{align}
\Delta E_{\rm NA}^{(\mu)}\ &
=\ \lambda N_c M M_\mu^2\int d^4 x\,I(x)\times \bigl(x^\mu\bigr)^2\ 
=\ \lambda N_c M M_\mu^2\Bigl(\tr\bigl(\Gamma^\mu\Gamma^\mu\bigr)\,+\,\frac12\,\tr\bigl(T\bigr)\Bigr)
\nonumber\qquad\\
&=\ \lambda N_c M M_\mu^2\left(
	\sum_{i=n}^A\Bigl(\bigl(\Gamma^\mu_{nn}\bigr)^2\,+\,\frac12\, T_{nn}\Bigr)\
	+\,\sum_{m\neq n}\bigl(\Gamma^\mu_{mn}\bigr)^2
	\right)\nonumber
\label{ENAprelim}\\
&=\ \lambda N_c M M_\mu^2\sum_n\Bigl(\bigl(X^\mu_n\bigr)^2\,+\,\frac12\, a_n^2\Bigr)\
	+\ \lambda N_c M M_\mu^2\sum_{m\neq n}\bigl(\alpha^\mu_{mn}\bigr)^2\,.
\end{align}
The above correction appears due to curvature $M_\mu^2$ in the direction $x^\mu$, \emph{cf.} Eq.~(\ref{GeneralCoupling}) for the five-dimensional coupling. Obviously, the first sum on the last line here is the sum of individual instantons' potential energies due to their radii and locations (relative to the $x^\mu=0$ hyperplane), while the second sum comprises the interactions between the instantons.

Irrespectively of the details of the instanton configuration, the small instanton expansion~(\ref{AlphaExpansion}) tells us that the interaction term in Eq.~(\ref{ENAprelim}) will contain only the two-body terms in the leading order in $a^2/D^2$. The leading order term is $O(\lambda N_c M^3 a^4/D^2)$, while the multi-body term will be $O(\lambda N_c M^3 a^6/D^4)$, or weaker. Specifically,
\begin{align}
\Delta E_{\rm NA}^{(\mu)}(\text{2-body};m,n)\ &
=\
\frac{\lambda N_c MM_\mu^2 a_m^2 a_n^2}{2|X_m-X_n|^2}
	\tr^2\Bigl( y_m^\dagger y_n\,(-i\vec N_{mn}\cdot\vec\tau)\Bigr),
\label{ENA2body}
\end{align}
where $\vec N_{mn}$ is the 3-vector, which contains all but $\mu$-th component of the unit 4-vector
\be
N^\nu_{mn}\ \equiv\ \bigl(N^1_{mn},N^2_{mn},N^3_{mn},N^4_{mn}\bigr)\
=\ \frac{X_n^\nu-X_m^\nu}{|X_n-X_m|}\,.
\label{Ndef}
\ee

Let us also quote the result for the Coulomb energy obtained through a \emph{tour de force} calculation in Ref.~\refcite{Kaplunovsky:2013iza}. In the multi-body expansion of the net Coulomb energy,
\be
E_{\rm C}^{\rm total}\
=\,\sum_n E_{\rm C}^{\rm 1\,body}(n)\ +\ \frac12\,\sum_{m\neq n} E_{\rm C}^{\rm 2\,body}(m,n)\ +\ \cdots,
\ee
the one-body and two-body terms are found to be
\begin{align}
E_{\rm C}^{\rm 1\,body}(n)\ &
=\  \frac{N_c}{5\lambda M}\times\frac{1}{a_n^2}\,,
\label{Eonebody}\\[5pt]
E_{\rm C}^{\rm 2\,body}(m,n)\ &
=\ \frac{N_c}{2\lambda M}\frac{1}{|X_m-X_n|^2}
\left(1\, +\,\frac{1}{5}\left(\frac{a_m^2}{a_n^2}+\frac{a_n^2}{a_m^2}\right)
	\bigl(\tr^2(y_m^\dagger y_n^{})-2\bigr)
	\right),\label{Etwobody}
\end{align}

For small instantons distant from each other, the one-body potential energies for the instanton radii $a_i$ are much larger than the two-body \emph{etc.} interactions between different instantons. Consequently, in the minimal-energy or near-minimal-energy configuration of the multi-instanton system, the instanton radii will be close to the equilibrium radius of a stand-alone instanton,
\be
a_n\ =\ a_0\ +\ O(a^3/D^2)z\,,
\ee
In particular, for the most general case here of non-zero $M_2$, $M_3$ and $M_4$, \emph{cf.} Eq.~(\ref{equiradius}),
\be
a_0^4\ =\ \frac{2/5}{\lambda^2 M^2(M_4^2+M_3^2+M_2^2)}\,,
\label{InstantonRadiusM}
\ee
Plugging this equilibrium radius into the combined non-abelian and Coulomb two-body interaction energy one obtains
\begin{multline}
\Delta E_{\rm net}^{(2)}(m,n)\
=\ \frac{2N_c}{5\lambda M}\,\frac{1}{|X_m-X_n|^2}\left(
	\frac12\,+\,\tr^2\bigl(y_m^\dagger y_n^{}\bigr)\right.
\\ \left.
	 +\sum_{\mu=2,3,4}C_\mu\,
		\Bigl(\eta^a_{\mu\nu} N^\nu_{mn}\,\tr\bigl( y_m^\dagger y_n^{}(-i\tau^a)\bigr)\Bigr)^2
	\right),
\label{E2bodyM}
\end{multline}
where
\be
C_\mu\ =\ \frac{M_\mu^2}{M_4^2+M_3^2+M_2^2}\,,\quad
C_4\,+\,C_3\,+\,C_2\ =\ 1.
\label{Cdef}
\ee
For the original holographic setting with $M_{1,2,3}=0$, while $M_4=M$, this formula simplifies to the Kim and Zahed formula:\cite{Kim:2008iy}
\be
E^{\rm 2\,body}(m,n)\
=\ \frac{2N_c}{5\lambda M}\frac{1}{|X_m-X_n|^2}\left[
	\frac12
	+ \tr^2\Bigl(y_m^\dagger y_n^{}\Bigr)
	+ \tr^2\Bigl(y_m^\dagger y_n^{}\,(-i\vec N_{mn}\cdot\vec\tau)\Bigr)
	\right].
\label{KZ}
\ee
Note that the expression inside the square brackets is always positive, so the two-body forces between the instantons are always repulsive, regardless of the instantons' $SU(2)$ orientations. However, the orientations do affect the strength of the repulsion: two instantons with similar orientations repel each other nine times stronger then the instantons at the same distance from each other but whose orientations differ by a $180^\circ$ rotation (in $SO(3)$ terms) around a suitable axis. This fact will be at the core of our analysis of instanton crystals in the remainder of this section.


\subsection{Phases and transitions}
\label{sec:phasediagram}

The perturbative analysis of instanton solutions and their interaction energies, based on the multi-body expansion discussed above, allows to vastly expand the search of available phases of one-dimensional lattices (both straight chains and zigzags). In particular, numerical algorithms can be implemented for a search of the ground state for a given set of parameters. Indeed we do not know \emph{a priori} which choice of instantons' moduli parameters, such as positions and orientations, will minimize the net energy. Actual realization of the exact solutions presented in section~\ref{sec:exactsolns} so far can only be conjectured. In view of a large number of possibilities the right strategy seems to be to start from a random values of the moduli and let them evolve and relax numerically to a local energy minimum. Comparing the energies of the local minima a true ground state can be selected.

The strategy outlined here was partially realized in Ref.~\refcite{Kaplunovsky:2013iza}, where the positions of instantons was assumed to fill a one--dimensional chain, or a zigzag, but the orientations were allowed to take any values in the numerical simulations. The numerical relaxation have indicated a set of minimum energy configurations, from which the true ground states were selected and preliminary phase diagrams were obtained. Given the ground states, the analytical ansatzes were constructed for a more accurate detection of the location of the phase boundaries. Let us review the main results of that investigation.

Following Ref.~\refcite{Kaplunovsky:2013iza}, we assume that the holographic direction $z\equiv x^4$, and there is an additional anisotropic curvature in the directions $x^2$ and $x^3$ so that the instantons are stabilized in a one--dimensional array along the direction $x^1$. This corresponds to the choice of five--dimensional gauge coupling as in Eq.~(\ref{GeneralCoupling}) with $M_4\equiv M$. We will also assume $M_2\ll M_4$, while $M_3$ will be allowed to take any values $M_2\leq M_3\leq M_4$. In such a setup the instantons will line--up along $x^1$ at low densities and the leading instability will be the zigzag deformation in the $\pm x^2$ direction. We will also see that the lowest energy configuration will be sensitive to the anisotropy ration $M_3/M_4$.

First, let us review various possibilities for the straight chain,
\be
X_n^\mu\ =\ (nD,0,0,0),\quad n\in\Z\,,
\label{chain1D}
\ee
where we will assume that the lattice spacing $D$ is much larger than the instanton radius $a$. The curvature creates the following effective one-body potential for the instanton centers $X_n$,
\be
\Delta E^{(1)}(n)\ =\ \lambda N_c M\left(M_4^2(X_n^4)^2\
+\ M_3^2\,(X_n^3)^2\ +\ M_2^2\,(X_n^2)^2\right)
+\ O(N_c\lambda M^5 X^4).
\label{ChainPotential}
\ee

First, consider the regime $M_2\lesssim M_3\ll M_4$. In this limit, the $M_2^2$ and $M_3^2$ parameters give rise to the $x_2^2$ and $x_3^2$ terms in the one-body instanton potential~(\ref{ChainPotential}), but their effect on the instanton radius $a$ or the two-body forces between the instantons may be neglected (compared to the effect of the $M^2_4x_4^2$ term). Therefore, the net two-body forces between the instantons remain approximately as
in Eq.~(\ref{KZ}), which for the one--dimensional lattice geometry~(\ref{chain1D}) of the instanton centers $|X_m-X_n|^2=D^2(m-n)^2$ and for $\vec{N}_{mn}\equiv(N_{mn}^1,N_{mn}^2,N_{mn}^3)=(\pm1,0,0)$ may be summarized as
\be
\Delta E_{\rm int}^{(2)}\ =\ \frac{N_c}{5\lambda M D^2}\sum_{m\neq n} \frac1{(m-n)^2}\left[ \frac12\, +\,\tr^2\bigl(y_m^\dagger y_n^{}\bigr)\,
	+\,\tr^2\bigl(y_m^\dagger y_n^{}(-i\tau_1)\bigr)
	\right]\,.
\label{Echain}
\ee

As we noted in the end of section~\ref{sec:twobody}, to minimize this energy, each pair of instantons $m$ and $n$ wants to have the relative orientation being a rotation through a $180^\circ$ angle, in $SO(3)$ terms, around some axis perpendicular to $x^1$. In other words it wants $y_m^\dagger y_n^{}$ to be a linear combination of $i\tau_2$ and $i\tau_3$. However, this cannot be achieved for all instanton pairs at once: minimizing the energies of the $(n,m)$ and the $(n,\ell)$ pairs will not produce a minimum of $(m,\ell)$. Instead, we will first minimize the energies of the nearest--neighbors and then look what happens with the less expensive pairs of instantons.

Thus, we want
\be
\label{Condition}
\forall m:\quad
y_m^\dagger y_{m+1}^{}\ =\ \cos\psi_m\,(i\tau_3)\ +\ \sin\psi_m\,(i\tau_2)\quad
\text{for some angle}\ \psi_m\,,
\ee
which has the following most general solution (modulo a common $SU(2)$ symmetry):
\be
\label{BigFamily}
y_n\ =\ \exp\bigl(i\phi_n\tau_1\bigr)\,(i\tau_3)^n\
=\begin{cases}
	\pm[\cos\phi_n\,\mathbbm{1}\,+\,\sin\phi_n\,(i\tau_1)] &
	\text{for even}\ n,\\
	\pm[\cos\phi_n\,(i\tau_3)\,+\,\sin\phi_n\,(i\tau_2)] &
	\text{for odd}\ n,\\
	\end{cases}
\ee
for some  angles $\phi_n$. In particular, $\psi_m=(-1)^m(\phi_{m+1}-\phi_m)$.

The big family of solutions~(\ref{BigFamily}) turn out to be completely degenerate! Indeed, for all sets of $\phi_n$, all instanton pairs $(m,n)$ with odd $m-n$ have minimal energies, while pairs with even $m-n$ have maximal energies. Consequently, regardless of the angles $\phi_m$, the net energy~(\ref{Echain}) per instanton of the chain is
\be
\frac{E_{\rm int}}{(\text{1-inst})}\ =\ \frac{N_c}{5\lambda MD^2}\left(
	\frac12\sum_{{\rm odd}\,\ell}\frac1{\ell^2}\,
	+\,\frac92\sum_{{\rm even}\,\ell\neq0}\frac1{\ell^2}
	\right)\ = \ \frac{\pi^2N_c}{10\lambda MD^2}\,.
\ee

In a generic lowest-energy configuration~(\ref{BigFamily}), the instanton orientations span an $SO(2)\times\Z_2$ subgroup of the $SO(3)\cong SU(2)\times \Z_2$ of the $y_n$, which corresponds to the rotational symmetries of a cylinder --- rotations through arbitrary angles around the $x_1$ axis, and $180^\circ$ rotations around axes perpendicular to $x_1$. The $y_n$ alternate between the two types of rotations, but apart from that they generically do not follow any regular patterns.

However, the family (\ref{BigFamily}) also contains some regular patterns in which the orientations $y_n$ (modulo sign) follow a repeating cycle of finite length $p$; moreover, the values of $y_n$  span a discrete subgroup of the cylindrical symmetry $SO(2)\times\Z_2$. Here are some examples:

\begin{itemize}
\item The \emph{anti-ferromagnetic chain}, with two alternating instanton orientations:
\be
y_{{\rm even}\,n}\ =\ \pm\mathbbm{1},\quad y_{{\rm odd}\,n}\ =\ \pm i\tau_3\,.
\label{AntiFerro}
\ee
In this configuration, for $\phi_n\equiv0$, the $y_n$ (modulo sign) span a $\Z_2$ subgroup of the $SO(2)\times\Z_2$.

\item Period--four configuration spanning the \emph{Klein group} of $180^\circ$ rotations around the three Cartesian axes:\footnote{%
	As a subgroup of $SO(3)$, the Klein group is abelian and isomorphic to $\Z_2\times \Z_2$.
	But its covering group in $SU(2)$ is non-abelian group and isomorphic
	to the group of unit quaternions $\pm 1,\pm i,\pm j,\pm k$.
	}
\be
\begin{array}{ccc}
y_{n\equiv0\!\!\!\pmod4}\ =\ \pm1,& \quad &
y_{n\equiv1\!\!\!\pmod4}\ =\ \pm\tau_3,\\
y_{n\equiv2\!\!\!\pmod4}\ =\ \pm\tau_1,& \quad &
y_{n\equiv3\!\!\!\pmod4}\ =\ \pm\tau_2.
\end{array}
\label{Klein}
\ee

\item Period--$2k$, $2k=6,8,10,\ldots$ configurations spanning \emph{prismatic groups} $\Z_k\times\Z_2$:
\be
\label{Prismatic}
\begin{split}
y_{{\rm even}\,n}\ &
=\ \cos\frac{\pi n}{2k}\,\mathbbm{1}\ +\ \sin\frac{\pi n}{2k}\,(i\tau_1),\\
y_{{\rm odd}\,n}\ &
=\ \cos\frac{\pi(n-1)}{2k}\,(i\tau_3)\ +\ \sin\frac{\pi(n-1)}{2k}\,(i\tau_2).
\end{split}
\ee

\item Period--$2k$, $2k=6,8,10,\ldots$ configurations spanning \emph{dihedral groups} $D_{2k}$, for $\phi_n=n\times(\pi/2k)$, \emph{i.e.}
\be
\label{Dihedral}
\begin{split}
y_{{\rm even}\,n}\ &
=\ \cos\frac{\pi n}{2k}\,\mathbbm{1}\ +\ \sin\frac{\pi n}{2k}\,(i\tau_1),\\
y_{{\rm odd}\,n}\ &
=\ \cos\frac{\pi n}{2k}\,(i\tau_3)\ +\ \sin\frac{\pi n}{2k}\,(i\tau_2).
\end{split}
\ee
\end{itemize}

There is a wider class of regular configurations, we shall call them link-periodic, in which the $y_n$ themselves are not periodic, but the relative rotations $y_n^\dagger y_{n+1}^{}$ between nearest neighbors follow a periodic pattern. In terms of the $\phi_n$ angles, this corresponds to periodic differences $\phi_{n+1}-\phi_n$, for example
\be
\phi_n\ =\ n\varphi\ -\ \frac{(-1)^n}2\,\theta\quad\Longrightarrow\quad
\phi_{n+1}\,-\,\phi_n\ =\begin{cases}
	\varphi+\theta & \text{for even}\ n,\\
	\varphi-\theta & \text{for even}\ n.
	\end{cases}
\label{LinkPeriodic}
\ee
For rational $\varphi/\pi$ this pattern produces a periodic array of instantons' orientations $y_n$, \emph{e.g.} the dihedral cycle for $\varphi=\pi/2k$ and $\theta=0$, or the prismatic cycle for $\varphi=\pi/2k$ and $\theta=-\varphi/2$. For irrational $\varphi$'s the orientations $y_n$ themselves do not have a finite period; instead, they wind irrationally around the cylinder group $SU(2)\times\Z_2$.

The large degeneracy of the one--dimensional configurations of instantons can be lifted if the condition $M_2\ll M_4$ is relaxed. For $M_2,M_3\sim M_4$, there is a unique lowest-energy straight chain configuration, namely the link-periodic array~(\ref{LinkPeriodic}), whose $\varphi$ and $\theta$ parameters depend on the $M_2/M_3$ ratio. To see this one can use Eqs.~(\ref{ENAprelim}) and~(\ref{E2bodyM}) for the one-body and two-body interaction energies in section~\ref{sec:twobody}.

Due to a larger number of parameters it is less trivial to minimize the interaction energy. In particular, it is not intuitively obvious how to balance the energy cost of nearest-neighbor interactions versus next-to-nearest neighbors and more distant instanton pairs. To find the ground state, in Ref.~\refcite{Kaplunovsky:2013iza} a numerical experiment was performed using a lattice of
200 $SU(2)$ matrices $y_n$. In each run, $y_n$'s were assigned some initial random values, and then they were alowed to evolve towards a
minimum of the energy function (\ref{E2bodyM}) via the relaxation method.

The bottom line of the numerical simulations in Ref.~\refcite{Kaplunovsky:2013iza} is that the ground state of the one--dimensional instanton lattice always has a link-periodic instanton orientations~(\ref{LinkPeriodic}) (up to a global symmetry, if any) for a periodicity angle $\varphi$ that depends on the ratio of parameters $M_2/M_3$ and not on the $M_4$ (as long as $M_4\ge M_3\ge M_2$).

Once the ground state is known the exact dependence of the angle $\varphi$ on the parameters $M_2$ and $M_3$ can be calculated analytically. Using Eq.~(\ref{LinkPeriodic}) for the $y_m^\dagger y_n$ and the expression for the energy~(\ref{E2bodyM}) one finds
\begin{multline}
\frac{E_{\rm int}}{(\text{1-inst})}\
= \ \frac{N_c}{5\lambda MD^2}\left(\frac{\pi^2}{2} \Bigl(1\,+\,C_3\,+\,C_2\,-\,(C_3-C_2)\cos(2\theta)\Bigr)\right.\\
\left. -\ 2\pi|\varphi| \Bigl( C_3\,+\,C_2\,-\,(C_3-C_2)\cos(2\theta)\Bigr)
	+\ 4\varphi^2 \left(C_3\,+\,C_2\right) \right).
\end{multline}
Minimizing this expression with respect to the $\varphi$ and $\theta$ produces four degenerate minima, namely
\be
\vcenter{\openup 1\jot \ialign{
	#\quad\hfil & $\displaystyle{\varphi\ =\ #\,}$,\quad\hfil & $\theta\ =\ #$,\hfil\cr
	(1) & +\frac\pi2\times\frac{C_2}{C_2+C_3} & 0\cr
	(2) & -\frac\pi2\times\frac{C_2}{C_2+C_3} & 0\cr
	(3) & -\frac\pi2\times\frac{C_3}{C_2+C_3} & \tfrac\pi2\cr
	(4) & +\frac\pi2\times\frac{C_3}{C_2+C_3} & \tfrac\pi2\cr
	}}
\label{BestPhi}
\ee
however the last two minima are physically equivalent to the first two. In agreement with the numerical experiment, the one-dimensional instanton lattice has two degenerate ground states related by the $\varphi\to-\varphi$ symmetry.  The value of $|\varphi|$ is also independent from $M_4$.

Two particularly interesting $M_2^2/M_3^2$ ratios need special handling, $M_2^2=M_3^2$ and $M_2^2\ll M_3^2$. For $M_2=M_3$ the background has a rotational symmetry in the $x^{2,3}$ plane. For the instanton chain, this translates into the $U(1)$ symmetry between the $i\tau^2$ and $i\tau^3$ directions in the $SU(2)$. Consequently, instead of two discrete ground states of Eq.~(\ref{BestPhi}) we have a continuous family:
\be
\varphi\ =\ \frac\pi2\,,\quad \theta\ =\ \rm anything.
\ee
For all configurations in this family, the instanton orientations $y_n$ (modulo signs) repeat with period four, while spanning the Klein groups $\Z_2\times\Z_2$ of $180^\circ$ rotations around three mutually perpendicular axes. For $\theta=-\frac\pi4$ the $y_n$ are spelled out in Eq.~(\ref{Klein}); for other values of $\theta$ we have similar cycles related by the $U(1)\subset SU(2)$ symmetry.

Finally, for the very asymmetric background with $M_2^2\ll M_3^2$, the two ground states of Eq.~(\ref{BestPhi}) become indistinguishable as $\varphi\to\pm0$. In this limit, the instanton orientations form the anti-ferromagnetic order~(\ref{AntiFerro}). This is indeed the case we have considered in section~\ref{sec:StraightChain} if we assume $M_3=M_4\equiv M'$ and $M_2\equiv M\ll M'$.

The analysis of the zigzag configuration proceeds similarly. We remind that the original chain was in the $x_1$ direction, while by virtue of the condition $M_2<M_3\leq M_4$ the most likely first step in the transition between one--dimensional and two--dimensional instanton lattices is the zigzag in the $\pm x_2$ direction (Fig.~\ref{fig:zigzag}). The ADHM ansatz for the zigzag appeared in Eqs.~(\ref{GammaZigzag1}), (\ref{SD3mat}) and~(\ref{dX3}). It is parameterized by two parameters $D$ and $\epsilon$; we shall refer to $D$ as the \emph{lattice spacing} and to $\epsilon$ as the \emph{zigzag amplitude}.

In the analysis of the zigzag transition we shall use the small--instanton approximation, $a\ll D$. It was demonstrated in section~\ref{sec:zigzagsoln} that for consistency one need to impose the condition $M_2\ll M_4z\equiv M$. Indeed in such a case the critical spacing for the zigzag transition, will respect the $a\ll D$ approximation, as in Eq.~(\ref{Dcrit}). The parameter $M_3$, instead, is allowed to take any values $M_2<M_3\leq M_4$. Thus, for the search of minimum energy configurations, one can evaluate the net interaction energy per instanton using Eq.~(\ref{E2bodyM}), with negligible $C_2\approx 0$. For a zigzag where all the instantons lie in the $x^{1,2}$ plane and hence all the $N_{mn}^\mu$ have form $N^\mu_{mn}=(*,*,0,0)$, Eq.~(\ref{E2bodyM}) can be rearranged as
\be
\label{E2D1}
\frac{\Delta E_{\rm int}^{\rm net}[{\rm zigzag}]}{(\text{1-inst})}\
=\ \frac{N_c}{5\lambda M}\sum_{m\neq n}\frac{Q_z(m,n)}{|X_m-X_n|^2}\,,
\ee
where
\begin{multline}
\label{E2D2}
Q_z(m,n)\
=\ \frac12\ +\ \tr^2\bigl(y_m^\dagger y_n^{}\bigr)\
+\ C_3\sum_{a=1,2} \tr^2\bigl(y_m^\dagger y_n^{}(-i\tau^a)\bigr)\
\\
+\ (1-2C_3)\,\tr^2\bigl(y_m^\dagger y_n^{}
	(-i\vec\tau\cdot\vec N_{mn})\bigr).
\end{multline}

Note that in general, that the force between two instantons is not central --- it depend not only on the distance between the instantons and their relative orientation $y_m^\dagger y_n^{}$ of their isospins, but also on the direction $\vec N_{mn}$ of their separation $\vec X_m-\vec X_n$ in space.
However, in the background with $M_3=M_4$, which was considered in Ref.~\refcite{Kaplunovsky:2012gb} (albeit in different notations) --- $C_3=1/2$, the last term in Eq.~(\ref{E2D2}) goes away, and the force becomes central (but isospin-dependent). As a result, for $M_3=M_4$, the only minimum energy configurations found in Ref.~\refcite{Kaplunovsky:2012gb}, corresponded to relative orientation of instantons generated by an abelian $U(1)$ in $SU(2)$. For $M_3\neq M_4$, as we shall see, the phase space is richer. Indeed, let us discuss the lowest-energy configuration of the orientations $y_n$ as a function of the $\epsilon/D$ and $M_3/M_4$ ratios.

\begin{figure}[bt]
\centerline{\psfig{file=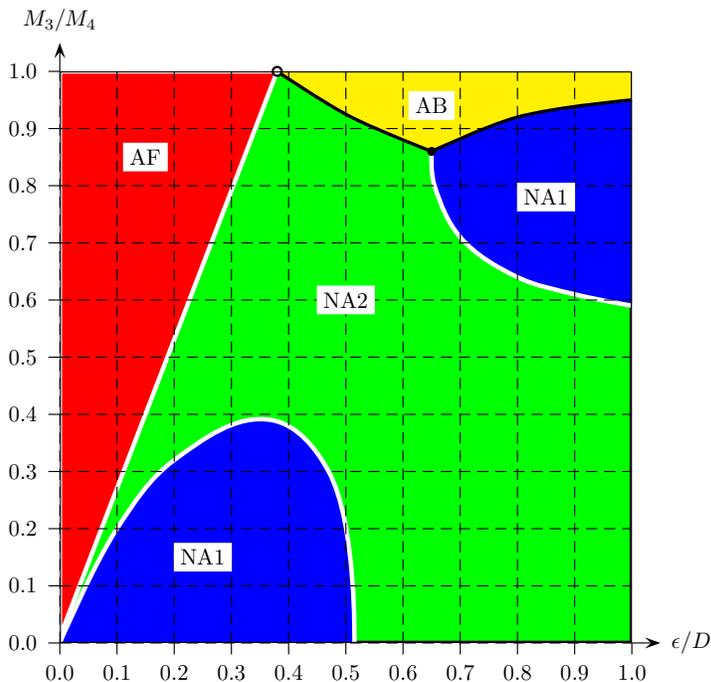,width=3.65in}}
\vspace*{8pt}
\caption{Phase diagram of the instanton orientation patterns in the zigzag parameter space.
	This diagram is obtained from the mostly-analytical calculation in Ref.~\protect\refcite{Kaplunovsky:2013iza}.
	The lines separating different phases indicate the order of the phase transition:
	a white line for a second-order transition and a black line for a first-order transition.}
\label{ZigzagDiagram}
\end{figure}

As in the case of the straight chain, it is hard to say \emph{a priori}, which particular pattern, or patterns the relative orientations $y_n^\dagger y_{n+1}$ of the neighbor instantons will follow. It is thus useful to start with a numerical simulation, starting with completely random $y_n$ and letting them evolve towards a minimum of the energy function~(\ref{E2D1}). In Ref.~\refcite{Kaplunovsky:2013iza} such a simulation for different combinations of the  $\epsilon/D$ and $M_3/M_4$ parameters was performed and the following four patterns of orientation was discovered:
\begin{itemize}
\item the anti-ferromagnetic pattern (AF) of instanton orientations, in which the nearest neighbors always differ by a $180^\circ$ rotation around
the third axis,
\be
\label{RedPattern}
y_n\ =\left.\begin{cases}
	\pm1 & {\rm for\ even}\ n,\\
	\pm i\tau_3 & {\rm for\ odd}\ n,
	\end{cases}\right\},\qquad
{\rm same}\ y_n^\dagger y_{n+1}^{}\ =\ i\tau_3\ {\rm for\ all}\ n\,;
\ee

\item another abelian pattern (AB), in which all nearest neighbors differ by the same $U(1)\subset SU(2)$ rotation, but now the rotation angle is less than $180^\circ$,
\be
\label{YellowPattern}
{\rm same}\ y_n^\dagger y_{n+1}^{}\
=\ \exp\bigl(\tfrac{i}{2}\phi\tau_3\bigr)\ {\rm for\ all}\ n,\quad
0<\phi<\pi\,;
\ee

\item a non-abelian link-periodic pattern (NA1), in which the relative rotation between nearest neighbors is always through a $180^\circ$ angle, but the direction of rotation alternates between two different axes in the $x^{1,2}$ plane: one axis for the odd-numbered instantons and the other for the even-numbered. In $SU(2)$ terms,
\be
\label{BluePattern}
\begin{split}
y_{2k}^\dagger y_{2k+1}^{}\ &
=\ \exp\left(\frac{i\pi}{2}\,\vec n_e\cdot\vec\tau\right)\
=\ i\vec n_e\cdot\vec\tau\
=\ +iA\tau_1\ +\ iB\tau_2\,,\\
y_{2k+1}^\dagger y_{2k+2}^{}\ &
=\ \exp\left(\frac{i\pi}{2}\,\vec n_o\cdot\vec\tau\right)\
=\ i\vec n_o\cdot\vec\tau\
=\ +iA\tau_1\ -\ iB\tau_2\,,
\end{split}
\ee
for some $A,B\neq0$ ($A^2+B^2=1$).

\item another non-abelian link-periodic pattern (NA2). Again, the relative rotation between nearest neighbors is always through a $180^\circ$ angle and the direction of rotation alternates between two different axes. However, this time the two axes no longer lie within the $x^{1,2}$ plane, thus
\be
\label{GreenPattern}
\begin{split}
y_{2k}^\dagger y_{2k+1}^{}\ &
=\ iA\tau_1\ +\ iB\tau_2\ +\ iC\tau_3\,,\\
y_{2k+1}^\dagger y_{2k+2}^{}\ &
=\ iA\tau_1\ -\ iB\tau_2\ -\ iC\tau_3\,,\\
\end{split}
\ee
where $A,B,C\neq0$ ($A^2+B^2+C^2=1$).

\end{itemize}
Figure~\ref{ZigzagDiagram} shows the distributions of the four phases in the parameter space.

Figure~\ref{ZigzagDiagram} also shows the order of the phase transitions between distinct phases. In particular, the transitions between the two non-abelian phases are second-order. (Notice, that the phase non-abelian NA1 appears in two disconnected regions of the diagram, separated by NA2.) Likewise, the transition between the non-abelian NA2 and the abelian anti--ferromagnetic phase AF is also second-order. On the other hand, the transitions between the other abelian AB phase and the phases NA1 and NA2 are first-order. Finally, the triple point at $(M_3/M_4)=1$, $(\epsilon/D)\approx0.38$ between the AF, NA2 and AB phases is \emph{critical}. At that point, the transitions between all three phases are second-order.

Thus far, we considered the zigzag amplitude $\epsilon$ as an independent input parameter. But physically $\epsilon$ is a dynamical modulus whose value follows from minimizing the net energy of the multi--instanton system. In fact, all instanton center coordinates $X_n^\mu$ are dynamical moduli, which raises two problems. First, in some situations the lowest-energy configuration of the instanton lattice may be more complicated than a zigzag or a straight chain. Second, for some lattice spacings $D$, a uniform lattice of any kind --- a straight chain, a zigzag, or anything else --- may be unstable against breaking into domains of different phases with different densities.

By way of analogy, consider a fluid governed by an equation of state such as Van-der-Waals. Formally, this equation allows a uniform fluid to have any density (up to some maximum). But in reality, at sub-critical temperatures one may have a low-density gas or a high-density liquid, but there are no uniform fluids with intermediate densities. If we constrain the overall volume $V$ of some amount of fluid such that its {\it average density}
would fall into the intermediate range, we would not get a uniform fluid; instead, part of the volume would be filled by the higher-density liquid while the other part by the lower-density gas. In the same way, if we fix the overall length $L$ of the $x^1$ axis occupied by some large number $A$ of instantons, their lowest-energy configuration is not necessarily a uniform lattice of some kind. Instead, for some average density $\rho=A/L$ we we would have $L$ split into domains of two different lattices of different densities.

To keep the fluid uniform, one should control its pressure $P$ rather than the volume $V$; consequently, the preferred phase follows from minimizing the free enthalpy $G=E-ST+PV$ rather than the free energy $F=E-ST$. Likewise, for the one-dimensional lattice, we should control the net compression force $\bf F$ along the $x^1$ axis, rather than the net length $L$ or the lattice spacing $D$. Also, we should minimize the free enthalpy of the lattice $G=E-ST+L\bf F$, but since we work at zero temperature all we need is the ordinary enthalpy $H=E+L\bf F$. Equivalently, we may minimize the \emph{non-relativistic chemical potential}%
\be
\hat\mu\ =\ \mu_{\rm rel}\ -\ M_{\rm baryon}\ =\ \frac{G_{\rm non-rel}}{A}\
\xrightarrow[{T=0}]{}\ \frac{E+L{\bf F}}{A}\ =\ \E\ +\ \frac{{\bf F}}{\rho}\,.
\label{ChemPot}
\ee
(We focus on the non-relativistic chemical potential $\hat\mu=\mu-M_{\rm baryon}$ because the relevant scale of $\hat\mu$ would be much smaller than the baryon mass.)

Thus, through the remainder of this section, we are going to impose a compression force $\bf F$ on the multi-instanton system, assuming that the instantons form a uniform zigzag of some lattice spacing $D$ and amplitude $\epsilon$, or a uniform straight chain for $\epsilon=0$, and vary $D$, $\epsilon$ and the orientation moduli of the zigzag to seek the minimum of the chemical potential~(\ref{ChemPot}) for any given combination of $\bf F$ and $M_3/M_4$.

It is convenient to introduce a new variable $\xi=\epsilon/D$ to write the chemical potental as
\be
\hat\mu\
=\ {\bf F}\times D\ +\ N_c\lambda MM_2^2\times D^2\xi^2\ +\ \frac{\pi^2 N_c}{20\lambda M}\times\frac{{\cal F}_m(\xi;M_3/M_4)}{D^2}\,,
\label{CPM}
\ee
where we have plugged the value of the function $\cal F$, computed in appendix~A of Ref.~\refcite{Kaplunovsky:2013iza}, minimized with respect to the orientation angles $\phi_n$:\footnote{In fact there is no explicit formula for the ${\cal F}_m(\xi;M_3/M_4)$ as the minimization of ${\cal{F}}$ with respect to the $\phi$ angle has to be done numerically. Nevertheless, it is fairly easy to calculate numerically both the ${\cal F}_m$ function itself and its derivative $\partial{\cal F}_m/\partial\xi$.\cite{Kaplunovsky:2013iza}}
\be
{\cal F}_m(\xi;M_3/M_4)\ =\ \min_{\phi_n}{\cal F}(\phi_n;(\xi=\epsilon/D),M_3/M_4)\,,
\ee
To find the local extrema of $\hat{\mu}$ with respect to $\xi$ and $D$ one has to solve the equations
\be
\label{Fextrema}
\left(\frac{\partial{\cal F}_m }{\partial D}\right)_\xi\ =\ 0\,,\quad
\left(\frac{\partial{\cal F}_m }{\partial \xi}\right)_D\ =\ 0\,.
\ee

As expected Eqs.~(\ref{Fextrema}) have two branches of solutions, corresponding to the straight chain $\epsilon=0$ and the zigzag $\epsilon>0$. However, for each branch, one may have more than one solution for the lattice moduli $(\xi,D)$. In such a case global minima must be selected upon the evaluation of $\hat{\mu}$ for each solution. Physically the presence of multiple solutions signals that the system is close to a first order phase transition. Indeed, varying the compression force ${\bf F}$ in this region one would typically find that the global minimum jumps from one lattice geometry to another.

\begin{figure}[bt]
\centerline{\psfig{file=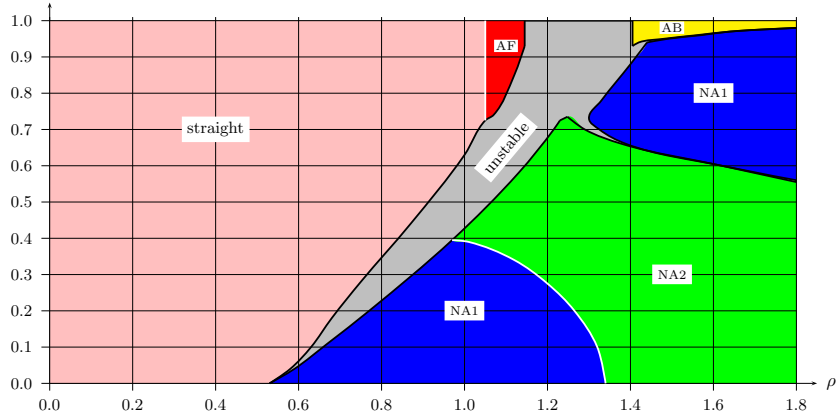,width=0.9\linewidth}}
\vspace*{8pt}
\caption{%
	Phase diagram of the zigzag and straight--chain phases in the linear density \emph{vs.}\ $M_3/M_4$ plane;
	the density $\rho=1/D$ is in units of $\sqrt{\lambda MM_2}$.
	The stable straight-chain phase is colored pink, while the stable zigzag phases are colored
	red, yellow, blue, or green according to the instanton orientation pattern.
	Finally, the gray color denotes densities at which a uniform zigzag or straight chain
	would be mechanically or thermodynamically unstable.
	}
\label{BigPhaseDiagramRho}
\end{figure}

Altogether, there is seven different transition sequences for different $M_3/M_4$ ratios. The transitions can be detected from figures~\ref{BigPhaseDiagramFM} and~\ref{BigPhaseDiagramRho}, which depict the phase diagram of all the zigzag and straight-chain phases in two different planes:\cite{Kaplunovsky:2013iza} the compression force $\bf F$ versus $M_3/M_4$ and the linear instanton density $\rho=1/D$ versus $M_3/M_4$. To complete this section let us summarize a few particularly noteworthy features of these diagrams:

\begin{figure}[bt]
\centerline{\psfig{file=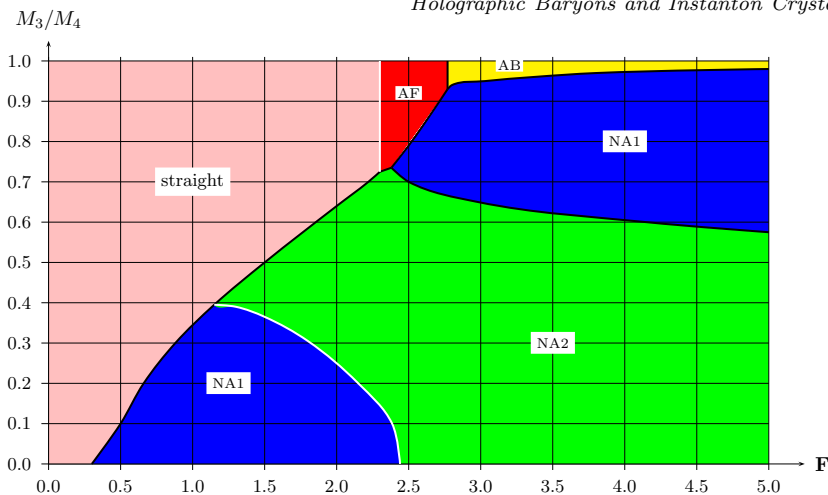,width=0.9\linewidth}}
\caption{Phase diagrams of the zigzag and straight-chain phases in the compression \emph{vs}.\ $M_3/M_4$ plane, $\bf F$ in units of $N_c\sqrt{\lambda MM_2^3}$). The straight-chain phase is colored pink, while four other colors --- red, yellow, blue, and green ---
	denote zigzag phases with different instanton orientation patterns.
	The first-order transition between phases are indicated by black lines, the second-order
	transitions by white lines.}
\label{BigPhaseDiagramFM}
\end{figure}

\begin{itemize}

\item Since we assume $M_2\ll M_3$, the straight-chain phase always has the anti-ferromagnetic order of the instantons' orientations.

\item The very first transition from the straight chain to a zigzag could be either first-order or second-order, depending on the $M_3/M_4$ ratio:
for $(M_3/M_4)<0.725$ the transition is first-order while for $(M_3/M_4)>0.725$ it's second-order. This difference is due to different orientation phases of the zigzag immediately after the transition: for $(M_3/M_4)>0.725$ the zigzag has the same anti-ferromagnetic order as the straight chain,
which allows a second-order transition; but for $(M_3/M_4)<0.725$ the zigzag has a different orientation pattern --- the non-abelian {NA1} or {NA2} --- so the transition is first-order.

\item The non-abelian phases {NA1} and {NA2} of the zigzag cover much larger areas of the phase diagrams on Figs.~\ref{BigPhaseDiagramFM} and~\ref{BigPhaseDiagramRho} than the abelian phases {AF} and {AB}. In particular, at larger compression forces $\bf F$ --- and hence larger chemical potentials $\hat\mu$, larger densities, and larger zigzag amplitudes --- the instanton orientations usually prefer the non-abelian patterns. Only the backgrounds with $M_3\approx M_4$ --- such as the model we have analyzed in Ref.~\refcite{Kaplunovsky:2012gb} --- favor the abelian orientations.

\item Figure~\ref{BigPhaseDiagramRho} has gray areas at which an instanton zigzag with a uniform lattice spacing $D=1/\rho$ and a uniform amplitude $\epsilon$ (or a uniform straight chain for $\epsilon=0$) would be unstable against instantons' motion along the $x^1$ axis (the long direction of the zigzag). If we put $A\gg1$ instantons into a box of fixed length $L=A/\rho$ and let them seek the lowest-energy configuration, they would organize themselves into domains of two different lattices with different lattice spacings and different amplitudes.

\item The {NA1} phase of the zigzag occupies two separate regions of the phase diagram separated by the region of the other non-abelian phase {NA2}.
The phase transition between the lower-left {NA1} region and the {NA2} region is second-order, while the transition between the upper-right {NA1} region and the {NA2} is weakly first-order: the lattice spacing and the zigzag amplitude are discontinuous across the transition,
but the discontinuity is very small and hard to see graphically on figure~\ref{BigPhaseDiagramRho}.

\item Likewise, the transition between the upper-right region of the {NA1} phase and the abelian {AB} phases of the zigzag is weakly first-order. On the other hand, the transitions between the anti--ferromagnetic phases of the straight chain or zigzag and all the other zigzag phases is strongly first-order, with largish discontinuities of the lattice spacing and even larger discontinuity of the zigzag amplitude.

\item However, for small $M_3/M_4$ the discontinuity becomes small; for $M_3/M_4=0$ it vanishes altogether and the phase transition between the straight chain and the {NA1} phase of the zigzag becomes second-order. This is OK because for $M_3/M_4=0$ --- or rather for $M_3=M_2\ll M_4$ --- the instantons' orientations in the straight-chain phase are no longer anti--ferromagnetic but form the period--four Klein-group pattern similar to the {NA1} pattern of the zigzag.

\end{itemize}


\subsection{Beyond Zigzag}
\label{sec:newresults}

Finally let us report some preliminary results of the study of crystalline structures beyond the zigzag. This section summarizes a research in progress, to be published soon.\cite{Kaplunovsky:toappear}

To study the two--dimensional instanton crystals beyond a single zigzag-shaped line, one can run a numeric simulation of a multi-instanton system, where the instantons' locations and orientations are both allowed to evolve seeking the minimum of the net energy. To obtain a two--dimensional crystal with two large dimensions we set $M_3\sim M_4$ but $M_2=0$, which allows the instantons to spread out in two dimensions $(x_1,x_2)$ due to two-body forces~(\ref{E2D1}) and~(\ref{E2D2}). Thanks to homogeneous scaling of these forces with distance the net energy scales like $\rho_2$ with two--dimensional density, so instead of varying the external pressure we have simply confined 1000+ instantons to a large fixed area of the simulator.

A priori, we did not know what kind of lattice the instantons would form: a square lattice, a triangular lattice (tiled by equilateral triangles), or something less symmetric. To our surprise, the simulations produces different lattices for different $M_3/M_4$ ratios:
\begin{itemize}

\item
For high ratios $0.80\lesssim (M_3/M_4)\le1$ the instantons form a square lattice with anti--ferromagnetic orientations, as shown in figure~\ref{fig:SquareAFLattice} (page~\pageref{fig:SquareAFLattice}).
\begin{figure}[bt]
\centerline{\psfig{file=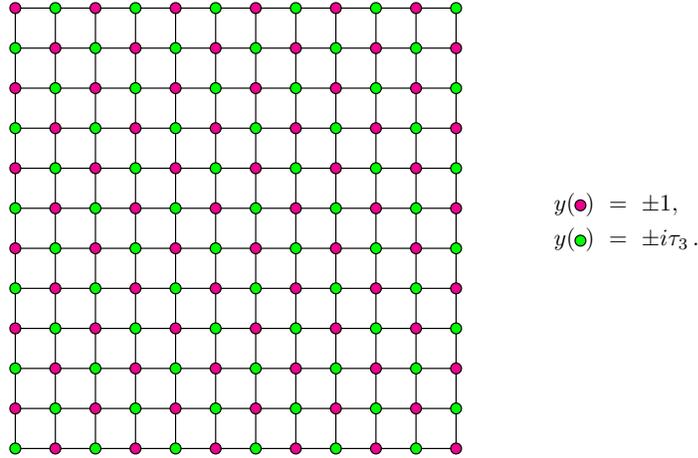,width=3.65in}}
\vspace*{8pt}
\caption{2D anti--ferromagnetic square lattice of instantons.}
\label{fig:SquareAFLattice}
\end{figure}

\item
For low ratios $0<(M_3/M_3)\lesssim0.55$ the instantons also form a square lattice, but with a very different orientation pattern shown in figure~\ref{fig:SquareNALattice} (page~\pageref{fig:SquareNALattice}). We call this orientation pattern \emph{non-abelian} since
unit translations of the lattice in different directions are accompanied by \emph{anticommuting} $SU(2)$ rotations of the instanton orientations.
\begin{figure}[bt]
\centerline{\psfig{file=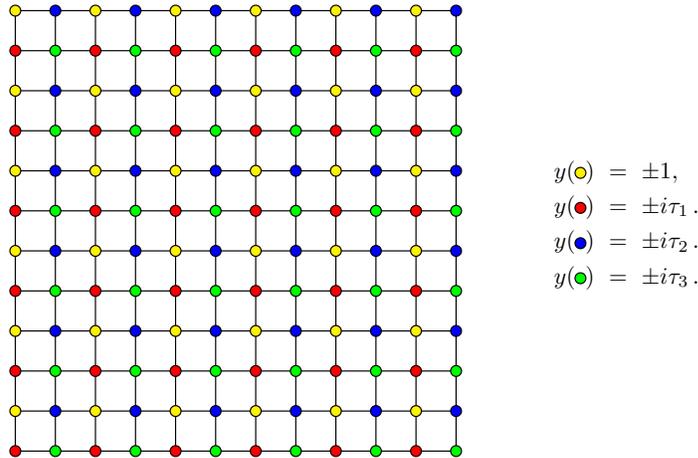,width=3.65in}}
\vspace*{8pt}
\caption{2D nonabelian square lattice of instantons.}
\label{fig:SquareNALattice}
\end{figure}

\item For the medium ratios $0.60\lesssim(M_3/M_4)\lesssim0.75$ the instantons form a triangular lattice (tiled by equilateral triangles), while the orientations form a non--abelian pattern shown in figure~\ref{fig:TriangularLattice} (page~\pageref{fig:TriangularLattice}).
\begin{figure}[bt]
\centerline{\psfig{file=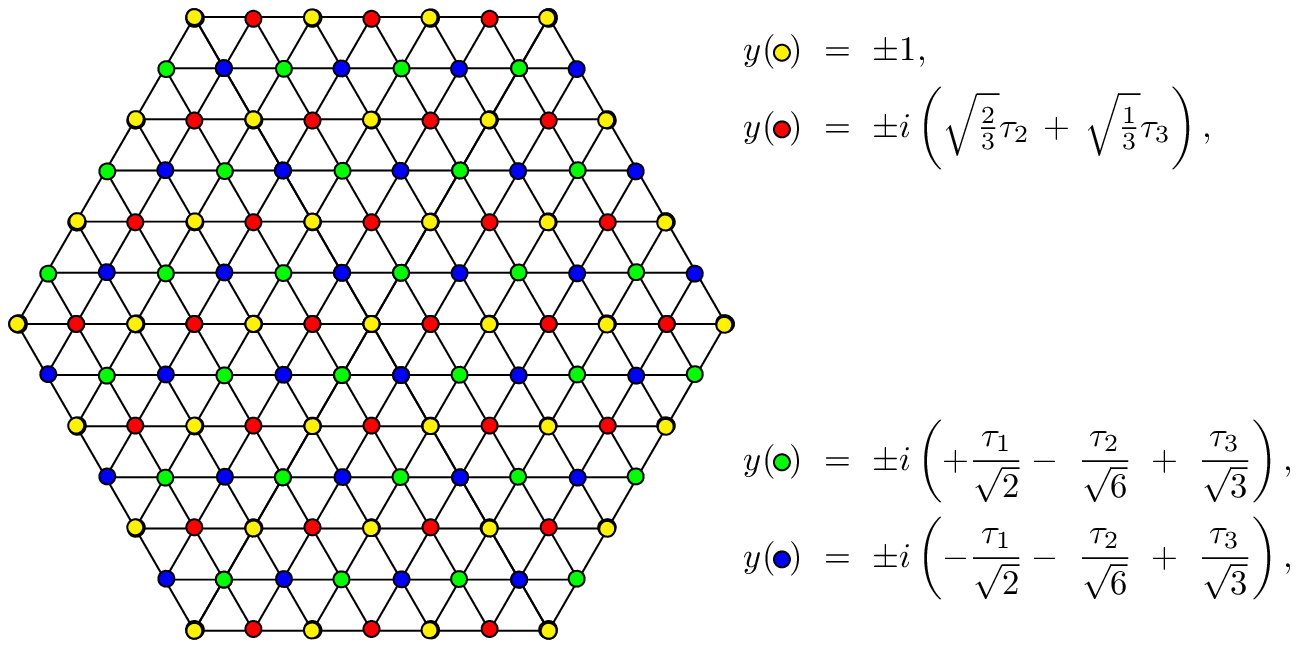,width=0.9\linewidth}}
\vspace*{8pt}
\caption{2D triangular lattice of instantons.}
\label{fig:TriangularLattice}
\end{figure}

\end{itemize}

To be precise, the numeric simulations did not produce clean monocrystalline lattices such as shown on figures~\ref{fig:SquareAFLattice}--\ref{fig:TriangularLattice} but rather polycrystals made of small randomly-oriented pieces of such lattices separated from each other by messy boundary layers. Fortunately, the nature of the ideal monocrystalline lattice was quite clear from such polycrystals for most values of the $(M_3/M_4)$ ratio. However, near the first-order phase transitions between the square and the triangular lattices, \emph{i.e.} for $0.55\lesssim(M_3/M_4)\lesssim0.60$ and $0.75\lesssim(M_3/M_4)\lesssim0.80$,  the simulations produce very confusing polycrystals with both square and
triangular domains. Strictly speaking, they might have also contained domains of some other types we could not identify.

To pinpoint the phase boundaries between the square and triangular lattices --- and also to check for other lattice types --- we have derived analytic formulae for energies (per instanton) of a rather large class of two--dimensional instanton crystals. Specifically, we allowed for simple lattices (one instanton per unit cell) of any geometry, where the orientations of the nearest neighbors are related by two fixed $SU(2)$ twists,
\be
y(m+1,n)\ =\ y(m,n)\times T_1\,\qquad \text{and},\qquad y(m,n+1)\ =\ y(m,n)\times T_2\,,
\ee
with the same $T_1,T_2\in SU(2)$ for all lattice sites $(m,n)$. This includes the abelian orientation patterns with $T_1T_2=+T_2T_1$ and the non-abelian ones with $T_1T_2=-T_2T_1$.

\begin{figure}[bt]
\centerline{\psfig{file=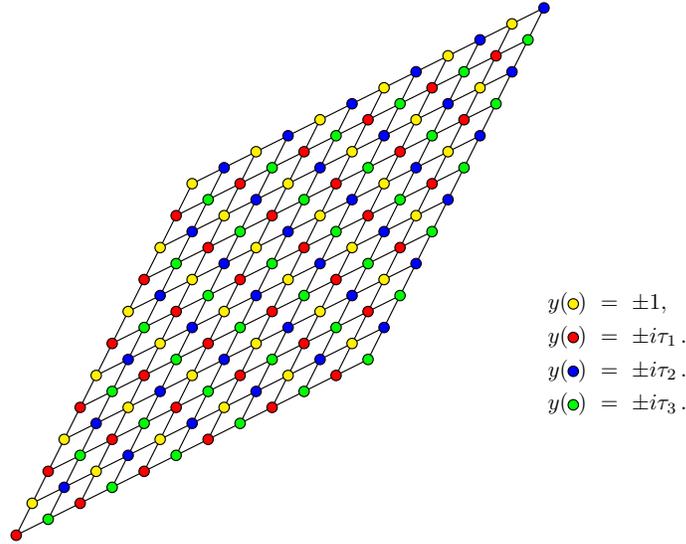,width=3.65in}}
\vspace*{8pt}
\caption{2D rhombic lattice of instantons.}
\label{fig:RhombicLattice}
\end{figure}

\begin{figure}[bt]
\centerline{\psfig{file=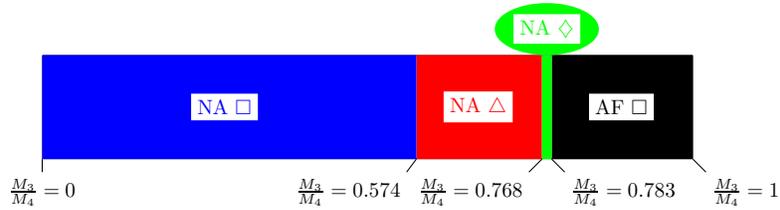,width=0.9\linewidth}}
\vspace*{8pt}
\caption{Phases of 2D crystals for different anisotropy ratios $M_3/M_4$.}
\label{fig:2Dphases}
\end{figure}

Minimizing the net energy per instanton as a function of lattice geometry and the orientation twists $T_1$ and $T_2$, we found \emph{four} distinct phases for different $M_3/M_4$ ratios: the three lattices we saw in the numerical simulations, plus the non-abelian rhombic phase which obtains in a narrow range of $0.768<(M_3/M_4)<0.783$. As shown on figure~\ref{fig:RhombicLattice} (page~\pageref{fig:RhombicLattice}), this lattice has similar instanton orientations to the non--abelian square lattice, but the unit cell is deformed from a square to a rhombus with axis ratio about 2.6.
The phase diagram for the four lattice types along the $(M_3/M_4)$ axis is shown in figure~\ref{fig:2Dphases}.

This completes our presentation of infinite two--dimensional crystals. The next subject concerns the \emph{thin instanton crystals}, which are infinitely long in the $x_1$ direction but have only a few layers in the $x_2$. The transition between a one- and a two--dimensional crystals goes through a sequence of such thin crystals, and back in Ref.~\refcite{Kaplunovsky:2013iza} and Sec.~\ref{sec:phasediagram} above, we have assumed that the first step in this sequence away from a straight one--dimensional chain was the zigzag. Obviously, we need to check this assumption and to investigate the other thin crystal phases beyond the zigzag.

The first tool for this task was again a numerical simulation, in which both instanton locations and their orientations were allowed to evolve as they would seek the net energy minimum. This time, we have turned on a small $M_2\neq 0$ (but $M_2\ll M_{3,4}$), so at low densities the instantons would form a one-dimensional chain along the $x_1$ axis, but at higher densities they would spread out in the $x_2$ direction. We did not prejudice the manner of such spread-out but let the instantons find the best way. We allowed the instantons to move in both $x_2$ and $x_1$ directions, but in the $x_1$ we put a movable wall subject to a controlled pressure. Indeed, as discussed in Sec.~\ref{sec:phasediagram}, controlling the one-dimensional pressure $\bf F$, rather than the one dimensional $L$, allows the instantons to form a uniform lattice for any $\bf F$.

A few thousands of such simulations for different pressures and $(M_3/M_4)$ ratios gave us a crude phase diagram of thin instanton crystals. Some features of this diagram were quite surprising:
\begin{itemize}
\item The abelian zigzag phase does not exist! Its appearance near the top of the phase diagrams in Ref.~\refcite{Kaplunovsky:2013iza} and in Figs.~\ref{ZigzagDiagram}--\ref{BigPhaseDiagramRho} in Sec.~\ref{sec:secondpaper} is an artefact of presuming the instanton centers form a zigzag when they actually prefer a different configuration --- a thin slice of a two--dimensional anti--ferromagnetic square lattice.

\item Even beyond the zigzag, there are no abelian phases with twist angles $\phi\neq\pi$. The only abelian orientation pattern for an instanton crystal --- thick or thin --- is the antiferromagnet.

\item While there is a whole series of anti--ferromagnetic thin crystals for $(M_3/M_4)\gtrsim0.8$, the anti--ferromagnetic zigzag phase does not exist!
Instead, there is anti--ferromagnetic period--three wave phase
$$
\centerline{\psfig{file=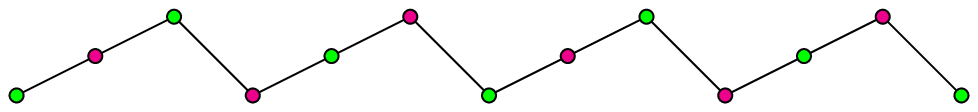,width=3.65in}}
$$
At $(M_3/M_4)\gtrsim0.7$, increasing pressure causes a second-order transition from a single straight line to this wave, followed by a first-order transition to two parallel lines of instantons with two-dimensional anti--ferromagnetic order of orientations
$$
\centerline{\psfig{file=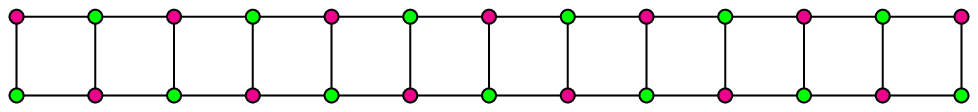,width=3.65in}}
$$

\item On the other hand, the non--abelian zigzag phases NA1 and NA2 do exist.

\end{itemize}

Not so surprisingly, at pressures beyond the AF wave or NA zigzag phases, the instantons form $n=2,3,4,5,\ldots$ parallel layers which act as a thin slice of an infinite two--dimensional lattice, with a similar pattern of orientations. For such slices of the AF square lattice, the square sides are always aligned with the long and short axes of the slice, although the squares themselves may be distorted into rectangles. On the other hand, for the NA square and triangular lattices, the squares or triangles do not look distorted, but the alignment of their sides is almost random, which makes for polycrystalline structures in the numeric simulations.

\begin{figure}[bt]
\centerline{\psfig{file=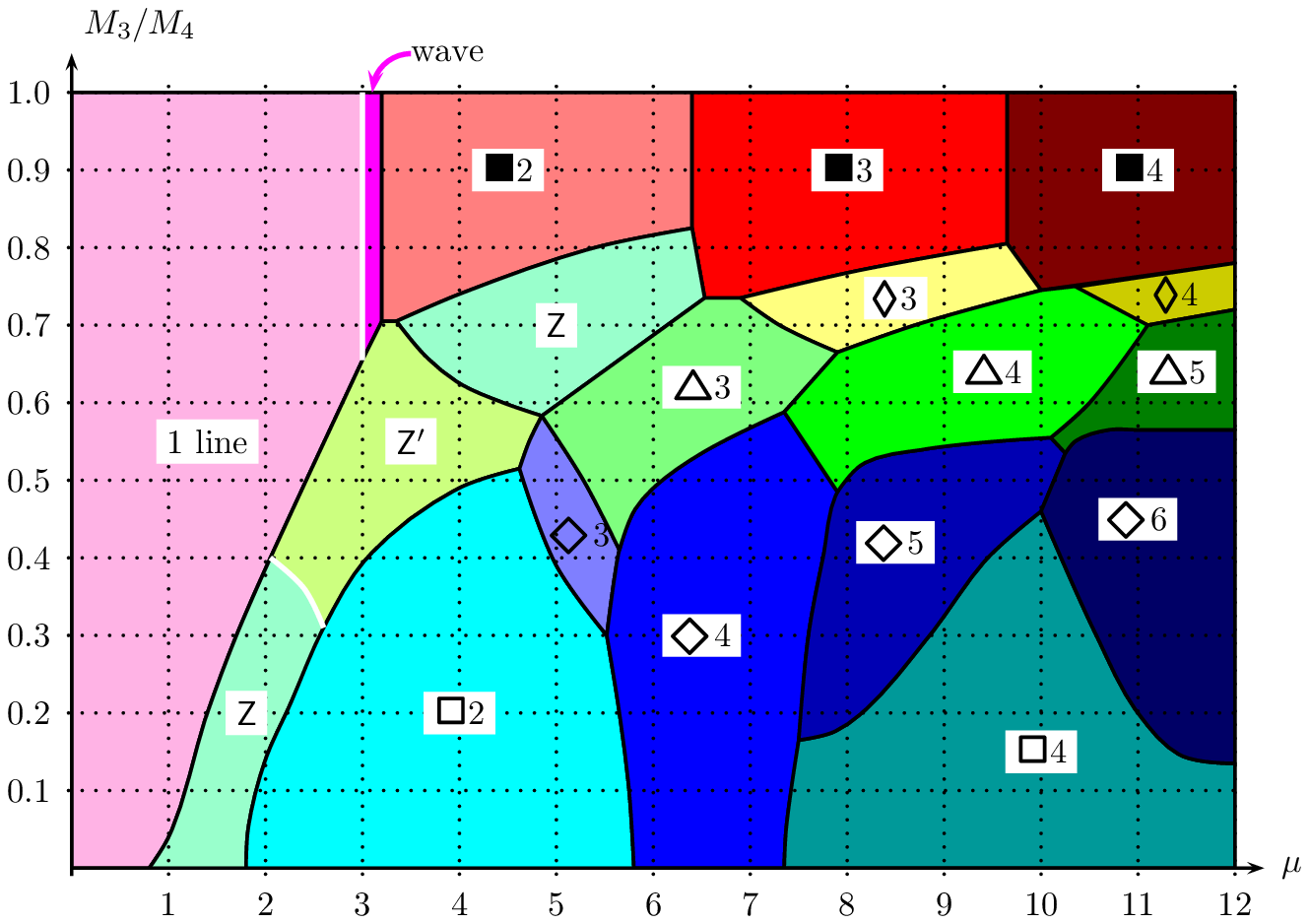,width=0.9\linewidth}}
\vspace*{8pt}
\caption{%
	Phase diagram of $\rm 1D\to2D$ instanton crystals, chemical potential $\mu$ (in units of $N_c M_2$) versus $M_3/M_4$ ratio. The phases are labeled as follows:
	}
\vspace*{4pt}
\centerline{\psfig{file=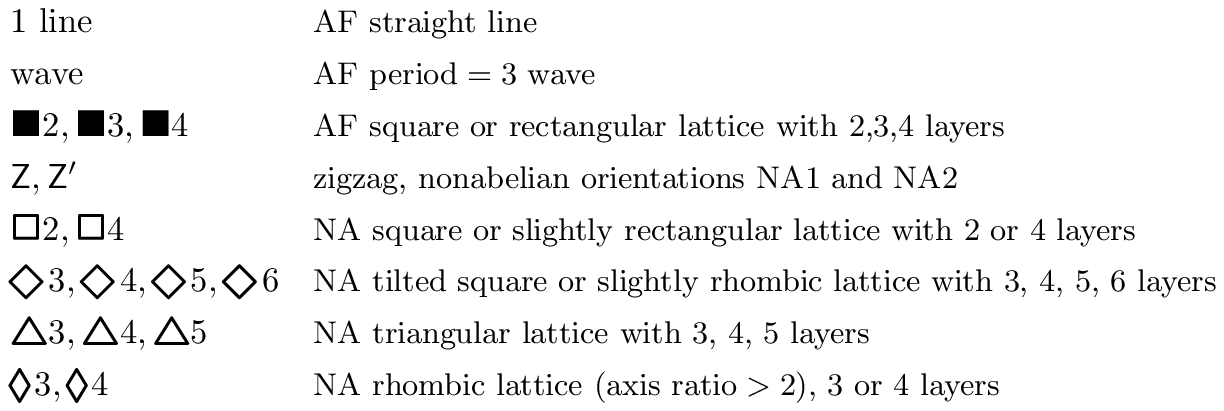,width=0.9\linewidth}}
\label{fig:ThinDiagram:mu}
\end{figure}

\begin{figure}[bt]
\centerline{\psfig{file=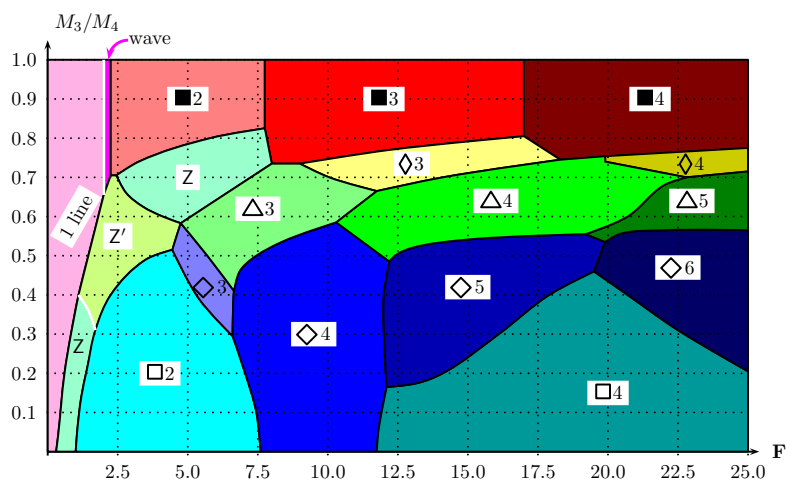,width=0.9\linewidth}}
\vspace*{8pt}
\caption{%
	Phase diagram of $\rm 1D\to2D$ instanton crystals,
	Pressure $\bf F$ along the $x_1$ axis (in units of $N_c\sqrt{\lambda MM_2^3}$)
	versus $M_3/M_4$ ratio.
	The phases are labeled similar to figure~\ref{fig:ThinDiagram:mu}.
	}
\label{fig:ThinDiagram:P}
\end{figure}

\begin{figure}[bt]
\centerline{\psfig{file=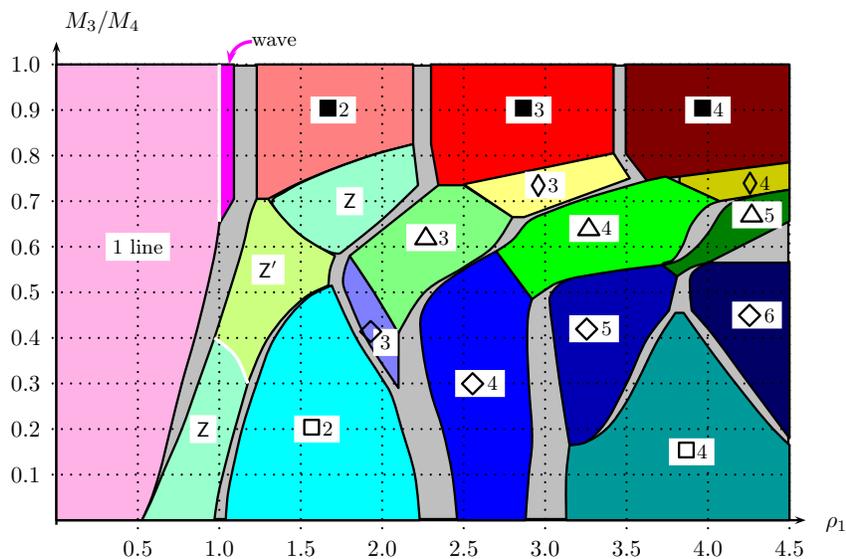,width=0.9\linewidth}}
\vspace*{8pt}
\caption{%
	Phase diagram of $\rm 1D\to2D$ instanton crystals,
	Linear instanton density along the $x_1$ axis
	(in units of $\sqrt{\lambda MM_2}$)
	versus $M_3/M_4$ ratio.
	The phases are labeled similar to figure~\ref{fig:ThinDiagram:mu}.
	The gray areas correspond to non--uniform (gas+liquid-like) mixtures of phases.
	}
\label{fig:newdiagram}
\end{figure}

To get a more precise phase diagram --- and to make sure we are not overlooking some phases due to poor convergence --- we have followed up
the numeric simulations with analytical calculations of net energies (per instanton) for all the phases we have seen or guessed. We have allowed some tunable parameters, such as square or rhombic deformations for square lattices, or deformation and global $SU(2)$ rotation of the orientation pattern
for the triangular lattice. Optimizing such parameters for each combination of pressure and $M_3/M_4$ ratio and always choosing the phase with the lowest \emph{enthalpy} $H=E+L{\bf F}$, and hence chemical potential $\hat\mu=H/N$ (at zero temperature), we have obtained the phase diagrams shown on figures
\ref{fig:ThinDiagram:mu}--\ref{fig:newdiagram} (pages \pageref{fig:ThinDiagram:mu}--\pageref{fig:newdiagram}).

To summarize, the preliminary analysis of the two--dimensional and quasi--one--dimensional lattices reported here demonstrates that the naive intuition about the preferred crystal structures at give densities does not always apply. Eventually one has to resort to numerical simulations to see which structures are actually taken by the instantons. Numerical simulations may also have convergence issues, especially close to the first--order phase transitions. Without appropriate care some crystalline phases may be overlooked.

%
%
%




\section{Summary}
\label{sec:summary}

In this review we have summarized the existing results on the description of the cold dense phase of holographic baryons as instanton lattices in the generalized Sakai--Sugimoto model. The main feature of a finite density system of instantons in such a setup is the competition between the two kind of forces: the Coulomb repulsion of instantons and the restoring (gravity--like) force transverse to the natural instantons' alignment. The latter force is the result of the existence of the transverse \emph{holographic} dimension that appears in the holographic description. However, in the effective description that we used here this force was merely an external force akin to the one in the electromagnetic traps used in the cold atom experiments to create one-- or two--dimensional arrays of atoms.

The outcome of the competition of the repulsive and restoring forces is the existence of different lattice structures depending on the density of instantons. The repulsion is minimized, when the instantons are far apart and the density is low. Meanwhile the instantons will occupy positions at the bottom of the potential well. As was demonstrated in a simple example in section~\ref{sec:pointcharge}, when the density is increased the repulsion energy may become so high that they will prefer to climb up the well to minimize the repulsion. The dimensionality of the original lattice changes in such a case. Overall, increasing the density will eventually make the lattice thicker and thicker in the transverse direction of the restoring force. The change of lattice structures will typically occur in a first order phase transition, as most of the lattice structures will not transform smoothly into each other.

The physical meaning of this kind of transitions should follow from the meaning of the transverse direction. Indeed, the holographic dimension has the interpretation of an energy scale. Growing size of the lattice in this dimension as the density is increased is resembling the emergence of the Fermi surface for finite density fermions.\cite{Rozali:2007rx} Note however, that there is no fermions in this picture. The baryons here are classical objects and the exclusion principle is realized by the hard Coulomb repulsion.

In this work we considered one-- and two--dimensional baryon lattices. Results for the two--dimensional lattices were presented in Sec.~\ref{sec:newresults}. They make part of a work in progress.\cite{Kaplunovsky:toappear}. The dependence of the lattice configuration on the anisotropy ratio parameter $M_3\to M_4$ was presented in Fig.~\ref{fig:2Dphases}. Three--dimensional lattices were only briefly discussed in the point--charge approximation in Sec.~\ref{sec:pointcharge}. A more detailed investigation of the three--dimensional lattices will appear elsewhere.

The details of the transitions between different phases depend very much on the details of the interaction forces. Those are quite non-trivial in the case of instantons. The leading order (in $1/\lambda$) solutions are the ADHM instantons parameterized by position, size and orientation in the $SU(N)$ space. The force between them is the NLO effect and it depends non-trivially on the orientations. Moreover, many--body interactions contribute to the net interaction force. As a result, calculating and minimizing the energy of a multi--instanton configuration is a complicated problem.

In section~\ref{sec:exactsolns} we reviewed the known exact solutions for the instanton lattices. To be precise we were only interested in the expressions for the instanton density, or even a related quantity $\det L$ defined by Eqs.~(\ref{Ldef}) and~(\ref{Idensity}). The solutions include the straight instanton chain~(\ref{BigFla}) and the abelian zigzag~(\ref{DetLmod}). As was demonstrated by a later study, only the anti-ferromagnetic chain still have chances to remain the ground state, in the isotropic setup $M_2\ll M_3=M_4$ (Sec.~\ref{sec:phasediagram}). As the preliminary results of Ref.~\refcite{Kaplunovsky:toappear} show, the abelian zigzag yields its place in the phase diagram to other phases (figure~\ref{fig:newdiagram} in section~\ref{sec:newresults}).

The solutions corresponding to the simplest configurations, such as the straight chain and the abelian zigzag are already sufficiently complicated. The formulae for the instanton density are so bulky, that there is not much sense presenting them in the paper. More elaborate configurations require even more resources. Therefore it seems no other efficient way than working in some approximation. In section~\ref{sec:twobody} we have reminded the reader that in the small--instanton approximation $a\ll D$, the forces between instantons can be reduced to just the two--body ones. This approximation allows to implement the numerical simulations in the search for lowest energy configurations. In Sec.~\ref{sec:phasediagram} we have demonstrated how the phase diagram looks like if we assume that the zigzag is the first non-trivial configuration at higher density. However, the preliminary results of the studies in Ref.~\refcite{Kaplunovsky:toappear} reported in Sec.~\ref{sec:newresults} show that for certain values of the anisotropy ratio, the zigzag phase is skipped and the chain transforms directly into another configuration (Figs~\ref{fig:ThinDiagram:mu}--\ref{fig:newdiagram}). This example shows that our naive intuition about the phase structure does not always work.

Rather than fixing a given pattern of the instanton positions, like a zigzag, or any other fixed geometry, a more accurate method would be to leave the positions as moduli and then minimize the net energy to find a preferred configuration. It is hopeless to achieve this analytically, but a numerical simulation, which allows instantons to select their equilibrium positions, would do. Such simulations have already been performed for (quasi--) one-- and two--dimensional lattices. The results were reported in this review. The transition from the infinite three-dimensional to four--dimensional lattices is a subject of a future work. We hope that the results of that work will allow to compare the phase diagram with that of the skyrmion lattices.

\section*{Acknowledgements}

DM would like to thank the hospitality of the Particle Theory group of Tel Aviv University, where this work was completed. The work was partially supported by the US National Science Foundation (grant \#PHY--1417366 (VK), by the Brazilian Ministry of Science, Technology and Innovation and, specifically in the scope of the Science without Borders program --- CNPq and of the MIT--IIP exchange program, by the foundation FUNPEC of the Federal University of Rio Grande do Norte, by the Russian RFBR grant \#14-02-00627 and by the grant for support of Scientific Schools NSh 1500.2014. (DM), by the Israel Science Foundation (grant 1989/14), the US-Israel bi-national fund (BSF) grant \#2012383 and the Germany--Israel bi-national fund GIF grant \#I-244-303.7-2013 (JS).


\vspace*{3pt}

\begin{thebibliography}{0}

\bibitem{Skyrme}
  T.~H.~R.~Skyrme,
  ``A Nonlinear field theory,''
  Proc.\ Roy.\ Soc.\ Lond.\  A {\bf 260} (1961) 127.

\bibitem{Witten:1979kh}
  E.~Witten,
  ``Baryons in the 1/n Expansion,''
  Nucl.\ Phys.\ B {\bf 160} (1979) 57.

\bibitem{Balachandran:1982dw}
  A.~P.~Balachandran, V.~P.~Nair, S.~G.~Rajeev and A.~Stern,
  ``Exotic Levels from Topology in the QCD Effective Lagrangian,''
  Phys.\ Rev.\ Lett.\  {\bf 49} (1982) 1124
   [Erratum-ibid.\  {\bf 50} (1983) 1630].

\bibitem{Adkins:1983ya}
  G.~S.~Adkins, C.~R.~Nappi and E.~Witten,
  ``Static Properties of Nucleons in the Skyrme Model,''
  Nucl.\ Phys.\ B {\bf 228} (1983) 552.

\bibitem{Balachandran:1985pa}
  A.~P.~Balachandran,
  ``Skyrmions,'' in {\it HIGH ENERGY PHYSICS 1985: proceedings}. ed.~M.~J.~Boswick and F.~Gursey, World Scientific, Singapore, 1985.

\bibitem{Meissner:1985qs}
  U.~G.~Meissner and I.~Zahed,
  ``Skyrmions in Nuclear Physics,''
  Adv.\ Nucl.\ Phys.\  {\bf 17} (1986) 143.

\bibitem{Brown:2010} {\it The Multifaceted Skyrmion -- Selected Papers},
ed.~G.~E.~Brown and~M.~Rho, World Scientific, Singapore, 2010.


\bibitem{Klebanov:1985qi}
  I.~R.~Klebanov,
  ``Nuclear Matter in the Skyrme Model,''
  Nucl.\ Phys.\ B {\bf 262} (1985) 133.

\bibitem{KuglerShtrikman}
  M.~Kugler and S.~Shtrikman,
  ``A new skyrmion crystal,''
  Phys.\ Lett.\  B {\bf 208}, 491 (1988).

\bibitem{Kugler:1989uc}
  M.~Kugler and S.~Shtrikman,
  ``Skyrmion Crystals and Their Symmetries,''
  Phys.\ Rev.\ D {\bf 40} (1989) 3421.

\bibitem{Goldhaber:1987pb}
  A.~S.~Goldhaber and N.~S.~Manton,
  ``Maximal Symmetry Of The Skyrme Crystal,''
  Phys.\ Lett.\  B {\bf 198}, 231 (1987).

\bibitem{McLerran:2007qj}
  L.~McLerran and R.~D.~Pisarski,
  ``Phases of cold, dense quarks at large N(c),''
  Nucl.\ Phys.\ A {\bf 796} (2007) 83
  [arXiv:0706.2191 [hep-ph]].

\bibitem{Maldacena:1997re}
  J.~M.~Maldacena,
  ``The Large N limit of superconformal field theories and supergravity,''
  Int.\ J.\ Theor.\ Phys.\  {\bf 38} (1999) 1113
   [Adv.\ Theor.\ Math.\ Phys.\  {\bf 2} (1998) 231]
  [hep-th/9711200].


\bibitem{WittenBaryons}
  E.~Witten,
  ``Baryons and branes in anti-de Sitter space,''
  JHEP {\bf 9807} (1998) 006
  [hep-th/9805112].

\bibitem{Brandhuber:1998xy}
  A.~Brandhuber, N.~Itzhaki, J.~Sonnenschein and S.~Yankielowicz,
  ``Baryons from supergravity,''
  JHEP {\bf 9807} (1998) 020
  [hep-th/9806158].

\bibitem{Callan:1999zf}
  C.~G.~Callan, Jr., A.~Guijosa, K.~G.~Savvidy and O.~Tafjord,
  ``Baryons and flux tubes in confining gauge theories from brane actions,''
  Nucl.\ Phys.\ B {\bf 555} (1999) 183
  [hep-th/9902197].

\bibitem{Sakai:2004cn}
  T.~Sakai and S.~Sugimoto,
  ``Low energy hadron physics in holographic QCD,''
  Prog.\ Theor.\ Phys.\  {\bf 113} (2005) 843
  [hep-th/0412141].

\bibitem{Witten:1998zw}
  E.~Witten,
  ``Anti-de Sitter space, thermal phase transition, and confinement in gauge theories,''
  Adv.\ Theor.\ Math.\ Phys.\  {\bf 2} (1998) 505
  [hep-th/9803131].

\bibitem{Karch:2002sh}
  A.~Karch and E.~Katz,
  ``Adding flavor to AdS / CFT,''
  JHEP {\bf 0206} (2002) 043
  [hep-th/0205236].

\bibitem{Sakai:2003wu}
  T.~Sakai and J.~Sonnenschein,
  ``Probing flavored mesons of confining gauge theories by supergravity,''
  JHEP {\bf 0309} (2003) 047
  [hep-th/0305049].

\bibitem{Aharony:2006da}
  O.~Aharony, J.~Sonnenschein and S.~Yankielowicz,
  ``A Holographic model of deconfinement and chiral symmetry restoration,''
  Annals Phys.\  {\bf 322} (2007) 1420
  [hep-th/0604161].

\bibitem{Mintakevich:2008mm}
  O.~Mintakevich and J.~Sonnenschein,
  ``On the spectra of scalar mesons from HQCD models,''
  JHEP {\bf 0808} (2008) 082
  [arXiv:0806.0152 [hep-th]].

\bibitem{Sonnenschein:2014jwa}
  J.~Sonnenschein and D.~Weissman,
  ``Rotating strings confronting PDG mesons,''
  JHEP {\bf 1408} (2014) 013
  [arXiv:1402.5603 [hep-ph]].

\bibitem{Seki:2008mu}
  S.~Seki and J.~Sonnenschein,
  ``Comments on Baryons in Holographic QCD,''
  JHEP {\bf 0901} (2009) 053
  [arXiv:0810.1633 [hep-th]].

\bibitem{Hata:2007mb}
  H.~Hata, T.~Sakai, S.~Sugimoto and S.~Yamato,
  ``Baryons from instantons in holographic QCD,''
  Prog.\ Theor.\ Phys.\  {\bf 117} (2007) 1157
  [hep-th/0701280 [HEP-TH]].

\bibitem{Hashimoto:2008zw}
  K.~Hashimoto, T.~Sakai and S.~Sugimoto,
  ``Holographic Baryons: Static Properties and Form Factors from Gauge/String Duality,''
  Prog.\ Theor.\ Phys.\  {\bf 120} (2008) 1093
  [arXiv:0806.3122 [hep-th]].

\bibitem{AtiyahManton} M.~F.~Atiyah and N.~S.~Manton,
  ``Skyrmions from instantons,''
  Phys.\ Lett.\  B {\bf 222} (1989) 438.

\bibitem{Kaplunovsky:2010eh}
  V.~Kaplunovsky and J.~Sonnenschein,
  ``Searching for an Attractive Force in Holographic Nuclear Physics,''
  JHEP {\bf 1105} (2011) 058
  [arXiv:1003.2621 [hep-th]].

\bibitem{Kaplunovsky:2012gb}
  V.~Kaplunovsky, D.~Melnikov and J.~Sonnenschein,
  ``Baryonic Popcorn,''
  JHEP {\bf 1211}, 047 (2012)
  [arXiv:1201.1331 [hep-th]].

\bibitem{Kaplunovsky:2013iza}
  V.~Kaplunovsky and J.~Sonnenschein,
  ``Dimension Changing Phase Transitions in Instanton Crystals,''
  JHEP {\bf 1404}, 022 (2014)
  [arXiv:1304.7540 [hep-th]].

\bibitem{Kim:2006gp}
  K.~Y.~Kim, S.~J.~Sin and I.~Zahed,
  ``Dense hadronic matter in holographic QCD,''
  J.\ Korean Phys.\ Soc.\  {\bf 63} (2013) 1515
  [hep-th/0608046].

\bibitem{Bergman}
  O.~Bergman, G.~Lifschytz and M.~Lippert,
  ``Holographic Nuclear Physics,''
  JHEP {\bf 0711} (2007) 056
  [arXiv:0708.0326 [hep-th]].

\bibitem{Rozali:2007rx}
  M.~Rozali, H.~H.~Shieh, M.~Van Raamsdonk and J.~Wu,
  ``Cold Nuclear Matter In Holographic QCD,''
  JHEP {\bf 0801}, 053 (2008)
  [arXiv:0708.1322 [hep-th]].

\bibitem{Kim:2007vd}
  K.~Y.~Kim, S.~J.~Sin and I.~Zahed,
  ``Dense holographic QCD in the Wigner-Seitz approximation,''
  JHEP {\bf 0809} (2008) 001
  [arXiv:0712.1582 [hep-th]].

\bibitem{Rho:2009ym}
  M.~Rho, S.~J.~Sin and I.~Zahed,
  ``Dense QCD: A Holographic Dyonic Salt,''
  Phys.\ Lett.\ B {\bf 689} (2010) 23
  [arXiv:0910.3774 [hep-th]].

\bibitem{DKS} A.~Dymarsky, S.~Kuperstein and J.~Sonnenschein,
  ``Chiral Symmetry Breaking with non-SUSY D7-branes in ISD backgrounds,''
  JHEP {\bf 0908} (2009) 005
  [arXiv:0904.0988 [hep-th]].

\bibitem{Dymarsky:2010ci}
  A.~Dymarsky, D.~Melnikov and J.~Sonnenschein,
  ``Attractive Holographic Baryons,''
  JHEP {\bf 1106} (2011) 145
  [arXiv:1012.1616 [hep-th]].

\bibitem{Kuperstein:2004yf}
  S.~Kuperstein and J.~Sonnenschein,
  ``Non-critical, near extremal AdS(6) background as a holographic laboratory
  of four dimensional YM theory,''
  JHEP {\bf 0411}, 026 (2004)
  [arXiv:hep-th/0411009].

\bibitem{Witten:1994tz}
  E.~Witten,
  ``Sigma models and the ADHM construction of instantons,''
  J.\ Geom.\ Phys.\  {\bf 15} (1995) 215
  [hep-th/9410052].

\bibitem{Douglas:1995bn}
  M.~R.~Douglas,
  ``Branes within branes,'' in {\it Cargese 1997, Strings, branes and dualities}, pp 267-275
  [hep-th/9512077].

\bibitem{Douglas:1996uz}
  M.~R.~Douglas,
  ``Gauge fields and D-branes,''
  J.\ Geom.\ Phys.\  {\bf 28} (1998) 255
  [hep-th/9604198].

\bibitem{Douglas:1996sw}
  M.~R.~Douglas and G.~W.~Moore,
  ``D-branes, quivers, and ALE instantons,''
  hep-th/9603167.

\bibitem{Lee:1997vp}
  K.~M.~Lee and P.~Yi,
  ``Monopoles and instantons on partially compactified D-branes,''
  Phys.\ Rev.\ D {\bf 56} (1997) 3711
  [hep-th/9702107].

\bibitem{Blake:2012dp}
  M.~Blake and A.~Cherman,
  ``Large $N_c$ Equivalence and Baryons,''
  Phys.\ Rev.\ D {\bf 86} (2012) 065006
  [arXiv:1204.5691 [hep-th]].

\bibitem{Casher:1979vw}
  A.~Casher,
  ``Chiral Symmetry Breaking in Quark Confining Theories,''
  Phys.\ Lett.\ B {\bf 83} (1979) 395.

\bibitem{Banks:1979yr}
  T.~Banks and A.~Casher,
 ``Chiral Symmetry Breaking in Confining Theories,''
  Nucl.\ Phys.\ B {\bf 169} (1980) 103.

\bibitem{Kaplan:1996rk}
  D.~B.~Kaplan and A.~V.~Manohar,
  ``The Nucleon-nucleon potential in the 1/N(c) expansion,''
  Phys.\ Rev.\ C {\bf 56} (1997) 76
  [nucl-th/9612021].

\bibitem{Rubakov} D.~V.~Deryagin, D.~Y.~Grigoriev and V.~A.~Rubakov,
  ``Standing wave ground state in high density, zero temperature QCD at large
  N(c),''
  Int.\ J.\ Mod.\ Phys.\  A {\bf 7}, 659 (1992).

\bibitem{SonShuster} E.~Shuster and D.~T.~Son,
  ``On finite density QCD at large N(c),''
  Nucl.\ Phys.\  B {\bf 573}, 434 (2000)
  [arXiv:hep-ph/9905448].


\bibitem{Horigome:2006xu}
  N.~Horigome and Y.~Tanii,
  ``Holographic chiral phase transition with chemical potential,''
  JHEP {\bf 0701} (2007) 072
  [arXiv:hep-th/0608198].

\bibitem{Nakamura:2006}
S.~Nakamura, Y.~Seo, S.~-J.~Sin and K.~P.~Yogendran,
  ``A New Phase at Finite Quark Density from AdS/CFT,''
  J.\ Korean Phys.\ Soc.\  {\bf 52} (2008) 1734
  [hep-th/0611021].

\bibitem{Yamada:2007} D.~Yamada,
  ``Sakai-Sugimoto model at high density,''
  JHEP {\bf 0810} (2008) 020
  [arXiv:0707.0101 [hep-th]].

\bibitem{ADHM}  M.~F.~Atiyah, N.~J.~Hitchin, V.~G.~Drinfeld and Yu.~I.~Manin,
  ``Construction of instantons,''
  Phys.\ Lett.\  A {\bf 65} (1978) 185.

\bibitem{Corrigan:1983sv}
  E.~Corrigan and P.~Goddard,
  ``Construction of Instanton and Monopole Solutions and Reciprocity,''
  Annals Phys.\  {\bf 154} (1984) 253.

\bibitem{Kim:2008iy}
  K.~Y.~Kim and I.~Zahed,
  ``Nucleon-Nucleon Potential from Holography,''
  JHEP {\bf 0903} (2009) 131
  [arXiv:0901.0012 [hep-th]].

\bibitem{Hashimoto:2009ys}
  K.~Hashimoto, T.~Sakai and S.~Sugimoto,
  ``Nuclear Force from String Theory,''
  Prog.\ Theor.\ Phys.\  {\bf 122} (2009) 427
  [arXiv:0901.4449 [hep-th]].

\bibitem{Hashimoto:2009as}
  K.~Hashimoto, N.~Iizuka and T.~Nakatsukasa,
  ``N-Body Nuclear Forces at Short Distances in Holographic QCD,''
  Phys.\ Rev.\ D {\bf 81} (2010) 106003
  [arXiv:0911.1035 [hep-th]].

\bibitem{Hashimoto:2010ue}
  K.~Hashimoto and N.~Iizuka,
  ``Three-Body Nuclear Forces from a Matrix Model,''
  JHEP {\bf 1011} (2010) 058
  [arXiv:1005.4412 [hep-th]].

\bibitem{Harrington:1978ve}
  B.~J.~Harrington and H.~K.~Shepard,
  ``Periodic Euclidean Solutions And The Finite Temperature Yang-Mills Gas,''
  Phys.\ Rev.\  D {\bf 17} (1978) 2122.

\bibitem{Kraan:1998pm}
  T.~C.~Kraan and P.~van Baal,
  ``Periodic instantons with non-trivial holonomy,''
  Nucl.\ Phys.\  B {\bf 533}, 627 (1998)
  [arXiv:hep-th/9805168].

\bibitem{Osborn} H.~Osborn,
  ``Calculation Of Multi - Instanton Determinants,''
  Nucl.\ Phys.\  B {\bf 159} (1979) 497.

\bibitem{Nahm} W.~Nahm,
  ``A Simple Formalism For The Bps Monopole,''
  Phys.\ Lett.\  B {\bf 90}, 413 (1980).

\bibitem{Kaplunovsky:toappear} V.~Kaplunovsky and J.~Sonnenschein, to appear.


\end{thebibliography}
\end{document}